\documentclass[12pt,a4paper]{report}
\usepackage[cp1250]{inputenc}
\usepackage{wrapfig}

\usepackage{amsmath}
\usepackage{graphicx}
\usepackage{amssymb}

\newcommand{\titl}[1]{{\centering\Large\bf #1\par}\bigskip}
\newcommand{\name}[1]{{\centering\rm\normalsize #1\par}\bigskip}

\newcommand{\adr}[1]{{\it \normalsize #1\par}\medskip}

\begin{document}
\begin{titlepage}
\begin{center}

\includegraphics[width=0.8\textwidth]{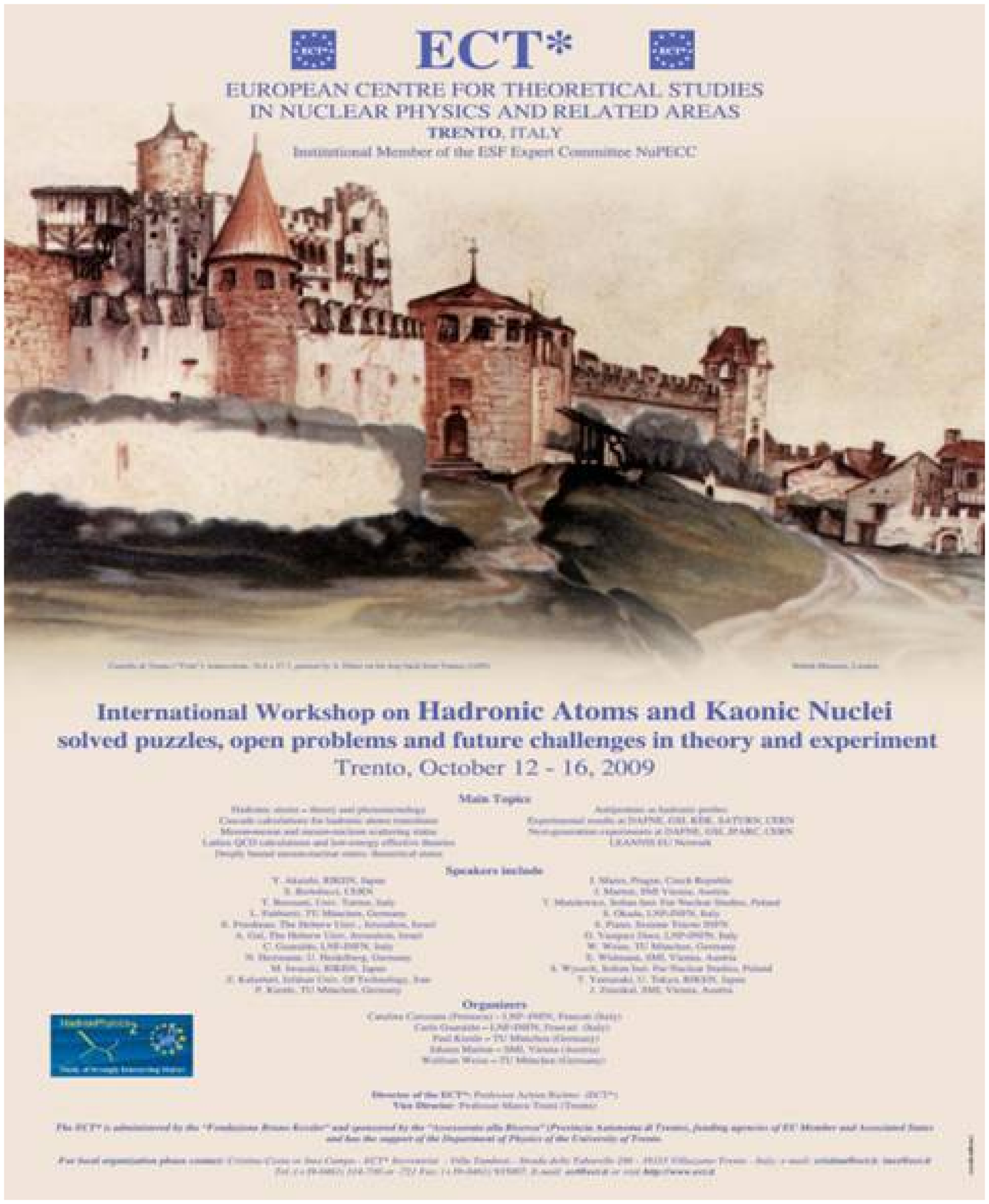}\\[2cm]
\textsc{\Large Mini-Proceedings}\\[0.5cm]
\textsc{\Large ECT* Workshop}\\[1.5cm]
\textsc{\LARGE Hadronic Atoms and Kaonic Nuclei}\\[3cm]
\text{Eds. C. Curceanu (INFN-LNF/Frascati) 
and J. Marton (SMI/Vienna)}\\[1.5cm]

\end{center}
\end{titlepage}

\tableofcontents
 
\clearpage
\addcontentsline{toc}{section}{
{\bf The fascinating world of strangeness - introductory remarks} \\
Catalina Curceanu, Johann Marton}
%
%
 
 
\titl{The fascinating world of strangeness - introductory remarks}
 
\name{ 
Catalina Curceanu$^{1}$, Johann Marton$^{2}$
} 
 
\adr{ 
$^1$ LNF-INFN. Frascati (Roma), Italy\\
$^2$ SMI-Vienna, Austria\\
} 
 
Experts and young researchers in strangeness hadronic and nuclear physics from all
over the world participated in the International Workshop ``Hadronic atoms and nuclei 
- solved puzzles, open problems and future challenges in theory and experiment'' 
held  from 12 to 16 October, 2009 at the European Center for Theoretical Studies in 
Nuclear Physics and related areas, ECT*, at Trento, Italy.

What is an hadronic atom and why there is growing interest in their study?\\
An exotic hadronic atom is formed whenever a hadron (pion, kaon, antiproton) from a beam enters
a target, is stopped inside and replaces an orbiting electron. Such an exotic atom is usually formed
in a highly excited state; a process of de-excitation through the respective atomic levels then
follows. The X-ray transitions to the lowest orbits ($1s$) are affected by the presence of
the strong interaction between the nucleus and the hadron, which is translated into a shift of
the $1s$ level with respect to the purely electromagnetic-based calculated value and by
a limited lifetime (width) of the level. Extracting these quantities, via the measurement 
of the X-ray transitions, provides fundamental information on the low-energy hadron-hadron
and hadron-nuclear interactions, impossible to obtain by other methods. Quantities such as
kaon-nucleon scattering lengths turn out to be directly accessible by measuring the properties
of exotic atoms. Furthermore, these quantities are important milestones to deal - 
in a unique way - with  important aspects of the low-energy QCD in the strangeness sector,
such as the chiral symmetry breaking. Although the field of exotic atoms has a long history,
it lives presently a second youth (renaissance), both from the experimental and from the 
theoretical point of view. On the experimental side, new ``hadronic'' beams became available
recently (kaons at DA$\Phi$NE) or will become available in 2010 (J-PARC facility), while new 
detectors, with improved performance (better energy resolution, stability, efficiency, 
trigger capability,) are starting to operate. On the theoretical side the field has advanced
significantly through recent developments in chiral effective field theories and their 
applications to hadron-nuclear systems.

The kaonic hydrogen was measured by the DEAR Collaboration (on DA$\Phi$NE) with an unprecedented
precision, leading to a lively debate on the kaon-proton scattering length extraction 
procedure, and on the compatibility with existent kaon-nucleon scattering data. The 
SIDDHARTA experiment is performing in 2009 an even more precise measurement and will 
complement it with an exploratory measurement of the kaonic deuterium; the results were
presented and discussed at the Workshop. Kaonic helium was measured at KEK (E570 experiment)
and SIDDHARTA - solving the ``kaonic helium puzzle'', while new experiments are planned at
J-PARC in the near future (E17) to measure X-ray spectrum of kaonic-$^3$He at highest precision.
Other experiments are already planned at existent and/or future machines (GSI, J-PARC, DA$\Phi$NE).
The future of exotic hadronic atoms will reach new horizons - in precision and in dealing 
with new types of exotic atoms, never measured before, such as kaonic deuterium, or sigmonic
atoms.

Another hot item intensively discussed at the Workshop deals with the recently studied
``Kbar -mediated bound nuclear systems''. It was originally suggested that in the few-body
nuclear systems the (strongly attractive) isospin $I=0 \bar{K}N$ interaction can favour the 
formation of  discrete and narrow $bar{K}$-nuclear bound states with large binding energy 
(100 MeV or even more).  Recent theoretical works suggest, however, that such deeply 
bound kaonic nuclear states do not exist: antikaon-nuclear systems might only weakly 
bound and very short-lived. Different interpretations for the existing experimental 
results are given, based for example on the interaction of negative kaons with two 
or more nucleons. This topic is related to a new puzzle in the physics of kaon-nucleon
interaction: the nature of the  $\Lambda$(1405) - single or double pole structure? Long 
discussions  were dedicated to this item in the Workshop. All these topics have important
consequences in astrophysics as well - in the physics of the neutron stars for example. 

The presently available experimental results were reviewed (KEK experiments. FINUDA at
DA$\Phi$NE, FOPI, BNL, OBELIX, Dubna experiments, DISTO at Saturne), together with a critical
discussion of current theories/models. Future perspectives at J-PARC (E15, E17); 
GSI (upgrade of FOPI and HADES) and DA$\Phi$NE (AMADEUS experiment), together with the
possibility to use antiprotons to create single and double strangeness nuclei 
(CERN, J-PARC or FLAIR) were discussed in the framework of  an integrated strategy
in which complementary facilities should bring together the various pieces of the puzzle. 

The field of {\it Hadronic atoms and kaonic nuclei} is a very active one, as was fully
proven during this Workshop. There are indeed {\it solved puzzles}, as kaonic 
hydrogen and kaonic helium ones - which were understood, both due to the newer experiments
(E570 and DEAR and SIDDHARTA at DA$\Phi$NE) and to theoretical interpretation, but many {\it open
problems} are still present. Important questions were targeted and formulated, but 
they still need both experimental results and deeper theoretical understanding. 
{\it So future challenges in experimental and theoretical sectors} are many and were,
for the first time, focussed and formulated in a unitary framework.
 
Last but not least, is to be underlined the strong participation of young researchers,
who were more than 50\% of participants.

We decided to collect in the form of present mini-proceedings short (1-page)
contributions from participants, in order to leave a testimony of the status of
the field by the end of 2009, since we are convinced progress will be quick and rather
dramatic in the coming years.

Full details of this workshop can be found at {\small http://www.smi.oeaw.ac.at/ect\_star/.}

\vfill  
 
 
\setcounter{equation}{0} 
\setcounter{figure}{0}
\clearpage

\addcontentsline{toc}{section}{
{\bf Structure of $K^-pp$ and single-pole nature of $\Lambda (1405)$} \\
 Y. Akaishi, T. Yamazaki}

%





\titl{Structure of $K^-pp$ and single-pole nature of $\Lambda (1405)$}

\name{ Y. Akaishi$^{1,2}$, 
T. Yamazaki$^{1,3}$
}

\adr{
$^1$ RIKEN Nishina Center, Wako, Saitama 351-0198, Japan \\
$^2$ College of Science and Technology, Nihon University, Funabashi, 
         Chiba 274-8501, Japan \\
$^3$ Department of Physics, University of Tokyo, Tokyo 113-0033, Japan 
} 


~~~We have predicted few-body deeply bound kaonic states [1,2], starting from the Ansatz that $\Lambda (1405)$ is a $K^-p$ quasi-bound state, $\Lambda^*$, of single pole. The structure of $K^- pp$ reveals a molecular feature, namely, the $K^-$ in $\Lambda^*$ as an ``atomic center" plays a key role in producing strong covalent bonding with the other proton [3]. Thus, it is proposed that strongly bound $\bar{K}$ nuclear systems are formed by ``super-strong" nuclear force due to migrating real boson, $\bar{K}$.

~~~Figure 1 shows that the $\Lambda (1405)$ is of single-pole nature in spite of the claim of a double-pole structure based on chiral SU(3) dynamics [4]. The experimental evidence for the two-pole structure of $\Lambda (1405)$ [5] is logically impossible to stand, since the 2nd pole cannot produce any peak structure in the relevant mass region. It is urgently important to distinguish the $\Lambda^*$ mass, 1405 or 1420 MeV/$c^2$. A bubble-chamber $\Sigma \pi$ invariant-mass data from stopped $K^-$ on $^4$He supports the $\Lambda(1405)$ Ansatz [6]. A stopped $K^-$ on $D$ experiment [7] would provide a decisive datum for the mass discrimination. 

\begin{figure}[h]
\centering
  \includegraphics[height=.29\textheight]{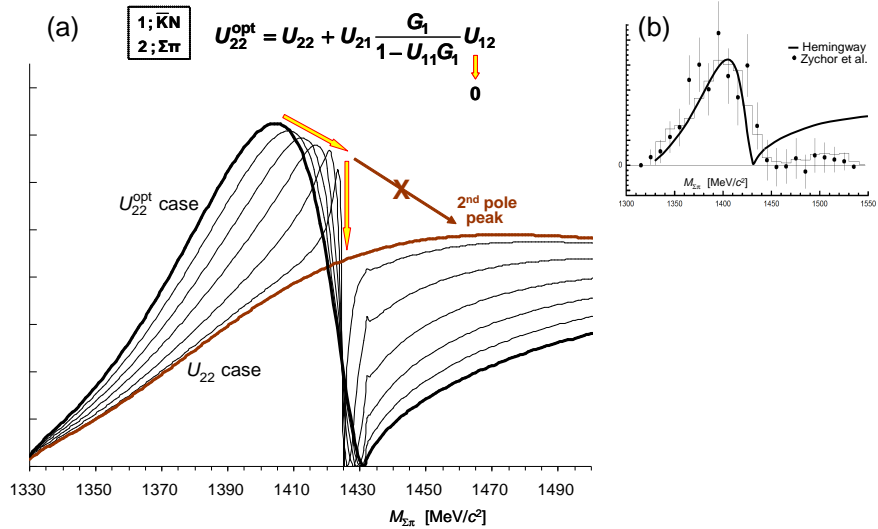}
  \caption{$\Sigma \pi$ invariant-mass spectrum by two-channel treatment of chiral SU(3) dynamics. (a) When the coupling to $\bar KN$ is reduced, the peak goes to disappear at the 1st pole position. (b) The spectrum is compared with experimental data: a large discrepancy is seen above the $\bar K$+$N$ threshold. }
\end{figure}

\vfill  

\noindent{\bf References }
\begin{description}
\setlength\itemsep{-3pt}

\item{[1]} Y. Akaishi and T. Yamazaki, 
                    Phys. Rev. C {\bf 65} (2002) 044005.
\item{[2]} T. Yamazaki and Y. Akaishi,
                    Phys. Lett. B {\bf 535} (2002) 70.
\item{[3]} T. Yamazaki and Y. Akaishi,
                    Proc. Jpn. Acad. B {\bf 83} (2007) 144. 
\item{[4]} T. Hyodo and W. Weise, Phys. Rev. C {\bf 77} (2008) 035204.
\item{[5]} V.K. Magas, E. Oset and A. Ramos, 
                    Phys. Rev. Lett. {\bf 95} (2005) 052301.
\item{[6]} J. Esmaili, Y. Akaishi and T. Yamazaki, 
                    arXiv: 0906.0505v1, 0909.2573v1 [nucl-th].
\item{[7]} T. Suzuki {\it et al.}, this Workshop.
                    
\end{description}


\setcounter{equation}{0} 
\setcounter{figure}{0}
\clearpage

\addcontentsline{toc}{section}{
{\bf The study of $\Lambda \pi^+$, $\Lambda \pi^-$,
$\Lambda\gamma$, $\Lambda p$ and $\Lambda p p$ spectra from p+C
interactions at momentum of 10 GeV/c} \\
P.Zh.Aslanyan, V.N. Yemelyanenko}

%





\titl{The study of $\Lambda \pi^+$, $\Lambda \pi^-$,
$\Lambda\gamma$, $\Lambda p$ and $\Lambda p p$ spectra from p+C
interactions at momentum of 10 GeV/c.}

\name{P.Zh.Aslanyan, V.N. Yemelyanenko 
  }

\adr{ $^1$ Joint Institute for Nuclear Research, LHEP }


The observed well-known resonances $\Sigma^0$, $\Sigma^{*+}$(1385)
and  $K^{*\pm}$(892) [1-3] from PDG are good tests for this method.
 Exotic strange multibaryon states have been observed in
the effective mass spectra of: $\Lambda \pi^+$, $\Lambda
\pi^-$(Fig.1),  $\Lambda \gamma$(Fig.2), $\Lambda p$(Fig.3) and
$\Lambda p p$ subsystems[1-3]. Fig.3 has shown $\Lambda p$ spectrum
in momentum range of 0.14$<Pp <$ 0.30 GeV/c with bin size 12
MeV/$c^2$ for 4669 combinations from primary p+C interactions. There
are significant enhancements in the mass regions of 2090(4.6
$\sigma$), 2144(6.2$\sigma$), 2215(4.3$\sigma$) and small
enhancements in maas regions 2290,There are new small enhancements
for total $\Lambda$p spectrum[3] ( from primary and secondary
protons interactions)in mass range of 2900, 3070 and 3210 MeV/$c^2$,
when applied geometrical weights of $\Lambda$. The enhancement
production for all registered hyperons than calculated geometrical
cross sections are observed [1-3]. The mass of exited
$\Sigma^{*-}(1385)$ is shifted to M(1370) which interpreted as
contribution from low momentum $\pi^-$[1](Fig.1). The mean value of
mass for $\Sigma^{*+}(1385)$ from secondary interactions is also
shifted in mass range of 1370 MeV/$c^2$ too[3]. The width of
$\Sigma^{*-}(1370)$ two time larger in medium of carbon than PDG
data. Such kind of behavior for width of $\Sigma^{*-}(1385)$
resonance is interpreted as the sum of contributions from stopped
$\Xi^-\to\Lambda\pi^-$, $\Sigma^{*-}(1385)$ and reflection from
$\Lambda^{*0}(1405)$ resonances. The width of $\Sigma^0$ is
$\approx$ 2 times larger than the experimental error when was
applied the sum of geometrical weights for $\Lambda$ and
$\gamma$[3](Fig.2).  It can interpret as reflection  of interactions
from $\Lambda$ and $\Sigma^0$ hyperons in medium of carbon nucleus.
The reflection from $\Xi^0$, $\Sigma^{*0}(1385)$, $\Lambda^*(1420)$
hyperons in ($\Lambda \gamma$) spectrum with total geometrical
weights are observed.

\begin{figure}[h]
\centering
{\includegraphics[height=.18\textheight]{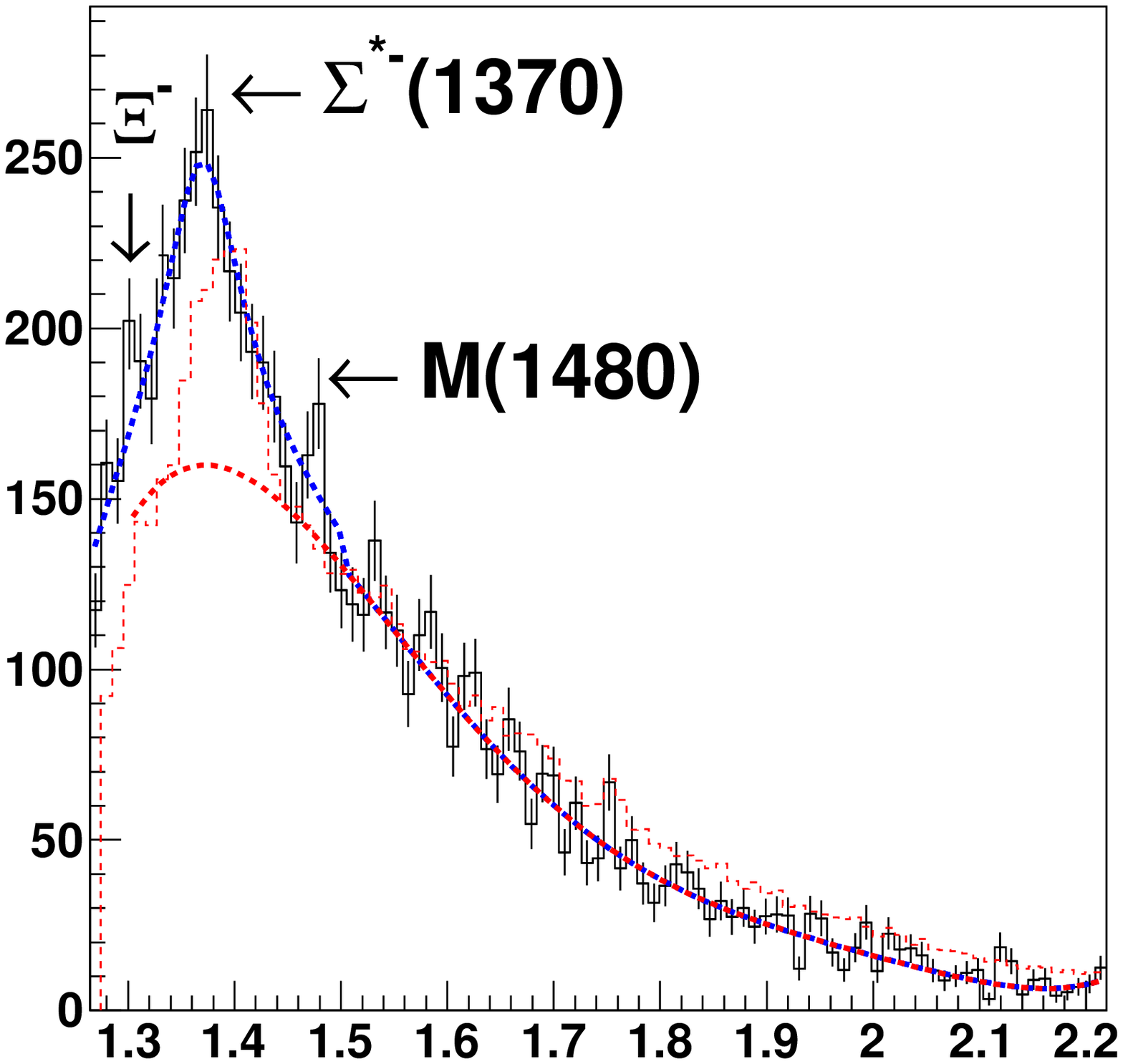}a)}
{\includegraphics[height=.18\textheight]{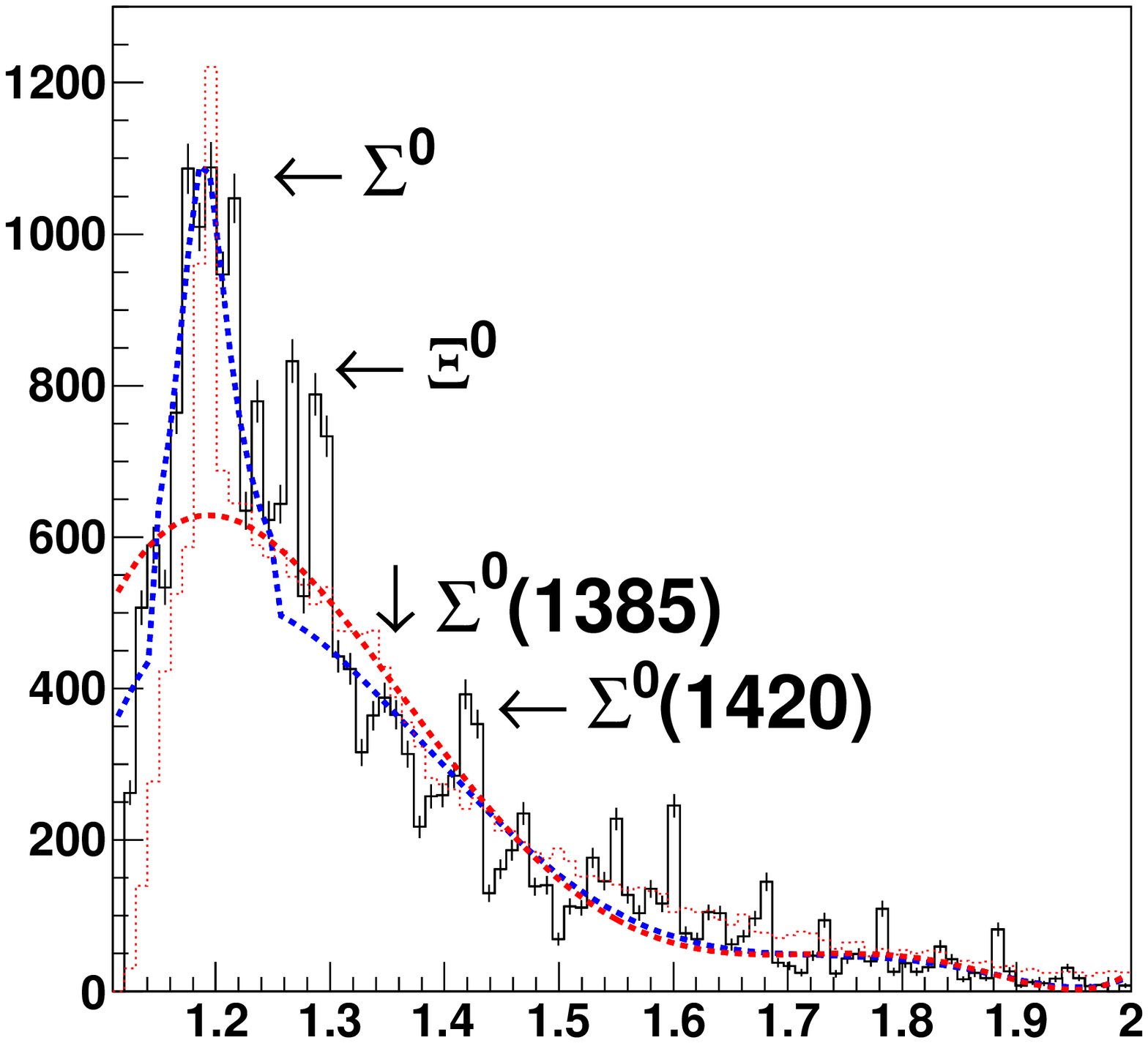}b)}
{\includegraphics[height=.18\textheight]{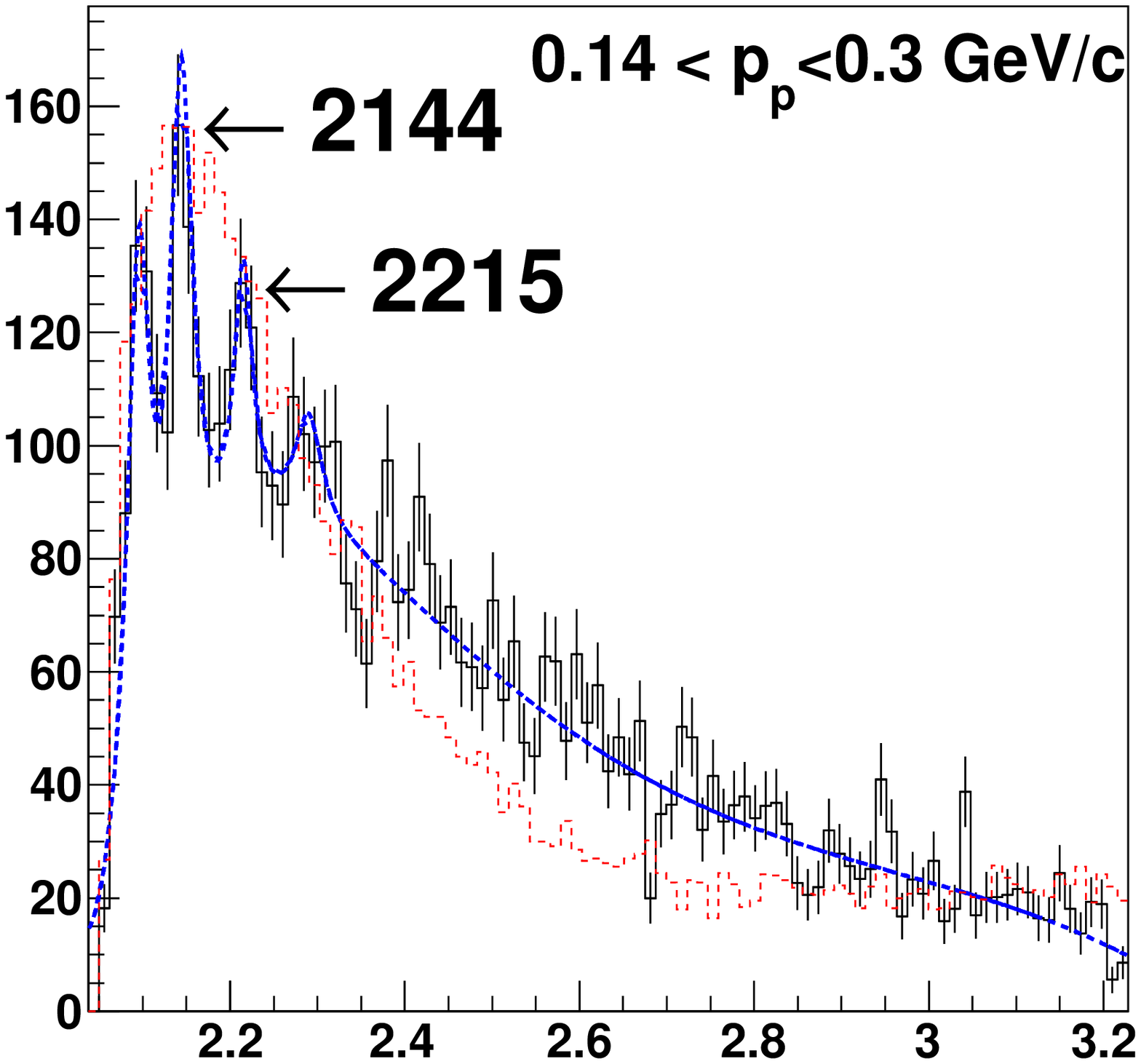}c)}
 \caption{a)The $\Lambda \pi^-$ - spectrum for all combinations
 with a bin size of 11 MeV/$c^2$;  b)the $\Lambda\gamma$ spectrum with total geometrical
weight and bin size 10 MeV/$c^2$; c)the $\Lambda p$ spectrum
 with stopped protons in the momentum range of
0.14$<P_p<$0.30 GeV/$c$ from primary protons with bin size 12
MeV/$c^2$;. The dashed curve is the experimental background fitted
by polynomial function. The dashed histogram is the  simulated
events by FRITIOF. \label{lpip}}
\end{figure}

\vfill  

\noindent{\bf References }
\begin{description}
\setlength\itemsep{-3pt}
\item{[1]}P.Z. Aslanyan, Physics of Particles and Nuclei, Vol. 40, No. 4,
 pp. 525  557, 2009.
\item{[2]}P.Z. Aslanyan, Proc. IUTP'09, Schladming, Austria, 29-6 March,
2009.
\item{[3]}P.Zh.Aslanyan, Proc. Int. Conf., Windows on the
Universe, 16-21 June, Blois, 2009.
\end{description}

 
\setcounter{equation}{0} 
\setcounter{figure}{0}
\clearpage
\addcontentsline{toc}{section}{
{\bf $\bar{K}N$ interactions at and near threshold} \\
A.~Ciepl\'{y}, J.~Smejkal}





\titl{$\bar{K}N$ interactions at and near threshold
\footnote{The work was supported by Grant Agency of the Czech Republic, grant 202/09/1441.  
}}

\name{
A.~Ciepl\'{y}$^{1}$, J.~Smejkal$^{2}$
}

\adr{
$^1$ Nuclear Physics Institute, 250 68 \v{R}e\v{z}, Czech Republic \\
$^2$ Czech Technical University, Horsk\'{a} 3a/22, 128~00~Praha~2, 
Czech Republic
}

We have constructed effective separable meson-baryon potentials to match 
the equivalent chiral amplitudes up to the second order in external meson 
momenta. The potentials were used in the standard coupled channel Lippman-Schwinger 
equations to compute the low energy $K^{-}p$ cross sections and branching 
ratios from the resulting transition amplitudes. At the same time 
the potential was also used to solve the $K^{-}$-proton bound state problem. 
This way the characteristics of the 1s level in kaonic hydrogen are obtained 
by direct calculation, not by relating them to the $K^{-}p$ scattering length 
by means of the Deser-Trueman formula. A direct calculation of the kaonic 
hydrogen characteristics is essential in view of the expected precision 
of the current $K$-atomic SIDDHARTA experiments on hydrogen and deuterium.

The model parameters were fitted to the three precisely measured threshold 
branching ratios $\gamma$, $R_c$ and $R_n$, the low energy $K^{-}p$ cross 
sections, the DEAR data on kaonic hydrogen atom and to the peak position 
of the $\pi \Sigma$ mass spectrum. In the Table 1 we show the results of our 
fits for four different choices of the pion-nucleon sigma term $\sigma_{\pi N}$.  
The strong interaction energy shift $\Delta E_N$ of the 1s level in kaonic 
hydrogen is reproduced well but we were not able to get a satisfactory fit 
of the 1s level decay width $\Gamma$ as our results are significantly 
larger than the experimental value. 

\begin{table}[h]
\caption{The fitted $\bar{K}N$ threshold data}
\begin{center}
\begin{tabular}{ccccccc}
$\sigma_{\pi N}$ [MeV] & $\chi^{2}/N$ & $\Delta E_{N}$ [eV] 
                       & $\Gamma$ [eV]& $\gamma$ & $R_c$ & $R_n$ 
                       \\[.333ex] \hline
 20           & 1.33  & 214     & 718      & 2.368   & 0.653 & 0.189 \\
 30           & 1.29  & 260     & 692      & 2.366   & 0.655 & 0.188 \\
 40           & 1.35  & 195     & 763      & 2.370   & 0.654 & 0.191 \\
 50           & 1.37  & 289     & 664      & 2.366   & 0.658 & 0.192 \\ \hline
experiment    &   -   & 193(43) & 249(150) & 2.36(4) & 0.664(11) & 0.189(15)
\end{tabular}
\end{center}
\label{fits}
\end{table} 

\vskip -1.5ex
The $\bar{K}N$ dynamics at low energies is strongly affected by the 
$\Lambda(1405)$ resonance observed in the $\pi \Sigma$ mass spectrum, 
just below the $\bar{K}N$ threshold. 
The coupled channel meson-baryon models based on chiral symmetry generate 
the resonance dynamically and it appears that there are two poles in 
the complex energy plane that may contribute to its spectrum. 
In our model, the position of one pole is more or less stable and does not depend 
much on the choice of the parameter set. Its complex energy 
\mbox{$z \approx (1400 - {\rm i}\:45)$ MeV} 
can clearly be associated with the observed 
$\pi \Sigma$ mass spectrum. The second pole is located further from the real 
axis and its position vary with the chosen parameter set. It is intriguing 
that neither of the poles is so close to the real axis as other authors claim. 

More details on our work including a discussion of the $K^{-}n$ elastic 
scattering amplitude are given in Ref.~[1].  

\vfill  

\vskip 1.25ex
\noindent{\bf References }
\begin{description}
\setlength\itemsep{-3pt}
\item{[1]}   A.~Ciepl\'{y} and J.~Smejkal, to appear in Eur.~Phys.~J. {\bf A},
  {\tt arXiv:0910.1822 [nucl-th]}
\end{description}

 
\setcounter{equation}{0} 
\setcounter{figure}{0}
\clearpage
\addcontentsline{toc}{section}{
{\bf Faddeev calculation of $K^{-}d$ scattering length}\\
J.~Donoval, N.V.~Shevchenko, A.~Ciepl\'{y}, J.~Mare\v{s}}

%

%



\titl{Faddeev calculation of $K^{-}d$ scattering length}

\name{
J.~Donoval$^1$, N.V.~Shevchenko$^1$, A.~Ciepl\'{y}$^1$, J.~Mare\v{s}$^1$}

\adr{$^1$ Nuclear Physics Institute, 250 68 \v{R}e\v{z}, Czech Republic}


Plausible values of the scattering lengths of $K^{-}d$ and $K^{-}p$
systems are essential for extracting the $K^{-}n$ scattering length,
better understanding of low-energy $\bar{K}N$ interaction and
extrapolating into $\bar{K}$-nuclear systems.

The $K^{-}d$ scattering length was calculated within the Faddeev
equations in the AGS form~[1]. Dealing with scattering at
low-energies, we work in s-wave approximation and we neglect
relativistic corrections. For simplicity, we assume that the isospin
symmetry is not broken. Our calculations are performed in the
momentum and isospin basis.

For the $NN$ interaction, we have considered the PEST potential [2]
and the energy dependent potential by Garcilazo [3], which we have
modified according to more recent experimental data. We have used
two separable $\bar{K}N$ potentials. The first one is our own
phenomenological potential with the Yamaguchi formfactors. The
parameters of the potential were determined from the $K^{-}p$
scattering lengths [4,5] and the properties of the $\Lambda(1405)$
resonance, which is usually assumed a $\bar{K}N$ quasibound state
and a resonance in the $\pi\Sigma$ channel. The second one is the
chirally motivated potential [6]. The effective chiral model was
based on the s-wave meson-baryon lagrangian up to the second order
and reduced to an equivalent single-channel complex potential for
our purposes.

\begin{table}[h]
\caption{The $K^{-}d$ scattering length calculated using various
two-body inputs (in $\mathrm{f\!m}$).} \centering
\begin{tabular}{ccc}
\hline\hline & PEST & E-dep $NN$ \\
\hline & & \\[-12pt]
phenomenological $\bar{K}N$& $-1.51+\mathrm{i}\,0.79$ & $-1.46+\mathrm{i}\,0.78$\\
chiral $\bar{K}N$ (KEK fit)& $-1.78+\mathrm{i}\,1.84$ & $-1.62+\mathrm{i}\,1.57$\\
chiral $\bar{K}N$ (DEAR fit)& $-1.66+\mathrm{i}\,1.88$ & $-1.53+\mathrm{i}\,1.55$\\
\hline
\end{tabular}
\end{table}

Our calculations confirmed that $I=0$ $\bar{K}N$ interaction, which
is much stronger, is more important for the $K^{-}d$ system than
$I=1$ $\bar{K}N$ interaction. Moreover, the chirally based
potentials give much higher imaginary part of $a_{K^{-}d}$ than the
phenomenological one (see~Table 1). By using the Deser-type formula
in next-to-leading order in isospin breaking we have calculated the
kaonic deuterium level shift $\epsilon^d$ and width $\Gamma^d$. The
phenomenological potential gives
$\epsilon^d_{\mathrm{phen.}}\simeq730\,\mathrm{e\!V},\Gamma^d_{\mathrm{phen.}}\simeq470\,\mathrm{e\!V}$,
while the chiral potentials lead to
$\epsilon^d_{\mathrm{chiral}}\simeq1020\,\mathrm{e\!V},\Gamma^d_{\mathrm{chiral}}\simeq890\,\mathrm{e\!V}$.
\\
\\
This work was supported by the GA AVCR grant KJB100480801 and the
GACR grant 202/09/1441.

\vfill  

\noindent{\bf References }
\begin{description}
\setlength\itemsep{-3pt}
\item{[1]} E.O.Alt, P.Grassberger, W.Sandhas, Nucl. Phys. {\bf B2} (1967) 167.
\item{[2]} H.Zankel, W.Plessas, J.Haidenbauer, Phys. Rev. {\bf C28} (1983) 538.
\item{[3]} H.Garcilazo, Lett. Nuovo Cimento {\bf 28} (1980) 73.
\item{[4]} T.M.Ito {\it et al.}, Phys. Rev. {\bf C58} (1998) 2366.
\item{[5]} G.Beer {\it et al.}, Phys. Rev. Lett. {\bf 94} (2005) 212302.
\item{[6]} A.Cieplý, J.Smejkal, arxiv.org/abs/0910.1822v1.
\end{description}
 
\setcounter{equation}{0} 
\setcounter{figure}{0}

\clearpage
\addcontentsline{toc}{section}{
{\bf Investigation of the $\Lambda(1405)$-resonance with HADES}\\
E. Epple  for the HADES collaboration}

%





\titl{Investigation of the $\Lambda(1405)$-resonance with HADES}

\name{
E. Epple$^{1}$ for the HADES collaboration}

\adr{$^1$ Excellence Cluster Universe, TU-Munich, Germany \\}

\begin{figure}[h]
\begin{center}
\begin{minipage}[hbt]{6.5 cm}
\center
\includegraphics[width=6.38 cm]{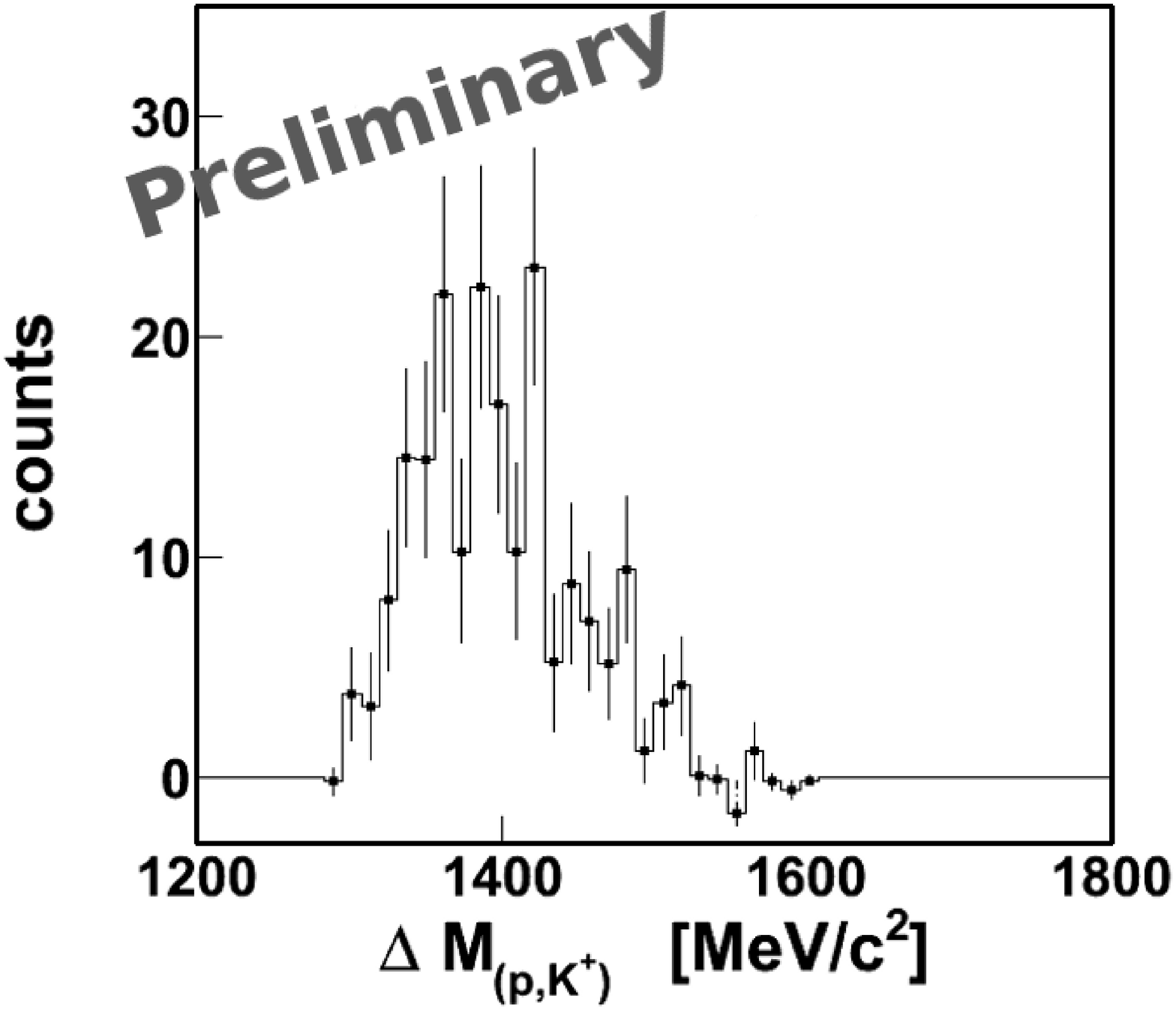}
\end{minipage}
\hspace{1 cm}
\begin{minipage}[hbt]{5.8 cm}
\center
\includegraphics[width=6 cm]{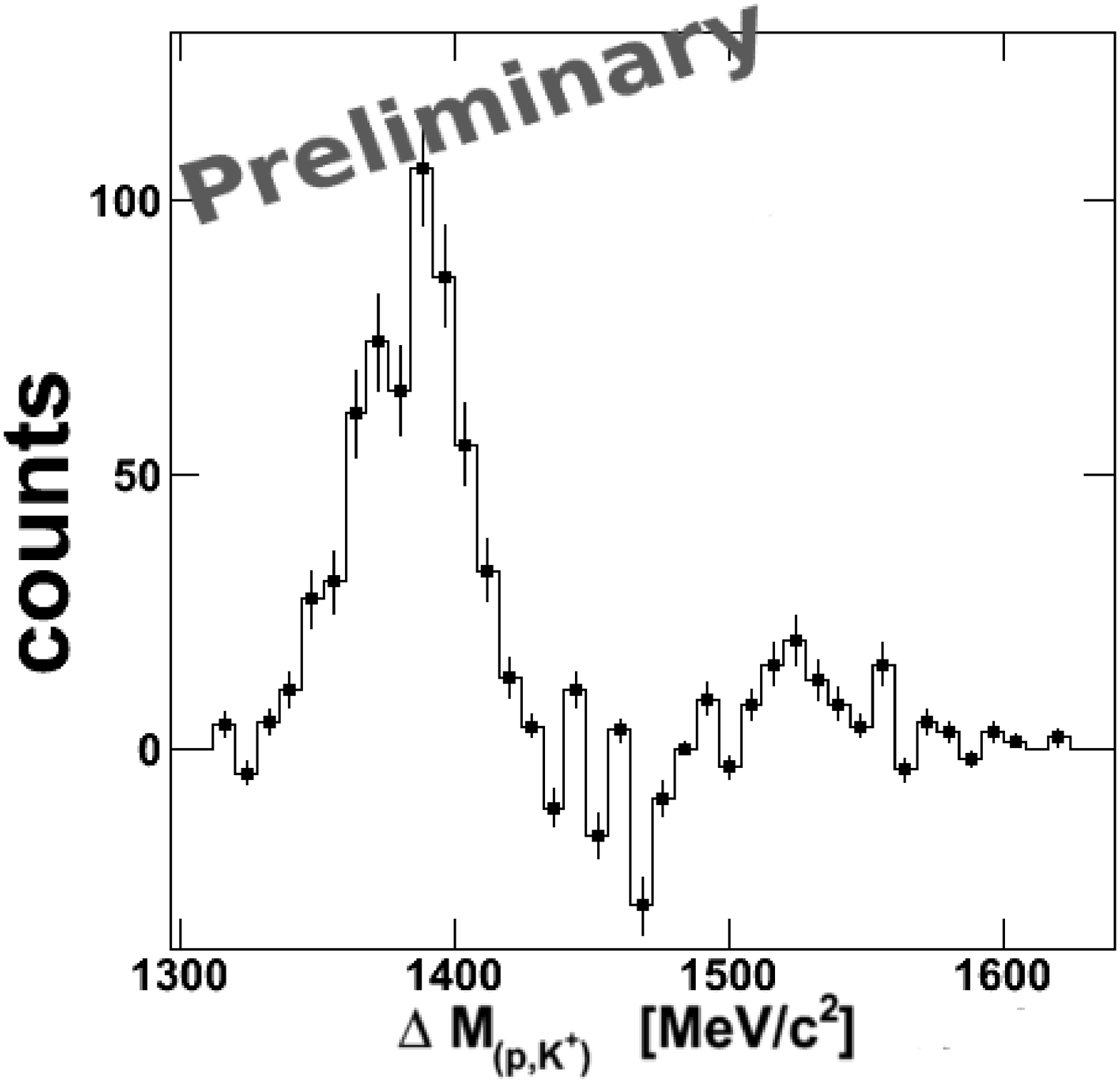}
\end{minipage}
\caption{Left: $\Delta M_{p,K^{+}}$ for the $\Sigma^{0}\pi^{0}$ decay channel. Right: $\Delta M_{p,K^{+}}$ for the two charged decay channels of the $\Lambda(1405)$ ($\Sigma^{+}\pi^{-}$ and $\Sigma^{-}\pi^{+}$) both only with statistical errors.}
\label{miss_mass_ALL}
\end{center}
\end{figure}

Although the $\Lambda(1405)$ resonance has been investigated since several decades there is still a lack of high statistic and quality data to verify the broad spectra of theories that have been developed on that field.
Data, extracted from kaon- and pion- induced reactions have been made available in the past [1,2].
Recently also data from p+p reactions have been published, which show the decay of $\Lambda(1405)\rightarrow \Sigma^{0}\pi^{0}$ [3].
We have also measured p+p at 3.5 GeV with the HADES spectrometer at GSI and attempted to reconstruct the $\Lambda(1405)$ line shape out of its decays in the $(\Sigma\pi)^{0}$ channels.
In p+p reactions the $\Lambda(1405)$ is produced together with a proton and a kaon, so that it can be reconstructed via the missing mass technique. The analysis however has to deal with the fact that a nearby resonance the $\Sigma(1385)^{0}$, that is produced in the same way like the $\Lambda(1405)$, is close in mass and broad, so that a separation in the analysis has to be done by the identification of their decay products. This however is only possible in the following decay, where a difference in the missing mass of p, $K^{+}$, p, $\pi^{-}$ for the both resonances can be found:
\begin{align}
\Lambda(1405)&\rightarrow \Sigma^{0}\pi^{0} \rightarrow (\Lambda\gamma)\pi^{0} \rightarrow (p\pi^{-})\gamma\pi^{0} \\
\Sigma(1385)^{0}&\rightarrow \Lambda\pi^{0} \rightarrow (p\pi^{-})\pi^{0}
\end{align}
The other decays of the $\Lambda(1405)$ are equal to the ones of the $\Sigma(1385)^{0}$, so that the contribution of the 
$\Sigma(1385)^{0}$ in the $\Lambda(1405)$ missing mass spectrum has to be subtracted. 
Fig. \ref{miss_mass_ALL} shows the background subtracted missing mass spectra of the $\Sigma^{0}\pi^{0}$ (left) and the 
$\Sigma^{\pm}\pi^{\mp}$ decay of the $\Lambda(1405)$ (right). The contamination of the $\Sigma(1385)^{0}$, which was estimated via simulations and found to be in the order of 15 \% was not subtracted yet. Furthermore the spectra are not efficiency and acceptance corrected.
\vfill  
\noindent{\bf References }
\begin{description}
\setlength\itemsep{-3pt}
\item{[1]} R. J. Hemingway {\it et al.}, Nucl. Physics {\bf B253} (1985) 742-752.
\item{[2]} D. W. Thomas, A. Engler {\it et al.}, Nucl. Physics {\bf B65} (1973) 15-45.
\item{[3]} I. Zychor {\it et al.}, Phys. Lett. {\bf B660} (2008) 167-171.
\end{description}

 
\setcounter{equation}{0} 
\setcounter{figure}{0}

\clearpage
\addcontentsline{toc}{section}{
{\bf Future Experiments with Pion Beams at GSI}\\
L.~Fabbietti, HADES and FOPI Collaborations}

%





\titl{Future Experiments with Pion Beams at GSI}

\name{
L.~Fabbietti$^{1}$ HADES and FOPI Collaborations.
}

\adr{
$^1$ Excellence Cluster "Universe", Technische Universit\"at M\"unchen \\
}

Pion beam with momenta varying from 1.0 to 1.7 GeV/c can be accelerated at the SIS18 facility at GSI Darmstadt and they can be exploited for fixed target experiments. As a result of the development carried out in the last years, currently secondary pion beams, with a maximal intensities of $10^4$ and $10^6$ particle per second for a beam momentum of 1.3 GeV/c can be delivered to the FOPI and HADES detector respectively. These two spectrometer can be exploited to perform complementary studies on the hadron properties in $\pi^-+A/ \pi^-+p$  reactions. 
The FOPI spectrometer, with its high geometrical acceptance for charged and neutral hadrons, can be exploited to study the production of $\phi$ mesons on nuclei and their decay into K$^+$K$^-$ pairs. Indeed, strangeness production in pion-induced reactions is well suited to study the kaon absorption in normal nuclear matter density.  Comparing the production rates in $\pi^-+p$ and $\pi^-+A$ absorption measurement for the $\phi$ meson for different A values the strength of the interaction between the K$^-$ and the nucleus can be determined [1].  On the other hand, in order to disentangle the intrinsic properties of the $\phi$ meson in the nucleus from the K$^-$ behavior, exclusive analysis can be carried out to tag the K$^-$ not coming from the $\phi$ decay. \\
If we consider the K$^-$ nuclear potential  U$_{KA}$, it consists of one part connected to the K$^-$-p scattering length and to a part dependent on the K$^-$ absorption in the nucleus.  The K$^-$ scattering length is known from kaonic atom experiments [2], while not very much is known on the absorption potential. The planned measurement should provide a valid contribution to determine the imaginary part of this potential.  \\
Similar measurements can be accomplished with the HADES spectrometer, with the difference that the geometrical acceptance is optimized for the mid-rapidity region and does not provide the full phase-space coverage.
On the other hand the performance of the data acquisition system (DAQ) of the HADES Spectrometer is a factor 30 higher than the maximal rate achieved with FOPI.
On the other hand, HADES can deliver a very precise measurement of the dilepton decay of vector mesons ( $\rho,~ \omega$) produced in the nucleus.  Since the $\omega$ and $\rho$ change their properties inside the nuclear matter, it is of interest to study how many of them decay inside the nucleus. According to simulations results, at the incident energy of 1.17GeV about 65\% of $\omega$'s  and 90\% of  $\rho$'s decay inside the nucleus. The modifications of the meson self-energy in the nucleus [3] can hence be studied.\\
Since the secondary pion beams are delivered with a momentum spread of about 8\%, accurate and radiation-hard devices are necessary to track the beam particles event by event.
Currently, we are working on the development of large area mono-crystallin diamonds [4] which are suited devices for our tracking purposes.


\vfill  

\noindent{\bf References }
\begin{description}
\setlength\itemsep{-3pt}
\item{[1]} Ye. S. Golubeva, L.A. Kondratyuk and W. Cassing, Nucl. Phys. A 625 (1997) 832.
\item{[2]} B. Borasoy et al. Phys. Rev. Lett. 94 (2005) 213401; M. Cargnelli et al. Int. J. Mod. Phys. A 20 (2005) 341.
\item{[3]} M. Urban et al., Nucl. Phys. A 641 (1998).
\item{[4]} J. Pietraszko et al., im Press in NIMA, arXiv/nucl-ex:0911.0337.
\end{description}


\setcounter{equation}{0} 
\setcounter{figure}{0}

\clearpage
\addcontentsline{toc}{section}{
{\bf Quantum--classical calculations of cascade transitions in hadronic
hydrogen atoms} \\
M.P.~Faifman, L.I.~Men'shikov}

%

%



\titl{Quantum--classical calculations of cascade transitions\\ in hadronic
hydrogen atoms}

\name{M.P.~Faifman, L.I.~Men'shikov 
 }

\adr{ Russian Research Center ``Kurchatov Institute'', 123182, Moscow, Russia }


The interpretation of experimental data and planning the high precision
measurements~[1] of the X-ray yields in the kaonic hydrogen $pK^{-}$ and
deuterium $dK^{-}$ require an additional theoretical study of cascade processes
in such atoms.

The attempts to create a universal code based on pure quantum mechanical
calculating the cascade characteristics encounter the problem of the lack of a
complete set of the collision de-excitation cross-sections (see [2,3] and
references therein). This reason has determined our choice to develop a
quantum--classical calculating method of a purely classical description of the
exotic atom collision with the hydrogen atom, whereas Auger processes are
considered in a semi-classical way, and radiative transitions are described
within the framework of quantum mechanics. Such ``a compromise'' provides a way
for {\it ab initio} calculation of the physical cascade characteristics with an
accuracy of $\sim25\%$.

The developed method has been applied to hadronic hydrogen atoms~[4] taking
into account the strong interaction between the kaon and the nucleus of
hydrogen isotope, and the yields for the K-series X-rays in $pK^{-}$ and
$dK^{-}$ atoms have been calculated at different hydrogen target densities
(Fig.1).

\begin{figure} [h]
\centering
\includegraphics[width=7.5cm]{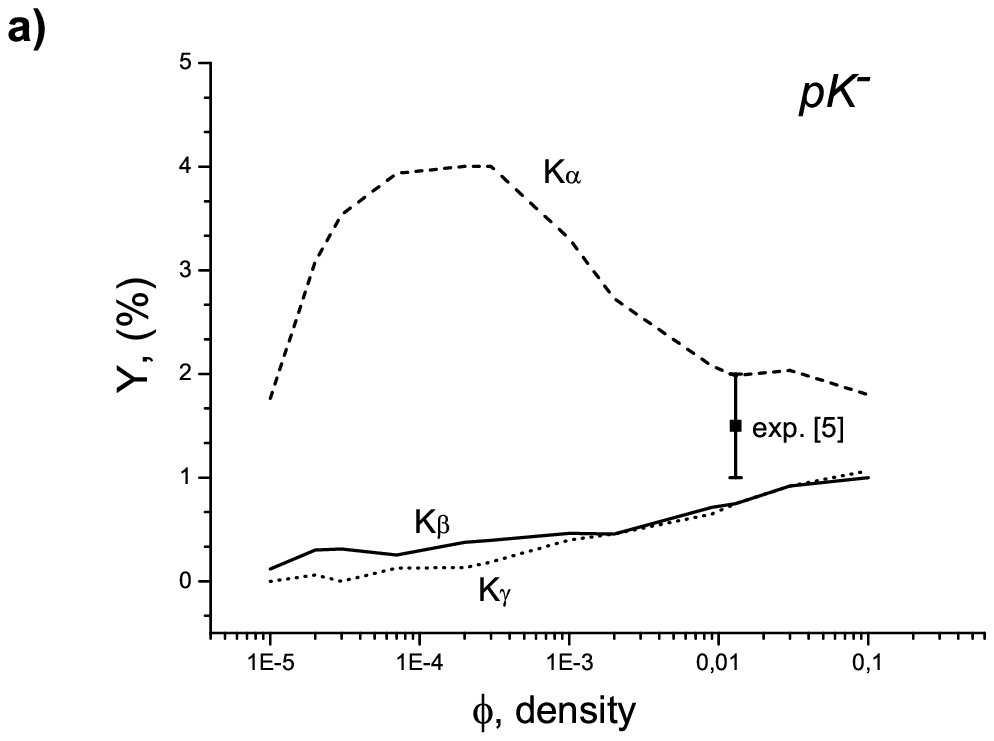}
\includegraphics[width=8.2cm]{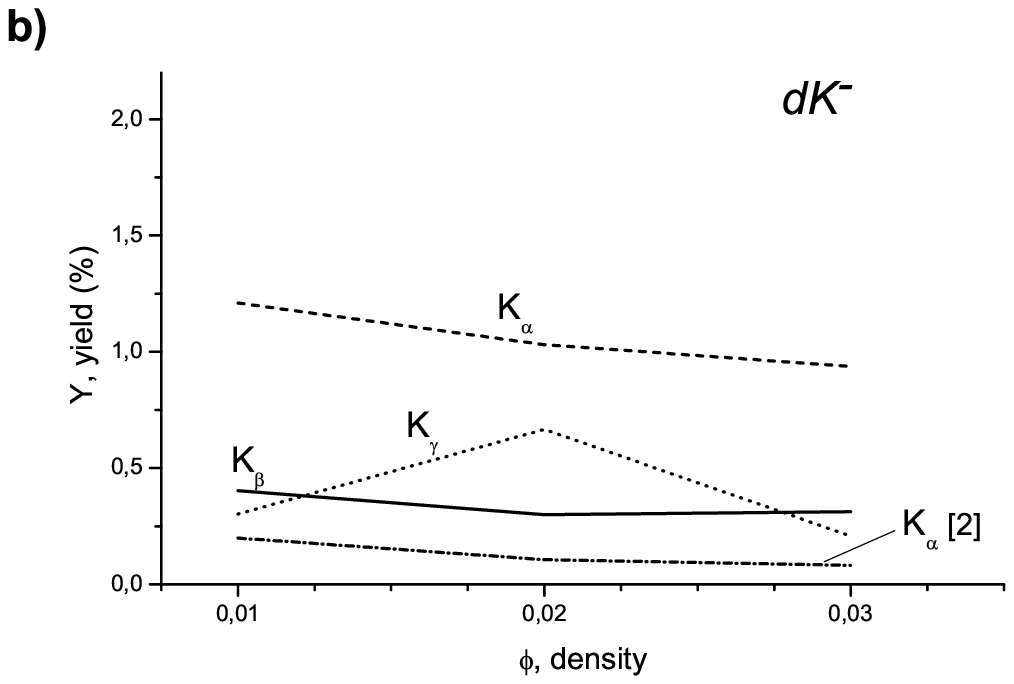}
\vspace{-0.2cm} \caption{The $K_\alpha$, $K_\beta$ and $K_\gamma$~-yields in
kaonic atoms as function of density reduced to the liquid hydrogen density.~{
a)}~The yields for $pK^{-}$ atoms calculated with the nuclear capture width in
the $2p$-state $\Gamma_{2p}=$2~meV~[4];~{ b)}~The $dK^{-}$ atom yields for the
nuclear capture width $\Gamma_{2p}=$4~meV [2].}
\end{figure}

Our calculating scheme allows to obtain as well the other basic cascade
characteristics which are needed for the detailed analysis of the
DEAR/SIDDHARTA experimental data.

\vspace{0.5cm}

\vfill  

\noindent{\bf References }
\begin{description}
\setlength\itemsep{-3pt}
\item{[1]} C. Curceanu et al., Eur. Phys. Journal {\bf A31}, 537 (2007); Hyp. Int. {\bf 193} (2009) 11.
\item{[2]}M.P.~Faifman et al., Frascati Phys. Series {\bf XVI} (1999) 637.
\item{[3]} M. Raeisi G.,  S. Z. Kalantari, Phys. Rev. {\bf A79}  (2009) 012510.
\item{[4]} M.P.~Faifman, L.I.~Men'shikov, Hyp. Int. {\bf 193} (2009) 141.
\item{[5]} T.M.~Ito et al, Phys. Rev. {\bf C58} (1998) 2366.
\end{description}

 
\setcounter{equation}{0} 
\setcounter{figure}{0}
\clearpage
\addcontentsline{toc}{section}{
{\bf Round table discussion: 
critical summary on reported BKS observations} \\
A. Filipp}

%





\titl{Round table discussion: \\ 
critical summary on reported BKS observations}

\name{
A. Filippi$^{1}$
}

\adr{
$^1$ INFN Torino, via P. Giuria, 1, 10125 Torino, Italy
}


It might
be useful to point out at possible
similarities of possible Bound Kaonic Systems (BKS) 
observations reported at this Workshop and
performed in different experimental environments, 
as a start for a thorough understanding of a few 
disagreements among some of them.

\medskip

{\it E549 vs FINUDA}.
Outa [1] reported the results of a missing mass analysis of the
$K^-{^4\mathrm{He}}\rightarrow \Lambda p X$ reaction in the E549 experiment.
The central component observed in the $(\Lambda p)$
invariant mass spectrum, peaked around 2230 MeV/$c^2$, could
be tentatively explained as due to the contribution of the two-body
absorption in $\Sigma N$, or due to the non-mesonic decay of a BKS.
Piano [2] showed the FINUDA $(\Lambda p)$ invariant mass spectrum for
$K^-{^6\mathrm{Li}}$ events. By means of the missing mass
technique the contributions from Q.F. $\Lambda p$
and $\Sigma^0 p$ two-nucleon absorption could be separated: 
the latter could explain just a small
part of the observed bump, centered at 2255 MeV/$c^2$. Moreover,
this signal could not be explained by $\Sigma N$ conversion reactions,
as the expected signature for such channels
did not comply to the observed strong $\Lambda p$
back-to-back angular correlation.
After a proper acceptance correction the positions of the peak observed by
E549 and by FINUDA agree, within the experimental errors.

\medskip

{\it FINUDA vs OBELIX}.
The missing mass spectrum of the $K^-\ ^6\mathrm{Li}\rightarrow \Lambda p X$
shown by Piano [2] exhibits a sharp peak in
correspondence to the opening of the
$K^-{^6\mathrm{Li}}\rightarrow \Lambda\pi^{0,\pm} p 4N$ reaction
threshold. This signal shows up as a narrow 
peak in the $(\Lambda p)$ invariant
mass spectrum at about 2200 MeV/$c^2$. The appearance of sharp peaks
at reaction thresholds is a well known kinematic (``cusp'') 
effect and had been observed several times right at this value
since the early bubble chamber
experiments [3]. 
A signal in the same invariant mass region and approximately the same width
was observed by OBELIX [4] in the
$(pp\pi^-)$ invariant mass spectrum, in the analysis of the
$\bar p ^4\mathrm{He}\rightarrow (p\pi^-)pK^0_SX$ annihilation reaction.
For this structure both the interpretation of BKS or of 
threshold effect can in principle be suggested.

\medskip

{\it FOPI vs early $K^- d$ observations}.
Herrmann reported the observations by FOPI of the $(\Lambda p)$ invariant
mass spectrum obtained in Ni+Ni 
and Al+Al
collisions [5]: they observed, in independent analyses,
a peak at 2135 MeV/$c^2$, a lower value as compared to the FINUDA peak.
This mass value is however in good agreement with the observations by
old bubble chamber experiments of $K^- d\rightarrow \Lambda p \pi^- X$
[6-7]: the 
signal was indicated to be due to 
a cusp effect related to the opening of the $\Sigma^0 n\pi^-$ channel, rather
than to a new resonance. 
Also in this case an alternative interpretation of the recently observed
signal can be suggested.

\vfill  

\noindent{\bf References }
\begin{description}
\setlength\itemsep{-3pt}
\item{[1]} E549 Experiment, H. Outa, {\it talk given at this Workshop}
\item{[2]} FINUDA Experiment, S. Piano, {\it talk given at this Workshop}
\item{[3]} T. Buran {\it et al.}, {\it Nucl. Phys.} {\it Phys. Lett.} {\bf 20}
(1966), 318
\item{[4]} G. Bendiscioli {\it et al.}, {\it Nucl. Phys.}{\bf A789} (2007),
222; {\it Eur. Phys. Jour.} {\bf A40} (2009), 11
\item{[5]} FOPI Experiment, N. Hermann, {\it talk given at this Workshop}
\item{[6]} D. Cline {\it et al.}, {\it Phys. Rev. Lett.} {\bf 20} (1968),
1452
\item{[7]} D. Eastwood {\it et al.}, {\it Phys. Rev.} {\bf D3} (1971), 2603
\end{description}


\setcounter{equation}{0} 
\setcounter{figure}{0}
\clearpage
\addcontentsline{toc}{section}{
{\bf Many facets of the kaonic atoms `puzzle'}\\
E. Friedman}

%







\titl{Many facets of the kaonic atoms `puzzle'}



\name{E. Friedman}

\adr{Racah Institute of Physics, Hebrew University, Jerusalem}


The expression `kaonic atoms puzzle'  refers to an apparent conflict between
phenomenological optical potentials obtained from fits to kaonic atom data
and the corresponding potentials constructed from more fundamental approaches
[1,2].  Whereas the best-fit phenomenological potentials
have for the real part typical depths of 180 MeV at nuclear-matter densities,
the corresponding depth of chiral-motivated potentials [3] is 30-40 MeV. 
The $\chi ^2$ values are 85 for 65 data points 
for the former as opposed to $\chi ^2$ of 260
for the latter. The question of how sensitive kaonic atom
observables are to the nuclear density was discussed recently in terms 
of a functional derivative approach [4]. It was shown that kaonic 
atom data are sensitive to
the surface region of nuclei up to almost the full nuclear density. 
Similar analysis of antiprotonic atoms show [2] that the sensitivity is only
at the extreme surface region where the density is $\sim$ 5\% of the 
central values. Indeed one can estimate a mean-free-path of a particle in
nuclear matter as $\lambda=1/ \rho \sigma $, with $\rho $ the density and
$\sigma $ the total cross section for particle-nucleon interaction.
Then $\lambda \sim $ 1 fm for $K^-$ mesons, $\lambda \sim $ 0.05 fm 
for antiprotons. The question of how sensitive kaonic atoms are to the
interior of nuclei is re-investigated here by analysing kaonic atoms
in parallel with antiprotonic atoms and scattering of low energy $\bar p$.

We have analyzed the LEAR data [5] on the
elastic scattering of 300 and of 600 MeV/c antiprotons by several nuclei.
We find these data to be well described by a potential very similar
to the global antiproton-nucleus potential of ref.[6] with typical $\chi ^2$
per point of 2.2 - 2.5. The potential uses a finite-range
$\bar p$-N interaction with rms radius of 1.4 fm and a fairly constant real
part of an effective scattering amplitude of 0.6 fm. The imaginary part
varies between 1.25 fm for atoms to 1.8 fm at 600 MeV/c. 
To clearly demonstrate the extreme peripheral nature of the $\bar p$-nucleus
interaction, we have repeated all the fits with a potential proportional to
the radial derivative of the nuclear density. Equally good fits are obtained
for the atoms as well as for scattering, where the potentials essentially 
vanish in the interior when using the derivative form. 
The non-penetration of the $\bar p$ is thus clearly demonstrated.

Returning to kaonic atoms, we have performed, in parallel to 
antiprotonic atoms, similar fits based on the derivative of the nuclear
density, but acceptable fits could not be achieved over very broad range
of parameter values. It thus demonstrates again, in addition to the 
functional-derivative analysis[4], that kaonic atom data are
sensitive to the nuclear density and the deep potentials which yield
the best fit between experiment and calculation very likely are 
a reliable representation of the $K^-$-nucleus interaction at close
to the full nuclear density.

\vfill  





\noindent{\bf References }

\begin{description}

\setlength\itemsep{-3pt}

\item{[1]}E.~Friedman, A.~Gal and C.J.~Batty, Nucl. Phys. A {\bf 579} (1994) 518. 

\item{[2]}E.~Friedman  and A.~Gal, Phys. Reports {\bf 452}, (2007) 89. 

\item{[3]}A.~Ramos and E.~Oset, Nucl. Phys. A {\bf 671} (2000) 481.

\item{[4]}N.~Barnea and E.~Friedman, Phys. Rev. C {\bf 75} (2007) 022202(R).

\item{[5]}S.~Janouin, Saclay report CEA-N-2453 (1985).

\item{[6]}E.~Friedman, A.~Gal and J.~Mare\v{s}, Nucl. Phys. A {\bf 761} (2005) 283.

\end{description}


\setcounter{equation}{0} 
\setcounter{figure}{0}
\clearpage
\addcontentsline{toc}{section}{
{\bf Multi-$\bar{K}$ (hyper)nuclei}\\
D.~Gazda, E.~Friedman, A.~Gal, J.~Mare\v{s}}

%

%



\titl{Multi-$\bar{K}$ (hyper)nuclei}

\name{
\underline{D.~Gazda}$^{1}$, E.~Friedman$^{2}$, A.~Gal$^{2}$, J.~Mare\v{s}$^{1}$
}

\adr{
$^1$ Nuclear Physics Institute, 25068 \v{R}e\v{z}, Czech Republic \\
$^2$ Racah Institute of Physics, The Hebrew University, Jerusalem 91904, Israel\\
}


In this contribution we report on relativistic mean-field (RMF) calculations of
baryonic systems containing several antikaons. We focus on the question whether
or not kaon condensation could occur in strong interaction self-bound nuclear
matter, thus whether the $K^-$ mesons could be the relevant degrees of freedom
of self-bound strange matter, as suggested recently in Ref.\ [1]. This scenario
requires that $\bar{K}$ separation energy exceeds 
\begin{wrapfigure}{l}{0.48\textwidth}
\centering
\includegraphics[width=0.48\textwidth]{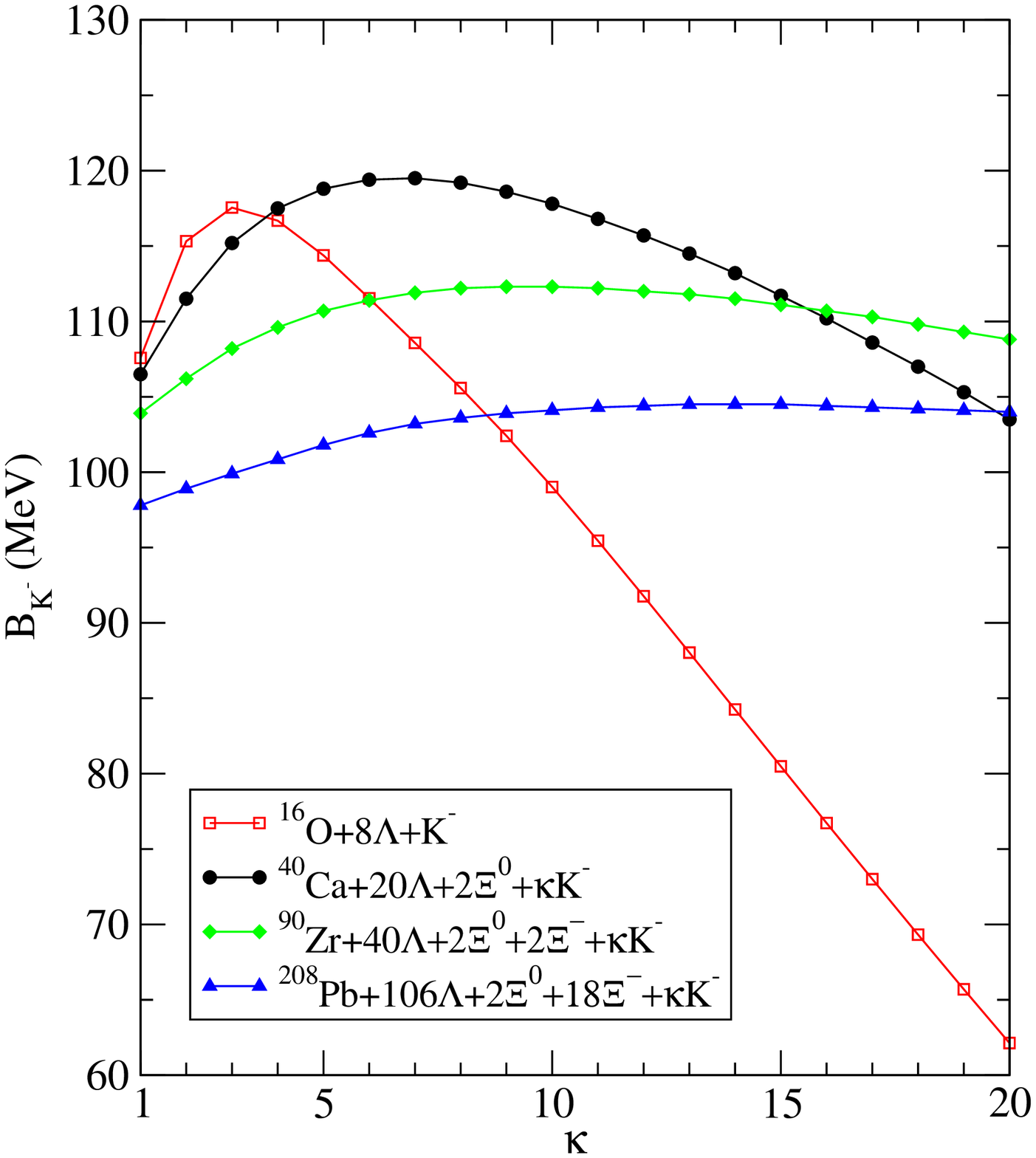}
\caption{The $K^-$ binding energy as a function of the number $\kappa$ of
antikaons in several hypernuclear configurations.}
\end{wrapfigure}
the threshold value of roughly $m_K+\mu_N-m_\Lambda\approx 320$~MeV, thus
allowing for the conversion $\Lambda \rightarrow \bar{K}N$ in matter. We
followed the observation of Refs.\ [2, 3], where the binding energy of $K^-$
mesons, as well as the associated nuclear and $K^-$ density distributions, were
found to \emph{saturate} upon increasing the number of the $K^-$ mesons embedded
in the nuclear medium. The saturation pattern was found to be a robust feature
of these multi-strange configurations. It is present across the entire periodic
table, independent of the RMF model used, and occurs for any mean field
composition containing the dominant $\omega$-meson exchange. In Ref.\ [4], we
generalized our calculations into the hypernuclear domain and studied whether
the presence of hyperons could alter our previous conclusions. In Fig.\ 1 we
present the $1s$ $K^-$ binding energies $B_{K^-}$ as a function of the number
$\kappa$ of antikaons bound in several particle-stable hypernuclear systems.
These configurations were built on top of selected core nuclei by adding first
the $\Lambda$ and then $\Xi$ hyperons until both conversions
$\Lambda\Lambda\leftrightarrow\Xi N$ were kinematically forbidden, thus the
resulting system being particle-stable against strong interactions (for details
see Ref.\ [4]). It is to be stressed that in all cases the $B_{K^-}$ does not
exceed 120~MeV in these multistrange configurations, which is considerably short
of the threshold value of $\approx$320~MeV necessary for the onset of kaon
condensation under laboratory conditions.

\bigskip
This work was supproted in part by GACR grant 202/09/1441 and by SPHERE within
the FP7 research grant system.

\vfill  

\noindent{\bf References }
\begin{description}
\setlength\itemsep{-3pt}
\item{[1]}  T.~Yamazaki, A.~Dot\'{e}, Y.~Akaishi, Phys.\ Lett.\ B 587, (2004) 167.
\item{[2]} D.~Gazda, E.~Friedman, A.~Gal, J.~Mare\v{s}, Phys.\ Rev.\ C 76 (2007) 055204.
\item{[3]} D.~Gazda, E.~Friedman, A.~Gal, J.~Mare\v{s}, Phys.\ Rev.\ C 77 (2008) 045206.
\item{[4]} D.~Gazda, E.~Friedman, A.~Gal, J.~Mare\v{s}, Phys.\ Rev.\ C 80 (2009) 035205.
\end{description}

 
\setcounter{equation}{0} 
\setcounter{figure}{0}
\clearpage
\addcontentsline{toc}{section}{
{\bf HadronPhysics2 and beyond}\\
C.~Guaraldo}

%





\titl{HadronPhysics2 and beyond}

\name{
C.~Guaraldo$^{1}$
}

\adr{
$^1$ INFN, Laboratori Nazionali di Frascati, CP 13,
 Via E. Fermi 40, I-00044, Frascati (Roma), Italy\\
}


$\it{HadronPhysics2 ~ is ~ an ~ Integrating ~ Activity ~ within ~ the ~ Seventh ~ Framework ~ Programme ~ of ~ EU ~ (FP7)}$
\begin{itemize}
\item{} The Project

Coordinator: INFN (Italy)~~~Project Coordinator: C. Guaraldo (LNF-INFN)\\
Other involved institutions: 104 Organizations\\
Total involved researchers: 2500 from 36 countries\\
European Commission budget: 10 million euro ~~~ Duration: 30 months

\item{} Project structure:

8 Networking Activities, 5 Transnational Access Activities, 14 Joint Research Activities, Management of the Consortium
\begin{itemize}
\item{} Networking Activities: 2 theoretical (QCD, TORIC), 3 experimental (PrimeNet, FAIRnet, ReteQuarkonii), 2 in the strangeness sector (SPHERE, LEANNIS), 
1 on the structure of the nucleon (TMDnet)
\item{} Transnational Access Activities:
3 with hadronic probes (FZJ-COSY, GSI, INFN-LNF), 1 with e.m. probes (UMainz-MAMI), 1 for theoretical studies (FBK-ECT*)
\item{} Joint Research Activities :
10 on detectors R\& D (CARAT, FPCC, FutureGas, DIRCs, SCiFi, HardEx, JointGEM, ULISI, JETCAL, SiPM), 2 on polarization (SPINMAP, PolAntiP), 1 on target development (FutureJet), 1 on lattice calculations (LatticeQCD)
\end{itemize}

\item{} Participation to the project
\begin{itemize}
\item{} Composition of the Consortium per country:\\
Germany: 18 organizations; 
United Kingdom: 4 organizations; 
France, Poland, Spain: 3 organizations;
Austria, Czech Republic, Italy, Sweden, The Netherlands: 2 organizations
\item{} Leaderships of activities per organization:\\
INFN: 6; GSI:4; FZJ, UGlasgow, CNRS: 2
\item{} Leaderships of activities per country:\\
Germany: 14; Italy: 7; Austria, France, United Kingdom: 2
\end{itemize}




\item{} Beyond HadronPhysics2

The strategy of the Commission to optimize the budget remaining in FP7 for Integrating Activities (about 350 Meuro) consists in a  $\it{targeted ~ approach}$ with selected topics, 
covering both projects funded in the past and new classes of the projects.
The Commission foresees to fund around 40 projects (up to of a maximum 10 Meuro) and to public around 80 topics in 3 Calls: 40 for potential follow-up projects, 40 for new projects.
HadronPhysics2 is among the follow up projects invited to the Call 8, to be published in July 2010, for an eventual HadronPhysics3.





\end{itemize}

\vfill  


 
\setcounter{equation}{0} 
\setcounter{figure}{0}
\clearpage
\addcontentsline{toc}{section}{
{\bf The Search for a K$^-$pp Bound State with FOPI}\\
O.N.~Hartmann on behalf of the FOPI Collaboration}

%

%



\titl{The Search for a K$^-$pp Bound State with FOPI.}

\name{
O.N.~Hartmann$^{1}$ on behalf of the FOPI Collaboration
}

\adr{\centering
$^1$ Stefan-Meyer-Institut f\"ur Subatomare Physik der \"OAW, Vienna, Austria \\
}

\vfill
The existence of bound states among antikaons and nucleons, with large binding energies and small widths, has been predicted by
T.~Yamazaki and Y.~Akaishi in [1]. In [2], the same authors propose to exploit proton-proton collisions to produce the simplest K$^-$-nucleon bound state K$^-$pp in the reaction

\begin{equation}
\label{eq1}
\text{p}+\text{p}\rightarrow \text{K}^-\text{pp} +  \text{K}^+.
\end{equation}

where a $\Lambda^{*}$ acts as a doorway state which together with another proton can form the bound state.

Furthermore, the maximal predicted cross section for reaction (\ref{eq1}) suggests a proton beam energy around 3 GeV [2]. A K$^-$pp could decay like (\ref{eq2}).

\begin{equation}
\label{eq2}
 \text{K}^-\text{pp}\rightarrow\Lambda + \text{p}.
\end{equation}

Making use of the $\Lambda\rightarrow\pi^-+\text{p}$ decay, an experimental approach for the identification of a K$^-$pp is to compute the invariant mass of $\Lambda+\text{p}$ (\ref{eq2}) and the K$^+$ missing mass (\ref{eq1}) from the charged final state particles.\\
An appropriate experiment had been proposed and has been accomplished in summer 2009 with the FOPI detector at the SIS accelerator of the GSI [3,4].
FOPI provides the necessary acceptance for the mentioned final state particles.\\ The detector setup was extended by a new segmented plastic scintillator as start detector, capable to deal with high beam rates, a liquid hydrogen target system with an upstream veto counter, and a trigger for $\Lambda$ hyperons.
The latter one is designed to allow for an online comparison of the multiplicity of charged particles. The bulk part of produced $\Lambda$ hyperons decay in between two layers of silicon strip detectors, hence the second layer sees more charges than the first one. The thresholds can be set hardwarewise. Employing this trigger, the non-strange background events can be suppressed significantly (approx. by a factor 10).\\
The second layer of the silicon detector provides a spatial information in $(x,y)$, too. This is used to improve the vertex reconstruction capability with tracks under small polar angles.\\
The liquid hydrogen target, a target cell with a kapton entrance and exit window, filled with liquid hydrogen, was placed in the vacuum of the beam pipe. \\
The new start detector delivered a time of flight resolution of $\approx130\text{ ps}$ (for one strip; q.v. [5] for details on FOPI's T.O.F. detectors).\\
The beamtime at the GSI took place in august and september 2009. In total 80 MEvents which fulfill the $\Lambda$ trigger condition were recorded. Currently, the data are being calibrated. From the first preliminary analysis a number of several 10$^4$ reconstructed $\Lambda$ hyperons in the forward hemisphere has been estimated. The invariant mass spectra of (p,$\pi^-$) and the K$^+$ missing mass spectra show the expected behaviour as seen in the simulation.


\vfill  

\noindent{\bf References }
\begin{description}
\setlength\itemsep{-3pt}
\item{[1]} Y. Akaishi and T. Yamazaki, Phys.Rev.C {\bf 65} (2002) 044005
\item{[2]} Y. Akaishi and T. Yamazaki, Phys.Rev.C {\bf 76} (2007) 045201
\item{[3]} http://www.gsi.de and http://www-fopi.gsi.de
\item{[4]} K. Suzuki et al, Nucl.Phys.A {\bf 827} (2009) 312c
\item{[5]} N. Herrmann, Contribution to this Workshop
\end{description}


\setcounter{equation}{0} 
\setcounter{figure}{0}
\clearpage
\addcontentsline{toc}{section}{
{\bf Kaonic $^4$He X-ray measurement with SIDDHARTA}\\
T. Ishiwatari {\it et al.}}

%

%



\titl{Kaonic $^4$He X-ray measurement with SIDDHARTA}

\name{
T. Ishiwatari$^{1}$,
M.~Bazzi$^{2}$,
G.~Beer$^{3}$
L.~Bombelli$^{4}$,
A.M.~Bragadireanu$^{4,5}$
M.~Cargnelli$^{1}$
G.~Corradi$^{2}$,
C.~ Curceanu~(Petrascu)$^{2}$,
A.~d'Uffizi$^{2}$,
C.~Fiorini$^{4}$,
T.~Frizzi$^{4}$,
F.~Ghio$^{6}$,
B.~Girolami$^{5}$,
C.~Guaraldo$^{2}$,
R.S.~Hayano$^{7}$,
M.~Iliescu$^{2,5}$,
M.~Iwasaki$^{8}$,
P.~Kienle$^{1,9}$,
P.~Levi~Sandri$^{2}$,
A.~Longoni$^{4}$,
V.~Lucherini$^{2}$,
J.~Marton$^{1}$,
S.~Okada$^{2}$,
D.~Pietreanu$^{2}$,
T.~Ponta$^{5}$,
A.~Rizzo$^{2}$,
A.~Romero Vidal$^{2}$,
A.~Scordo$^{2}$,
H. Shi$^{7}$,
D.L.~Sirghi$^{2,5}$,
F.~Sirghi$^{2,5}$,
H.~Tatsuno$^{7}$,
A.~Tudorache$^{5}$,
V.~Tudorache$^{5}$,
O.~Vazquez~Doce$^{2}$,
E.~Widmann$^{1}$,
J.~Zmeskal$^{1}$

}

\adr{
$^1$ Stefan-Meyer-Institut f\"{u}r subatomare Physik, Vienna, Austria\\
$^2$ INFN, Laboratori Nazionali di Frascati, Frascati (Roma), Italy \\
$^3$ Dep. of Phys. and Astro., Univ. of Victoria, Victoria B.C., Canada\\
$^4$ Politechno di Milano, Sez. di Elettronica, Milano, Italy\\
$^{5}$ IFIN-HH, Magurele, Bucharest, Romania\\
$^{6}$ INFN Sez. di Roma I and Inst. Superiore di Sanita, Roma, Italy\\
$^{7}$ Univ. of Tokyo, Tokyo, Japan\\
$^{8}$ RIKEN, The Inst. of Phys. and Chem. Research, Saitama, Japan\\
$^{9}$ Tech. Univ. M\"{u}nchen, Physik Dep., Garching, Germany
}

The SIDDHARTA experiment measured the kaonic $^4$He $3d \to 2p$ X-ray transition 
in a gaseous target, where Compton scattering in helium is negligible. 
The kaonic atom X-rays were detected with large-area Silicon Drift Detectors  
using the timing information of the $K^+K^-$ pairs produced by $\phi$ decays at the 
DA$\Phi$NE $e^+e^-$ collider. A new value of the strong interaction shift 
of the kaonic helium-4 $2p$ state was determined to 
be $ 0 \pm 6 \mbox{ (stat)} \pm 2 \mbox{ (syst)} \mbox{ eV}$ [1]. The resultant shift of 0 eV confirms the 
result by the E570 group [2]. Prior to the experiment by the E570 group, 
the average of the three previous results was $-43 \pm 8$ eV [3], 
while most of the theoretical calculations give $\sim 0$ eV [2]. 
This five-sigma discrepancy between theory and experiment was known as the ``kaonic
helium puzzle.'' A resolution of this long-standing puzzle provided by the E570 group is
now firmly established by the present work.

\medskip
\begin{tabular}{cc}
\begin{minipage}{0.4\hsize}
\begin{center}
\includegraphics[scale=0.32]{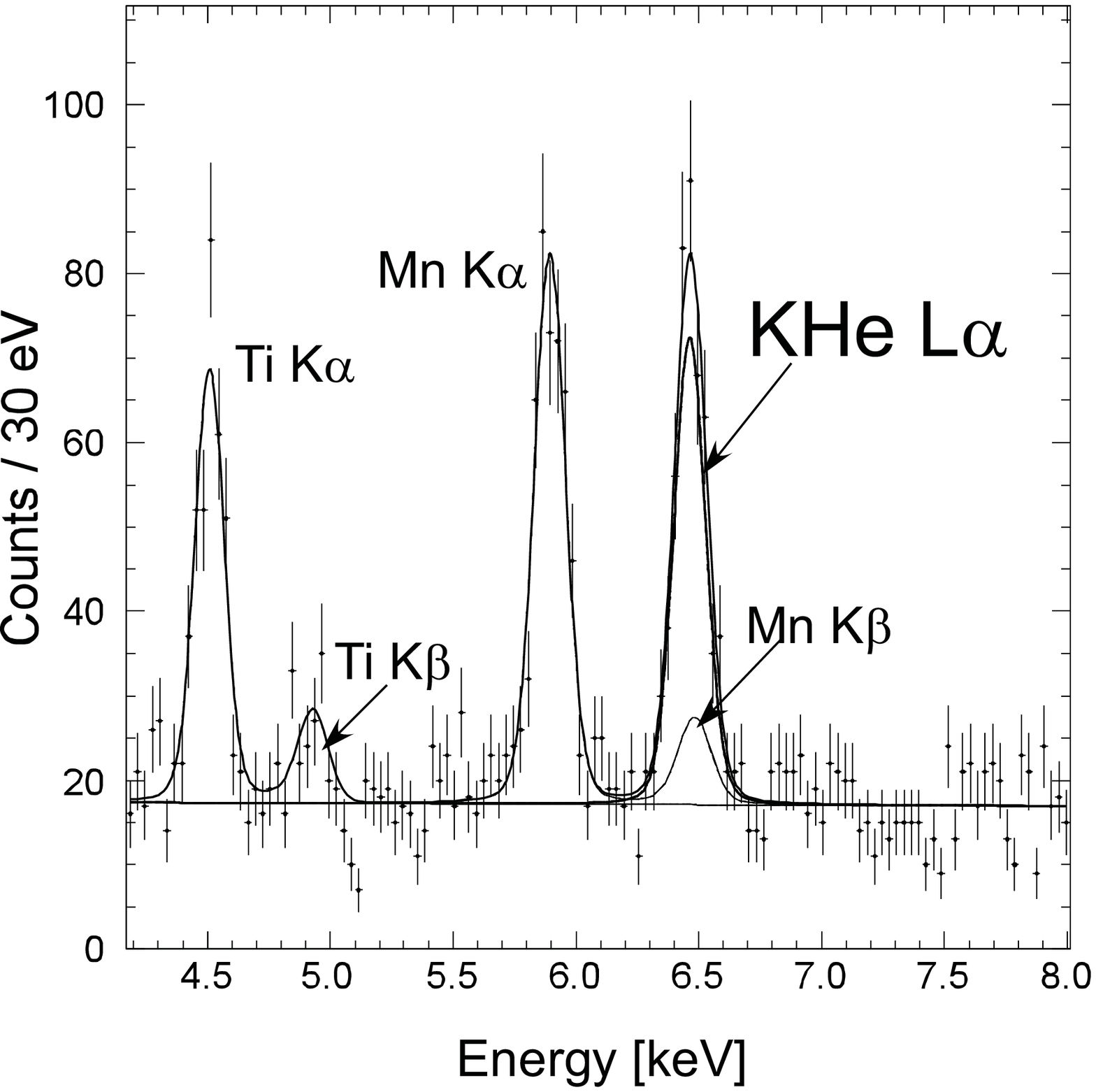}
\end{center}
\end{minipage}
&
\begin{minipage}{0.5\hsize}
\begin{center}
Energy shift of the kaonic helium $2p$ state\\
\begin{tabular}{ccc}
\hline \hline
$\Delta E_{2p}$ (eV) & Ref. \\
\hline
$-41 \pm 33$    &Wiegand {\it et al.} [3] \\
$-35 \pm 12$    &Batty {\it et al.}  [3] \\
$-50 \pm 12$    &Baird {\it et al.} [3] \\
\hline
$-43 \pm 8$     &Average of above [3] \\
\hline
$+2 \pm 2 \mbox{ (stat)} \pm 2$ \mbox { (syst)} &  Okada {\it et al.} [2] \\
\hline
$0 \pm 6 \mbox{ (stat)} \pm 2 \mbox{ (syst)}$&  This work [1]\\
\hline \hline
\end{tabular}
\end{center}
\end{minipage}\\
Energy spectrum of kaonic $^4$He X-rays\\
\end{tabular}



\vfill  

\noindent{\bf References }
\begin{description}
\setlength\itemsep{-3pt}
\item{[1]} M. Bazzi, {\it et al.}, (SIDDHARTA Collaboration), Phys. Lett. B  681 (2009) 310.
\item{[2]} S. Okada, {\it et al.},  Phys. Lett. B  653 (2007) 387.
\item{[3]} C. J. Batty, Nucl. Phys. A 508 (1990) 89c, and refrences therein.
\end{description}


\setcounter{equation}{0} 
\setcounter{figure}{0}
\clearpage
\addcontentsline{toc}{section}{
{\bf Molecule model for kaonic nuclear cluster $\bar{K}NN$}\\
M.~Faber, \underline{\rm A.~N.~Ivanov}, P.~Kienle,
J.~Marton, M.~Pitschmann}

%





\titl{Molecule model for kaonic nuclear cluster $\bar{K}NN$}

\name{M.~Faber$^1$, \underline{\rm A.~N.~Ivanov$^1$}, P.~Kienle$^{2,3}$,
J.~Marton$^2$, M.~Pitschmann$^1$}

\adr{ $^1$ Atominstitut der \"Osterreichischen Universit\"aten,
Technische Universit\"at Wien,\\ Wiedner Hauptstrasse 8-10, A-1040 Wien,
Austria\\ $^2$ Stefan Meyer Institut f\"ur subatomare Physik
\"Osterreichische Akademie der Wissen-\\schaften, Boltzmanngasse 3,
A-1090, Wien, Austria\\ $^3$ Excellence Cluster Universe Technische
Universit\"at M\"unchen, D-85748 Garching,\\ Germany }

The kaonic nuclear cluster (KNC) $\bar{K}NN$ (or ${^2_{\bar{K}}}{\rm
H}$) with quantum numbers $I(J^{\pi}) = \frac{1}{2}(0^-)$ and the
structure $N\otimes (\bar{K}N)_{I = 0}$ has been predicted by Akaishi
and Yamazaki [1] in a non--relativistic potential model approach with
the binding energy $B_{{^2_{\bar{K}}}{\rm H}} = 48\,{\rm MeV}$ and the
width $\Gamma^{(\pi)}_{{^2_{\bar{K}}}{\rm H}} = 61\,{\rm MeV}$, caused
by the pionic ${^2_{\bar{K}}}{\rm H} \to p \Sigma \pi$ decay modes
only.  According to [1], the simplest KNC is the isospin--singlet
state $(\bar{K}N)_{I = 0}$ (or ${^1_{\bar{K}}}{\rm H}$), identified
with the $\Lambda(1405)$ hyperon with quantum numbers $I(J^{\pi}) = 0(
\frac{1}{2}^-)$ and mass $M_{\Lambda^*} = 1405\,{\rm Mev}$.  We
propose to treat the KNCs ${^n_{\bar{K}}}{\rm H}$, where $n > 1$ is
the number of nucleons, as kaonic molecules [2]. In the center of mass
frame molecular states are defined by rotational and vibrational
degrees of freedom.  Since the angular momentum of the KNC
${^2_{\bar{K}}}{\rm H}$ is equal to zero, the properties of this state
are determined by the vibrational degrees of freedom only, which we
describe by trial harmonic oscillator wave functions. Such a choice
can be justified by the short--range character of forces producing
quasi--bound KNCs [1,2].  The later allows also to describe the
properties of the KNC ${^1_{\bar{K}}}{\rm H}$, treated as an
antikaonic $(\bar{K}N)_{I = 0}$ atom in the ground state, by a
harmonic oscillator wave function.  In our model the binding energies
and widths of KNCs ${^n_{\bar{K}}}{\rm H}$ are defined by the real and
imaginary parts of the diagonal matrix elements $\langle
{^n_{\bar{K}}}{\rm H}|\mathbb{T}|{^n_{\bar{K}}}{\rm H}\rangle$ of the
$\mathbb{T}$-- matrix [2]. This agrees with the definition of the
binding energies and the widths of strongly coupled nuclear systems
within the optical potential approach. The matrix elements of the
$\mathbb{T}$-- matrix are calculated in the heavy baryon
approximation, using chiral Lagrangians with derivative meson--baryon
couplings invariant under chiral $SU(3) \times SU(3)$ symmetry, which
are used in Chiral Perturbation theory (ChPT) and $SU(3)$
coupled--channels technique for the analysis of low--energy strong
interactions. This allows to calculate both non--pionic and pionic
decay modes of the KNC ${^2_{\bar{K}}}{\rm H}$, expressed in terms of
two input parameters $\Omega_{\Lambda^*}$, the frequency of the
$\bar{K}N$ oscillations in the KNC ${^1_{\bar{K}}}{\rm H}$, and
$g_{\Lambda^*}$, the coupling constant defining the strength of the
$\Sigma\pi$ decay modes of the KNC ${^1_{\bar{K}}}{\rm H}$. With the
assumption that the mass of the KNC ${^1_{\bar{K}}}{\rm H}$ has a
simple structure $M_{{^1_{\bar{K}}}{\rm H}} = m_K + m_N -
B_{{^1_{\bar{K}}}{\rm H}}$, where $m_K$ and $m_N$ are the kaon and
nucleon masses, and equal to $M_{{^1_{\bar{K}}}{\rm H}} = 1405\,{\rm
MeV}$ the calculated binding energy and width of the KNC
${^2_{\bar{K}}}{\rm H}$ agree well with the results obtained by
Akaishi and Yamazaki [1]. The explanation of the experimental data by
the DISTO Collaboration $B_{{^2_{\bar{K}}}{\rm H}} = 105(2)\,{\rm
MeV}$ and $\Gamma_{{^2_{\bar{K}}}{\rm H}} = 118(8)\,{\rm MeV}$ [3]
goes beyond the description of the KNC ${^1_{\bar{K}}}{\rm H}$
proposed in [2].  The analysis of the experimental data by the DISTO
Collaboration in the molecule model of kaonic nuclear clusters is
carried out in a forthcoming paper.

\vfill  

\noindent{\bf References }
\begin{description}
\setlength\itemsep{-3pt}
\item{[1]} Y. Akaishi and T. Yamazaki, Phys. Rev. C {\bf 65}, 044005
(2002); T. Yamazaki and Y. Akaishi, Phys. Lett. B {\bf 535}, 70
(2002); Phys. Rev. C {\bf 76}, 045201 (2007).
\item{[2]} M. Faber {\it et al.}, arXiv: 0912.2084 [nucl-th].
\item{[4]} M. Maggiora {\it et al.}, (DISTO Collaboration), arXiv: 0912.5116 [nucl--ex].
\end{description}


\setcounter{equation}{0} 
\setcounter{figure}{0}
\clearpage
\addcontentsline{toc}{section}{
{\bf Analysis of cascade dynamics and x-ray yields for $K^{-}p$
and $K^{-}d$ atoms by Monte-Carlo method}\\
S.Z. Kalantari and M. Raeisi G.}

%





\titl{Analysis of cascade dynamics and x-ray yields for $K^{-}p$
and $K^{-}d$ atoms by Monte-Carlo method}

\name{S.Z. Kalantari
and M. Raeisi G.}

\adr{Department of Physics, Isfahan University of
Technology,Isfahan, 84156-83111,Iran}


The cascade calculations presented here have led to a
 more detailed understanding of the atomic cascade in $K^{-}x$
atoms. The density dependence of the x-ray yields and
  number of nuclear absorption, nuclear reaction and Stark mixing in different excited states are calculated.
  We have analyzed our results to clear the answer of some important questions about the kaonic cascade
  dynamics. For example the results of forthcoming SIDDHARTA experiment [1] are predicted. For this purpose we have used Monte-Carlo method.
  In this presentation we review the cascade dynamics of kaonic
  atoms [2,3,4] and our Monte-Carlo simulation [5].  Our calculations show that two processes, nuclear
  absorption and Stark mixing, have an important role in atomic cascade at high density
  targets.\\
  \indent In order to compare our results with experimental values we should calculate the x-ray yields per incoming kaon
  to the target. The value of $\Gamma_{2p}$ ($2p$ strong interaction width of
$K^{-}p$ atoms) is determined by fitting our results with existing
experimental data in KEK [6] and LNF [7]. Our estimated value for
$\Gamma_{2p}$ is 0.105$\pm$ 0.002 meV. We suggest that the
SIDDHARTA experiment for $K^{-}d$ and $K^{-}p$ atoms are done in
different density. Because if it is possible, We can take
$\Gamma_{2p}$ as a free parameter then it can be determined more
precisely by fitting the simulated x-ray curves with experimental
results. We have also predicted the optimum range of the target
density to detect the higher x-ray yields for forthcoming
experiments on $K^{-}p$ and $K^{-}d$ atoms. The optimum range of
the target density are 0.03-0.06 of liquid hydrogen density (LHD)
for $K^{-}p$ atoms and 0.06-0.2 of LHD for $K^{-}d$ atoms.\\ We
have also investigated the kinetic energy distribution of $K^{-}p$
atoms and the role of Coulomb transition on x-ray yields. The
high kinetic energy component of $K^{-}p$ atoms appear as Doppler
broadening profile in the experimental x-ray lines. We have
calculated the average Doppler broadening contribution on the
observed width in $K_{\alpha}$ line by using the simulated
kinetic energy distribution of $K^{-}p$ atoms in $2p$-state at
the instant of $2p\rightarrow1s$ radiative transition. It should
be used to extract the $\Gamma_{1s}^{had}$ from experimental
results.


\vfill  

\noindent{\bf References }
\begin{description}
\setlength\itemsep{-3pt}
\item{[1]} C. Curceanu {\it et al.}, Eur. Phys. J. A {\bf 31} (2007)
537; A. Romero, this proceeding.
\item{[2]} M. Leon and H.A. Bethe, Phys. Rev. {\bf 127} (1962) 636.
\item{[3]} T.S. Jensen and V.E. Markushin, Lect. Notes Phys. {\bf 627} (2003) 37.
\item{[4]} M.Raeisi G. and S.Z. Kalantari, Phys. Rev. A {\bf 79}
(2009) 012510.
\item{[5]} S.Z. Kalantari and M.Raeisi G., submitted to Phys. Rev. C (2009).
\item{[6]} T.M. Ito {\it et al.}, Phys. Rev. C {\bf 58} (1998) 2366.
\item{[7]} G. Beer {\it et al.}, Phys. Rev. Lett. {\bf 94} (2005) 212302; and private discussion with one of the authors (C. Curceanu).
\end{description}


\setcounter{equation}{0} 
\setcounter{figure}{0}
\clearpage
\addcontentsline{toc}{section}{
{\bf Search for Double Antikaon Production in Nuclei by
 Stopped Antiproton Annihilation}\\
Paul~Kienle}

%





\titl{Search for Double Antikaon Production in Nuclei by Stopped Antiproton Annihilation}

\name{
Paul~Kienle$^{1}$
}

\adr{
$^1$ Excellence Cluster Universe, Technische Universit\"{a}t M\"{u}nchen, Germany
}


Recently antiproton annihilation in nuclei was proposed to search for deeply 
bound double-strange systems with S = -2 of light nuclei [1, 2]. A reanalysis of 
OBELIX data on stopped antiproton annihilation in $^4$He at LEAR-CERN showed 
a surprisingly large production probability of $\sim10^{-4}$ of strangeness
S = -2 channels [3]. The corresponding momentum transfer is large in such a reaction; 
thus the observed high formation probability indicates compact systems as products. 
Such a scenario became recently support by the observation of a deeply bound 
S = -1 di-baryon resonance with a binding energy of $\sim100$ MeV in high momentum
transfer $p-p$ collisions [4]. So there is reason to believe that one can produce
S = -2 di-baryon systems with still higher binding energies and densities using 
antiproton annihilation at rest. This led to letters of intent for searches of dense, 
S = -2 nuclear systems at CERN, FAIR and J-PARC [5, 6, 7] with dedicated detectors.

In this contribution we present details of experimental approaches to search for 
the most elementary S = -2 di-proton system using the antiproton annihilation 
reaction on $^3$He targets at rest
\begin{equation}
\bar{p} + ^3{\rm He} \to K^+ K^0 + X ({\rm S}=2)
\end{equation}
where $X$ denotes a di-proton with S = -2, which one can see as deeply 
bound $K^-K^-pp$ system with a binding energy of about 200 MeV 
(double as high as the observed 100 MeV of $K^-pp$ [4]). We furthermore assume that 
it is formed with a probability of $\sim10^{-4}$ per antiproton [3], and that it 
decays with an assumed 10\% probability into $\Lambda \Lambda$, the energetically 
most favored channel. For an exclusive experiment we need a large acceptance detector 
with the ability of identifying all charged particles, measuring their low momenta 
for determination the missing mass of $K^+K^0$ and their correlated $\Lambda \Lambda$
pairs; the $K^0 \to \pi^+ \pi^-$, and the $\Lambda \to p \pi^-$ decay being 
reconstructed via their respective invariant masses.

A detailed simulation for the $K^+K^0$ and $\Lambda \Lambda$ missing mass spectra
has been performed for two detector concepts with different magnetic field 
configurations (cylindrical and dipole magnets).  With an antiproton beam of 
0.65 GeV/c momentum, $3.4 \times 10^4$ antiprotons in a spill time of 3.5 s, 
a stopping efficiency of 3.9\%, and assuming a double strangeness production 
probability of $10^{-4}$, one expects double-strangeness production of 
$9.6 \times 10^4$ per month. Assuming a conservative $K^+ K^0 \Lambda \Lambda$
branching ratio of 10\% one produces $9.6 \times 10^3$ $K^+K^0 \Lambda \Lambda$ 
events per month.
Simulated $\Lambda \Lambda$ and $K^+K^0$ missing mass spectra to be measured in 
the two setups during one month show well resolved lines with about
50 and 20 events, respectively.

\vfill  

\noindent{\bf References }
\begin{description}
\setlength\itemsep{-3pt}
\item{[1]} W. Weise,  arXiv: 0507.058 (nucl-th) 
\item{[2]} P. Kienle, J. Mod. Phys. A22 (2007) 365, and J. Mod. Phys. E16 (2007) 905
\item{[3]} G. Bendiscioli, {\it et al.}, Nucl. Phys. A 797 (207) 109
\item{[4]} T. Yamazaki, P. Kienle, M. Maggiora, and K. Suzuki in behalf of the DISTO collaboration, 
              Hyp. Int., DOI:10.1007/s10751-009-9997-5
\item{[5]} J. Zmeskal, {\it et al.}, EXA/LEAP08 Proc. Hyp. Inter.
\item{[6]} J. Zmeskal, {\it et al.}, New opportunities in the physics landscape of CERN, May 2009
\item{[7]} M. Iwasaki, P. Kienle, H. Onishi, F.Sakuma, and J. Zmeskal, Lett. of Int. J-PARC, June 2009
\end{description}


\setcounter{equation}{0} 
\setcounter{figure}{0}
\clearpage
\addcontentsline{toc}{section}{
{\bf Hypernuclear production in ($K^-_{\rm stop}, \pi$) reactions}\\
V.~Krej\v{c}i\v{r}\'{i}k, A.~Ciepl\'{y}}

%

%
%


\titl{Hypernuclear production in ($K^-_{\rm stop}, \pi$) reactions}

\name{
V.~Krej\v{c}i\v{r}\'{i}k$^{1,2}$, A.~Ciepl\'{y}$^{1}$
}

\adr{
$^1$ Nuclear Physics Institute, ASCR, \v{R}e\v{z}, Czech Republic \\
$^2$ Faculty of Mathematics and Physics, Charles University, Prague, Czech Republic \\
}


We studied the $\Lambda$-hypernuclear production induced by the stopped kaon. 
The calculation was performed within a framework of the distorted wave impulse approximation (DWIA) as described in detail by
Gal and Klieb [1]. The formula for the capture rate $R_{\rm if}$ can be expressed as
a product of three terms:
\begin{equation}
R_{\rm if} = { { q_{\rm f} \omega_{\rm f} }\over{ \overline{q}_{\rm f} \overline{\omega}_{\rm f} } } \cdot R(K^- N\rightarrow \pi Y) \cdot R_{\rm if}/Y .
\label{CRperKaon}
\end{equation}
The first term  is a kinematic factor,
the second 
stands for the branching ratio of the elementary process and
the third represents the rate per hyperon that contains the overlap of the initial and the final state wave functions. 

We used an effective potential model based on chiral symmetry [2,3] to generate
the elementary branching ratios,
whereas previous authors took values derived from experiment. 
The pertinent branching ratios are $R(K^-n \rightarrow \pi^- \Lambda) = 10.72$, and $R(K^-p \rightarrow \pi^0 \Lambda) = 5.36 $. 

\begin{wrapfigure}{l}{6.5cm}
\vspace*{-1.1em}
\begin{center}
\includegraphics[width=5.5cm]{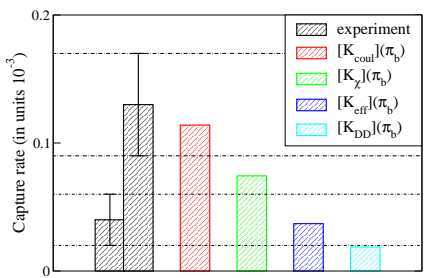}
\caption{The production of $^{16}_{~\Lambda}{\rm O}$ in the $1S_{\Lambda}$ state.}
\end{center}
\vspace*{-1em}
\end{wrapfigure}

We performed the calculation with three different pion wave functions: $(\pi_0)$-free wave, $(\pi_{\rm b})$ and $(\pi_{\rm c})$-generated
by two different optical potentials, which were fitted to pion scattering data.
We found that the capture rate is up to one order
higher if the pion distortion is neglected while the other two options give comparable capture rates. 

For the choice $(\pi_{\rm b})$, the sensitivity to the kaon wave function is demonstrated in Fig.1.
Four different potentials were used to describe the $K^-$-nucleus interaction: pure electromagnetic [$K_{\rm EM}$] ($V_0=0\,{\rm MeV}$),
chiral $[K_{\chi}]$ ($V_0\approx 50\,{\rm MeV}$), and two phenomenological ones $[K_{\rm eff}]$ ($V_0\approx 80\,{\rm MeV}$),
 and $[K_{\rm DD}]$($V_0\approx 190\,{\rm MeV}$). It appears
that the capture rate is a decreasing function of the central strong interaction potential depth which is denoted by~$V_0$.



In the figure, our results are also compared with experimental data [4,5] (the first two bars). Although the data from different experiments
do not match,
we can say that our results are in better agreement with experiment than results of previous authors.
Currently, we have results for capture rates of the production of $^{7}_{\Lambda}{\rm Li}$, $^{9}_{\Lambda}{\rm Be}$,
$^{12}_{~\Lambda}{\rm B}$, $^{12}_{~\Lambda}{\rm C}$, $^{13}_{~\Lambda}{\rm C}$, $^{16}_{~\Lambda}{\rm O}$,
and we also tested the sensitivity of
the model to various inputs.
A more compete acount of our work was presented in [6].

\vfill  

\noindent{\bf References }
\begin{description}
\setlength\itemsep{-3pt}
\item{[1]} Gal A., Klieb L.: Phys. Rev. C {\bf 34} (1986) 956.
\item{[2]} Kaiser N., Siegel P.B., Weise W.: Nucl. Phys. A {\bf 594}, (1995) 325.
\item{[3]} Ciepl\'{y} A., Smejkal J.: arXiv:0910.1822 (2009). 
\item{[4]} G. Bonomi, talk at HYP-X @ J-PARC (2009).
\item{[5]} Tamura H., et al.: Prog. Theor. Phys. Suppl. {\bf 117}, (1994) 1.
\item{[6]} Krej\v{c}i\v{r}\'{i}k, A.~Ciepl\'{y}: arXiv:0912.1505 (2009).
\end{description}


\setcounter{equation}{0} 
\setcounter{figure}{0}
\clearpage
\addcontentsline{toc}{section}{
{\bf The in-flight $ ^{12}C(K^-,p)$ reaction at KEK}\\
V.K. Magas, J. Yamagata-Sekihara, S. Hirenzaki, E. Oset, and A. Ramos}

%





\titl{The in-flight $ ^{12}C(K^-,p)$ reaction at KEK}

\name{
V.K. Magas$^1$, J. Yamagata-Sekihara$^{2,3}$, S. Hirenzaki$^4$, E. Oset$^3$, and A. Ramos$^1$
}

\adr{
$^1$ Departament d'Estructura i Constituents de la Mat\`eria, \\
Universitat de Barcelona, Diagonal 647, 08028 Barcelona, Spain\\
$^2$ Yukawa Institute for Theoretical Physics, Kyoto University, Kyoto 606-8502, Japan\\
$^3$ Departamento de F\'{\i}sica Te\'orica and IFIC Centro Mixto\\
Universidad de Valencia-CSIC, Institutos de Investigaci\'on de Paterna \\
Apdo. correos 22085, 46071, Valencia, Spain \\ 
$^4$ Department of Physics, Nara Women's University, Nara 630-8506, Japan
}


We study the $(K^-,p)$ reaction on $ ^{12}C$ with a kaon beam of 1 GeV/c momentum, paying a special attention at the region of emitted protons having kinetic energy above 600 MeV, which was used to claim the evidence of a deep kaon nucleus optical potential, $Re~V \approx -200~ \rho/\rho_0$ MeV [1]. 

We perform a Monte Carlo simulation of this reaction. The advantage of our method with respect to the Green's function method used in [1] is that it   
allows one to account not only for quasi-elastic $K^- p$ scattering, but also for  the other processes which contribute to the proton spectra. 
We investigated the effect of $K^-$ absorption by one and two nucleons ($K^- N \rightarrow \pi Y$ and $K^- NN \rightarrow YN$) followed by the weak decay of the hyperon into $\pi N$, which can also produce strength in the region of interest [2]. Our calculation is done within a local density approximation and considers  final state interactions in terms of multiple scattering of the $K^-$, $p$ and all other primary particles on their way out of the nucleus.
To compare our calculations with the published data [1]  we have to also simulate the experimental coincidence requirement of having, together with
the energetic proton, at least one charged particle detected in the decay counters surrounding the target.

We show that the above mentioned absorption mechanisms, not considered in the original publication [1], together with the coincidence simulation, allow us to explain the observed spectrum with "standard" shallow kaon nucleus optical potential, $Re~V \approx -60~ \rho/\rho_0$ MeV, obtained in chiral models  [2].   

Also, contrary to what was assumed in
Ref. [1], we clearly see that the
spectrum shape is affected by the required coincidence [2]. In fact, the distorsion
of the experimental spectrum due to the coincidence requirement can easily be
much bigger than the difference between results for shallow  and deep optical potentials.
Thus, we conclude that the experiment of
Ref. [1] is not appropriate for extracting information on
the kaon optical potential. The theoretical analysis of [1]
was based on the assumption that the shape of the spectrum does not change with
the coincidence requirement. Since we have shown this not to be case,  the
conclusions obtained there do not hold. Certainly, the experimental data 
without the coincidence requirement of [1] would be a much 
more useful observable.

\vfill  

\noindent{\bf References }
\begin{description}
\setlength\itemsep{-3pt}
\item{[1]} 
  T.~Kishimoto {\it et al.},
  Prog.\ Theor.\ Phys.\  {\bf 118} (2007) 181.

\item{[2]} 
  V.~K.~Magas, J. Yamagata-Sekihara, S. Hirenzaki, E. Oset,  A. Ramos,
  arXiv:0911.3614 [nucl-th];
%
  arXiv:0911.2091 [nucl-th];
A.~Ramos, V.~K.~Magas, J. Yamagata-Sekihara, S. Hirenzaki, E. Oset,
  arXiv:0911.4841 [nucl-th].

\end{description}


\setcounter{equation}{0} 
\setcounter{figure}{0}
\clearpage
\addcontentsline{toc}{section}{
{\bf Kaonic nuclei}\\
J.~Mare\v{s}}

%

%



\titl{Kaonic nuclei}

\name{
J.~Mare\v{s}$^{1}$}
\adr{
$^1$ Nuclear Physics Institiute, 250 68 \v{R}e\v{z}, Czech Republic}
We performed fully selfconsistent calculations of ${\bar K}$ nuclear bound states
within the relativistic mean field approach~[1,2] with the view of exploring   
dynamical aspects of the ${\bar K}$--nucleus interaction.  
We scanned on the ${\bar K}$ coupling constant in order to cover   
a wide range of ${\bar K}$ separation energies. 
  
The widths $\Gamma_{\bar K}$ of the ${\bar K}$ nuclear states are mostly determined 
by phase-space suppression on top of the increase provided by the compressed nuclear 
density. A lower limit $\Gamma_{\bar K} \sim 50\pm 10$~MeV was placed on the widths 
of the $1s$ ${\bar K}$ nuclear states expected for separation energies in the range 
$B_{\bar K} \sim 100 - 200$~MeV.  
Fig.~1 illustrates that replacing $\rho$ by $\rho^2$ for the density dependence of the 
$2N$ absorption modes and switching on the $\pi\Lambda$ decay mode adds further 
conversion width to the ${K^-}$ nuclear states.  

\begin{figure}[h]
\centering
\includegraphics[width=8cm]{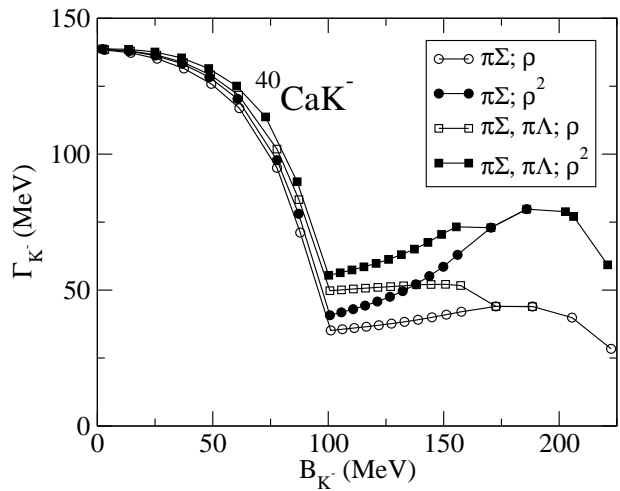}
  \caption{Width of the $1s$ $K^-$ nuclear state in $^{40}_{K^-}$Ca
as function of the $K^-$ separation energy, for absorption through
${\bar K}N \rightarrow \pi\Sigma$, with and without ${\bar K}N
\rightarrow \pi\Lambda$, and assuming $\rho$ or $\rho^2$
dependence for ${\bar K}NN \rightarrow {\Sigma}N$.
}
\end{figure}

Significant polarization of the core nucleus was found in light nuclei for deeply bound 
${K^-}$ nuclear states. The resulting nuclear central density is increased by a 
factor of 2. However, this enhancement, which follows closely the 
${K^-}$ density, is limited to a region $\leq 2$~fm about the origin. 
As a result, the calculated average nuclear density increases substantialy in light kaonic 
nuclei ($^{12}_{K^-}$C, $^{16}_{K^-}$O),  
while it remains almost constant as function $B_{K^-}$ in $^{208}_{K^-}$Pb. 

The dynamical calculation gives higher binding than the static calculation does. 
The gain in $B_{K^-}$ increases monotonically with the $K^-$ separation energy. It is 
worth noting that Im$V_{K^-}$ may safely be ignored in the calculation of $B_{K^-}$ 
above 100~MeV.  
 
\vspace*{1mm}  
We refer the interested reader to Refs.~[1,2] for more detailed discussion of the 
dynamical effects for ${\bar K}$ deeply bound nuclear states.
 
\vfill  
This work was supported in part by the GACR grant 202/09/1441.  

\noindent{\bf References }
\begin{description}
\setlength\itemsep{-3pt}
\item{[1]} J. Mare\v{s}, E. Friedman, A. Gal, Nucl. Phys. A {\bf 770} (2006) 84.
\item{[2]} D. Gazda, E. Friedman, A. Gal, J. Mare\v{s}, Phys. Rev. C {\bf 76} (2007) 055204.

\end{description}


\setcounter{equation}{0} 
\setcounter{figure}{0}
\clearpage
\addcontentsline{toc}{section}{
{\bf LEANNIS: a liason between theory and
experiment in antikaon physics}\\
J.~Marton}

%





\titl{LEANNIS: a liason between theory and
experiment in antikaon physics}

\name{
J.~Marton$^{1}$
}

\adr{
$^1$ Stefan Meyer Institute, 1090 Vienna, Austria \\
}

The Network LEANNIS (Low-energy Antikaon Nucleon and Nucleus Interaction) is devoted to the
current research in experiment as well as in theory on kaonic atoms and kaonic nuclei bound by the strong force.
It is a networking activity in the European project HadronPhysics2 of the 7$^{th}$ framework program.
12 institutes from 5 EU countries (Austria, Finland, Germany, Italy and Poland) and institutes from the associated country Japan participate in LEANNIS.
Many open questions are addressed in new experiment like SIDDHARTA at LNF [1] and FOPI [2]. For the interpretation of
the experimental data a close connection with theoreticians is essential and provided by the network.
The field of low-energy antikaon interaction has many facets. Even the study of the most simple kaonic atom - kaonic hydrogen - is a challenge in experiment and theory. The so-called kaonic hydrogen puzzle was solved by the experiment KpX
[3] and verified by DEAR [4] but more precise data of the observables (shift and width of the ground-state due to K$^-$-p interaction is main topic of the SIDDHARTA experiment taking advantage of new detectors (silicon drift detectors) for
x ray spectroscopy. On the other hand the corresponding observables of kaonic deuterium are a hot topic since they are
necessary to extract the isospin dependent K$^-$-nucleon scattering lengths. For the first time kaonic deuterium was studied by SIDDHARTA very recently. The extraction of the scattering lengths is subject of intense theoretical work [5] taking into account the corrections (e.g. isospin breaking) to the Deser-Trueman formula.\\
 Many questions about kaonic nuclei are still open: production mechanism, binding energy, decay widths etc. Experimental data claim the observation of kaonic nuclei in different reactions using stopped K$^-$, proton-induced reaction or antiproton absorption [6,7]. However, the overall picture is not coherent since the binding energies and decay width extracted from data with different production reactions vary in a broad range. On the theoretical side studies with effective field approaches, Faddeev calculations and phenomenological approaches also differ in a wide rang. Certainly this situation calls for new dedicated experiments (fully exclusive experiments) and
also new theoretical studies in order to design the experiments.
To promote the field the LEANNIS activities are concentrated on developing new strategies, both in experimental and theoretical sectors, to attack the still many open problems in the field. The development of new experimental methods and techniques will profit from this coordinated network. Furthermore, major European institutes working in this field are participating in this network, therefore a platform is created to strengthen and bundle the research efforts.

\vfill  

\noindent{\bf References }
\begin{description}
\setlength\itemsep{-3pt}
\item{[1]} http://www.lnf.infn.it/
\item{[2]} K. Suzuki et al., Nucl.Phys.{\bf A827} (2009)312C.
\item{[3]} M. Iwasaki et al., Phys. Rev. Lett. {\bf 78} (1997) 3067.
\item{[4]} G. Beer, et al., Phys. Rev. Lett. {\bf 94} (2005) 212302.
\item{[5]} M. Lage, U.-G. Mei{\ss}ner and A. Rusetzky, Hyp. Int. {\bf 193} (2009) 69. 
\item{[6]} M. Agnello et al., Phys. Rev. Lett. 94, 212303 (2005).
\item{[7]} T. Yamazaki et al.,Hyp. Int. {\bf 193} (2009) 181,  arXiv: 0810.5182 [nucl–ex].
\end{description}


\setcounter{equation}{0} 
\setcounter{figure}{0}
\clearpage
\addcontentsline{toc}{section}{
{\bf Recent results on $K^{-}$ absorption at rest 
on few nucleon systems with FINUDA}\\
S.~Piano}

%





\titl{Recent results on $K^{-}$ absorption at rest \\ 
on few nucleon systems with FINUDA}

\name{
S.~Piano$^{1}$
}

\adr{
$^1$ INFN Sezione di Trieste, via A. Valerio 2, 34127 Trieste, Italy
}


The FINUDA experiment is dedicated to the study of kaonic interactions
and is installed at the DA$\Phi$NE collider (INFN Laboratori
Nazionali di Frascati - Italy) [1]. The experimental
setup consists of a large acceptance ($> 2\pi$ sr) magnetic spectrometer 
with high momentum resolution and excellent particle mass identification.
In FINUDA, negative kaons are absorbed at rest in nuclear
targets, which are constituted by thin plates; thus, allowing for 
$A$ dependence studies, from $^6$Li to $^{51}$V. \\
While the $K^{-}_{stop} N\rightarrow Y\pi$ one-nucleon absorption is 
the basic mechanism for hypernuclear production, the negative kaon 
can initially interact also with two or more nucleons, 
$K^-(2N)\rightarrow YN,\, K^-(3N)\rightarrow Y(NN),\, K^-(4N)\rightarrow Y(NNN)$,
leading to hyperon-nucleons final states without pion emission. 
The experimental study of mesonless multinucleon absorption received a 
strong boost after the Akaishi and Yamazaki [2] predictions 
of the existence of Bound Kaonic Nuclear (BKN) states because
their formation can be viewed as the intermediate step 
of a $K^-$ many-body absorption.\\ 
The FINUDA studies of kaon absorption on few nucleon systems  
were carried out by examining $\Lambda-p(d,t)$ correlations [3] [4] [5].
Regardless of $A$, the $\Lambda-p(d,t)$ pairs are found to be preferentially 
emitted in opposite directions. 
The observation of a nearly constant production rate of back-to-back
$\Lambda-p(d,t)$ pairs suggests that the absorption
of $K^{-}$ at rest in nuclei may proceed through the formation
of intermediate BKN states. Moreover, the experiment could fruitfully 
apply the quasi-invariant mass spectroscopy technique, thanks to the capability 
to detect completely the full topology of the final states, also by means of
the missing mass information. 
In the $\Lambda p$ [3] and $\Lambda d$ [4] invariant 
mass spectra two strong signatures of BKN states were found 
and the binding energy and the width were provided. \\
The analysis of the FINUDA second data taking are in the final stages 
and will allow to obtain spectra from different targets. The A dependence may play 
a key role to understand  the negative kaon multinucleon absorption and to
disentangle the effect of Final State Interactions. \\

\vfill  

\noindent{\bf References }
\begin{description}
\setlength\itemsep{-3pt}
\item{[1]} FINUDA Collaboration, M. Agnello {\it et al.}, Phys. Lett. {\bf B622} (2005) 35

\item{[2]} Y. Akaishi, T. Yamazaki, {\it Phys. Rev.} {\bf C65} (2002), 044005 \\
T. Yamazaki, Y. Akaishi, {\it Nucl. Phys.} {\bf B535} (2002), 70 \\
Y. Akaishi, A. Dote, T. Yamazaki, {\it Phys. Lett.} {\bf B613} (2005), 140

\item{[3]} FINUDA Collaboration, M. Agnello {\it et al.}, Phys. Rev. Lett. {\bf 94} (2005), 212303
\item{[4]} FINUDA Collaboration, M. Agnello {\it et al.}, Phys. Lett. {\bf B654} (2007), 80 \\
FINUDA Collaboration, M. Agnello {\it et al.}, Eur. Phys. J. {\bf A33} (2007), 283
\item{[5]} FINUDA Collaboration, M. Agnello {\it et al.}, Phys. Lett. {\bf B669} (2008), 229
\end{description}


\setcounter{equation}{0} 
\setcounter{figure}{0}
\clearpage
\addcontentsline{toc}{section}{
{\bf SIDDHARTA recent results}\\
A. Romero Vidal on behalf of the SIDDHARTA Collaboration}

%





\titl{SIDDHARTA recent results}

\name{ A. Romero Vidal on behalf of the SIDDHARTA Collaboration
}

\adr{
 Laboratori Nazionale di Frascati-INFN, Frascati (Rome) \\
}


  The SIDDHARTA experiment at $DA\Phi NE$ studies 
the X-ray spectroscopy of exotic atoms, as kaonic helium ($He^3$ and $He^4$) and kaonic hydrogen and deuterium.
For the case of kaonic hydrogen and deuterium the strong interaction leads to a shift $\Delta E$ of the energy of the fundamental $1s$ level
compared to the value calculated from electromagnetic interaction only, and a broadening $\Gamma$ due
to the absorption of the hadron by the nucleus. These quantities can be obtained by comparing the measured 
energies of the X-ray transitions to the $1s$ level to the purely $QED$ calculated ones.

  A measurement of the $\Delta E$ and $\Gamma$ for the K-H and the K-d would 
provide very valuable information for a better comprehension of the low energy $QCD$, improving the
measurements done by $KpX$ [1] and $DEAR$ [2] experiments, and will allow to determine the isospin dependent kaon-nucleon scattering lengths [3].

 The SIDDHARTA setup is described in detail in [3]. It uses the low-energy kaon beam from
the $DA\Phi NE$ $e^+e^-$ collider , stopping kaons in a high-density cryogenic gas target and employs 144 novel Silicon Drift Detectors (SDDs) $1cm^2$ to measure the X-rays energy.
A system of two scintillators placed above (in coincidence with entrance window of setup) and
below the beam pipe are used to dectect the kaons coming from the $\Phi \rightarrow K^+ K^-$ reaction, and to deliver the trigger signal to the SDD detectors, eliminating in this way the background produced by losses from circulating beams (main source of background).

The data taking campaign of SIDDHARTA: a measurement of kaonic hydrogen, the first exploratory measurement of kaonic deuterium, precision measurements of $KHe^4$ and $KHe^3$  ended by middle November 2009. Presently the data analyses is undergoing, with very promising results. For preliminary analyses results see [5].



\vfill  

\noindent{\bf References }
\begin{description}
\setlength\itemsep{-3pt}
\item{[1]} M. Iwasaki et al., Phys. Rev. Lett. 78 (1997) 3067.
\item{[2]} G. Beer et al., Phys Rev. Lett. 94 (2005) 212302.
\item{[3]} J. Zmeskal, Progr. Part. Nucl. Phys. 61 (2008) 512.
\item{[4]} T. Ishiwatari, Nucl. Instr. Meth. Phys. Res. A 581 (2007) 326.
\item{[5]} http://agenda.infn.it/getFile.py/
\item{} access?contribId=1\&sessionId=1\&resId=0\&materialId=slides\&confId=1841

\end{description}


\setcounter{equation}{0} 
\setcounter{figure}{0}
\clearpage
\addcontentsline{toc}{section}{
{\bf QGP formation in antiproton annihilation at rest}\\
P.~Salvini, G.~Bendiscioli, T.~Bressani}

 
 
 
\titl{QGP formation in antiproton annihilation at rest} 
 
\name{ 
G.~Bendiscioli$^{1,3}$, T.~Bressani$^{2}$, P.~Salvini$^{3}$ 
} 
 
\adr{ 
$^1$ Dipartimento di Fisica Nucleare e Teorica dell'Universita' di Pavia,Pavia, Italy \\ 
$^2$ Dipartimento di Fisica Sperimentale, Universita'  di Torino, Torino, Italy  \\ 
$^3$ Istituto Nazionale di Fisica Nucleare, Sezione di Pavia, Italy  \\ 
} 
 

In this paper we report data on pion spectra reinforcing the hypothesis of quark-gluon plasma 
formation in $\overline p ^4He$ annihilation at rest previously reported [1]. According to some models
the transition from a hadronic phase to QGP is 
expected to occur above a critical temperature of the order of 150-200 MeV [2] and one of its expected 
signatures is a high production of strangeness [3]. We experimentally investigated the  $K^+$, $K^-$ production in the 
$\overline p$ annihilation at rest on $^1$H and $^4$He gas targets and found that in 
reaction channels selected as annihilations with involvement of more than one nucleon and without 
neutral meson production the ratio between the $K^+$ production  on $^4$He and that on single nucleon, i.e. $^1$H, 
raise up to 30 [1], while the strangeness increase observed for the global set of 
events was only about 2 times.
The data were obtained with the spectrometer Obelix working at the 
complex LEAR of CERN in the years 1992-1996.

\begin{figure}[h] 
\centering 
\includegraphics[height=.18\textheight]{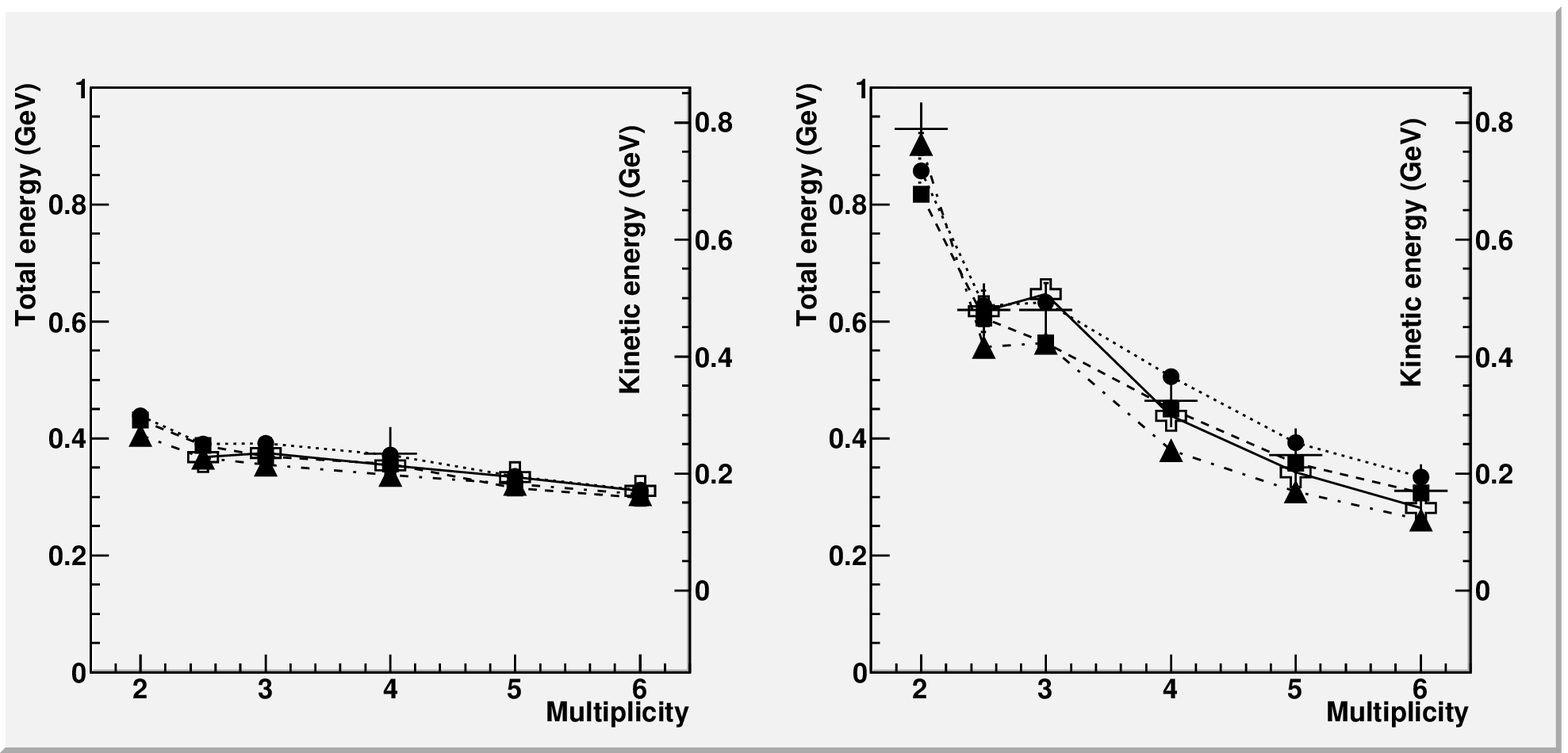} 
\caption{Pion mean energy E$_{\pi}$ vs. multiplicity for reaction channels with the following detected particles: 
(crosses) $\pi ^{\pm}$; (circles) $\pi ^{\pm}$1K$^{\pm}$; (triangles)$\pi ^{\pm}$p; (squares)$\pi ^{\pm}$K$^{\pm}$p. 
(a) All events; (b) events without neutral meson production
[1]. The E$_{\pi}$ for $^1$H ($+$) is
calculated as 2m$_{proton}$ divided by the $\pi ^{\pm}$ multiplicity.} \label{pinco} 
\end{figure} 
 
In Figs.\ref{pinco}a,b the $\pi^{\pm}$ mean kinetic energy  results to be equal or higher than
the critical temperature Tc $\approx$ 170 MeV predicted for the QGP formation. Taking into account
that the $q \overline q$ state masses are higher than the single quarks masses, from figs.\ref{pinco} the quarks kinetic energy 
(corresponding to the temperature in the statistical picture [4]) is high enough to be compatible with the occurrence of QGP. 
This means that at least two  conditions predicted for the QGP formation  are well satisfied in $\overline p$$^4$He annihilation at rest.

\vfill  
\noindent{\bf References } 
\begin{description} 
\setlength\itemsep{-3pt} 
\item{1} P.Salvini{\it et al.},Nucl.Phys.A760(2005)34, G.Bendiscioli{\it et al.},Nucl.Phys.A815(2009)67
\item{2} F.Karsch {\it et al.}, Nucl.Phys.B605(2001)579, J. Rafelski, Nucl. Phys. A418(1984)215c
\item{3} J. Rafelski and B. M\H{u}ller,  Phys. Rev. Let
\item{4} D. Evans {\it et al.}, J. Phys. G: Nucl. Phys. 25(1999)209
\end{description} 
 

\setcounter{equation}{0} 
\setcounter{figure}{0}
\clearpage
\addcontentsline{toc}{section}{
{\bf The trigger system for the AMADEUS experiment}\\
A.Scordo, {\it et al.}}

%





\titl{The trigger system for the AMADEUS experiment}

\name{A.Scordo,
 M.Bazzi, G.Corradi, A.Romero Vidal,  D.Tagnani, O.Vazquez Doce}

\adr{
 Laboratori Nazionali di Frascati, INFN, Via E. Fermi 40, 00044, Frascati, IT \\
}


Multi-Pixel Photon Counters (MPPC) consist of hundreds of micro silicon Avalanche PhotoDiodes (APD) working in Geiger mode. The high gain, low noise and low voltage values needed for operating of these relatively new devices, together with their good behaviour in magnetic fields make them ideal for the readout of scintillating fibers as front-end detectors, as planned for the trigger system of the AMADEUS experiment. A first test of a prototype of MPPC+Sci-Fi setup was performed on the DA$\Phi$NE collider at LNF-Frascati. 5 scintillating fibers ($\sim 10\, cm\, length$) were coupled at both sizes with MPPCs and were placed few centimiters under the lower scintillator of SIDDHARTA's Kaon Monitor [1] wich is located $6\, cm$ below the interaction point. 
The Kaon Monitor consists in 2 scintillators mounted above and below the beam pipe in order to detect a K-/K+ couple produced ind the $\Phi$ decay in back-to-back configuration and also MIPs coming from the interaction point; KM signal is the coincidence between upper and lower scintillator. Radiofrequency of the machine is taken as reference time; this signal is provided each time a collision between e+ and e- bunches occurs ($\simeq 340\, MHz$).
Timing separation between MiPs and Kaons in the KM is $\simeq 1\,ns$
Triggering the acquisition with the SIDDHARTA's Kaon Monitor and selecting signals coming from Kaons or MIPs, charge and timing information were collected. 

\begin{figure}[htbp]
\centering
\begin{tabular}{lll}
\mbox{\includegraphics[height=.27\textheight]{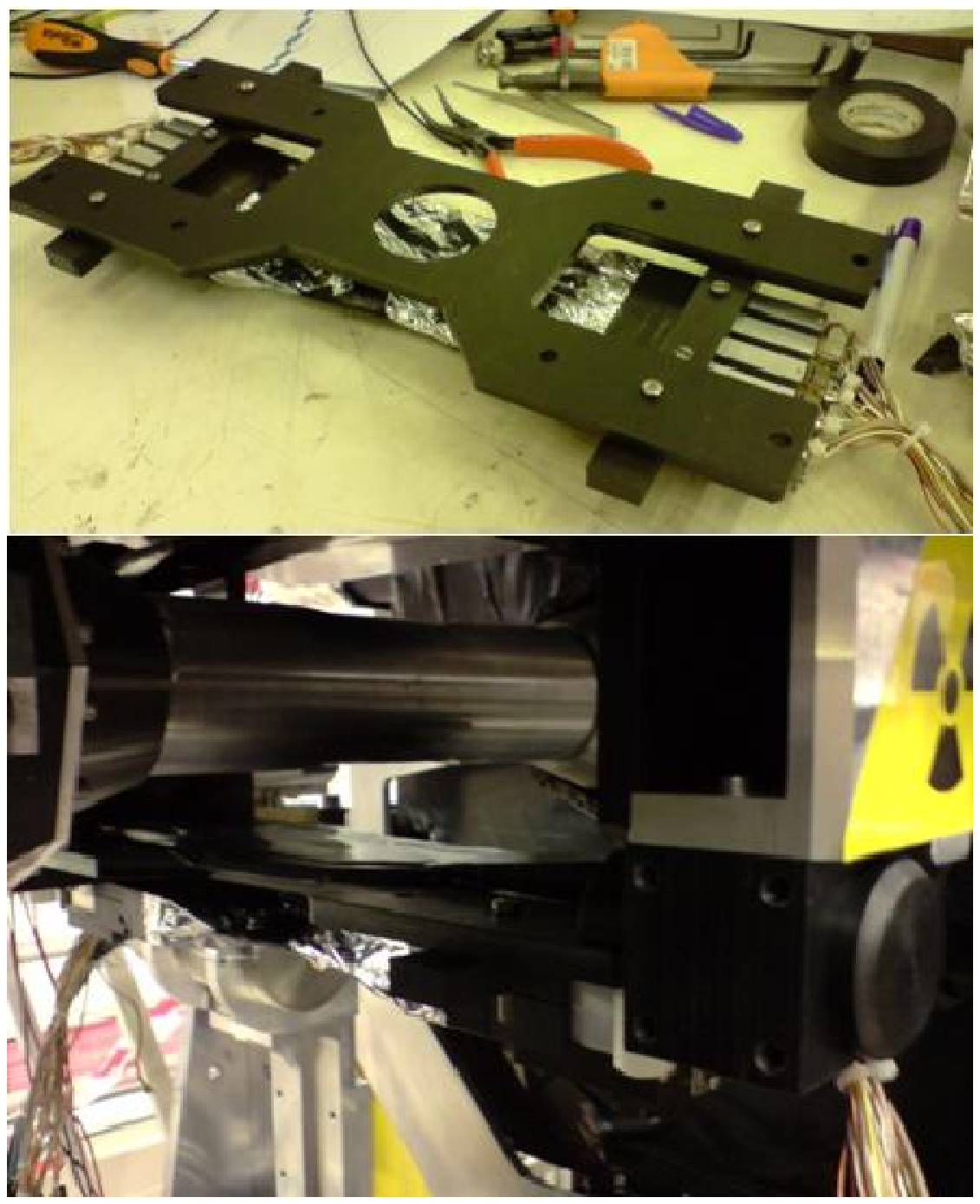}}
\mbox{\includegraphics[height=.27\textheight]{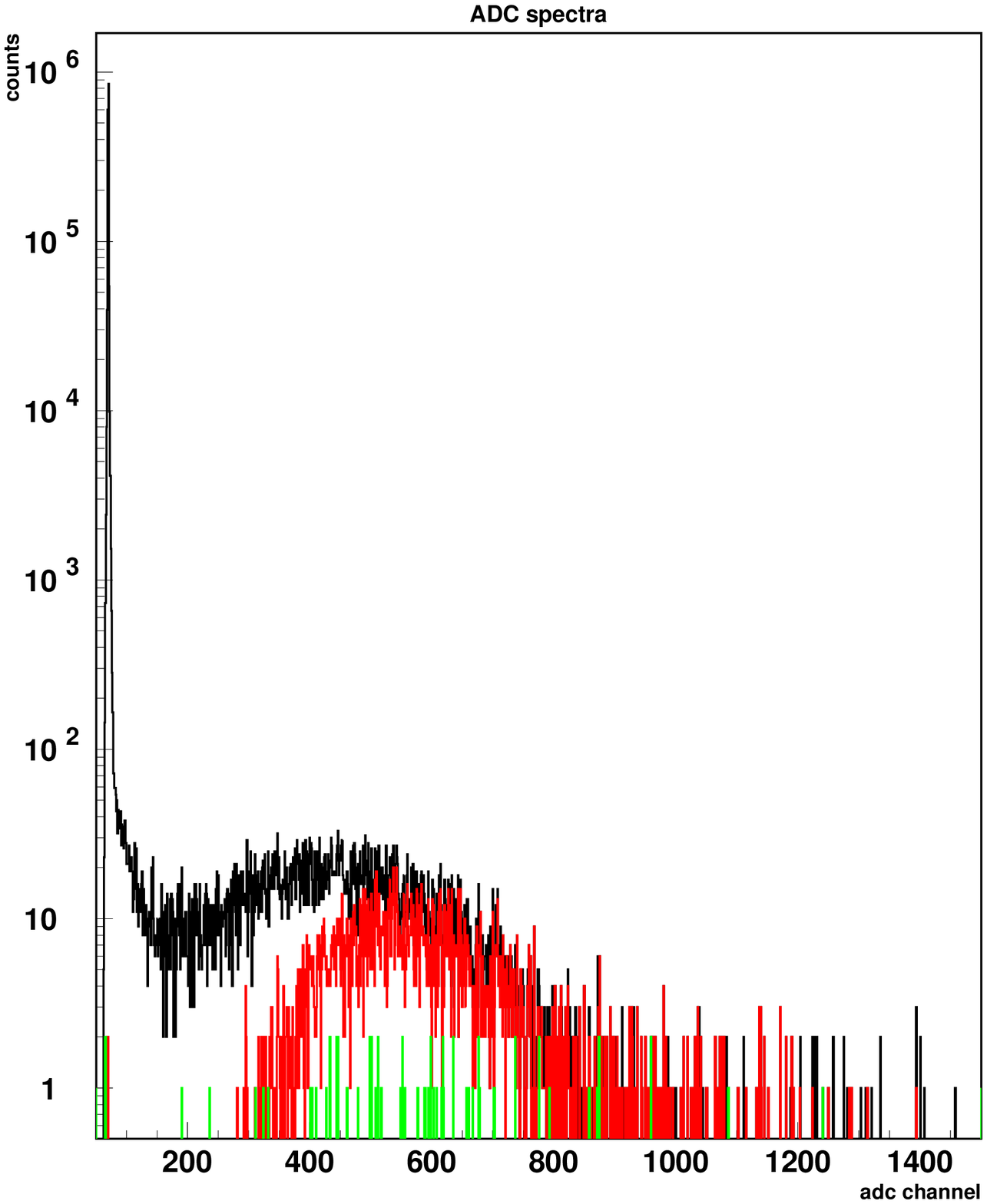}}
\mbox{\includegraphics[height=.27\textheight]{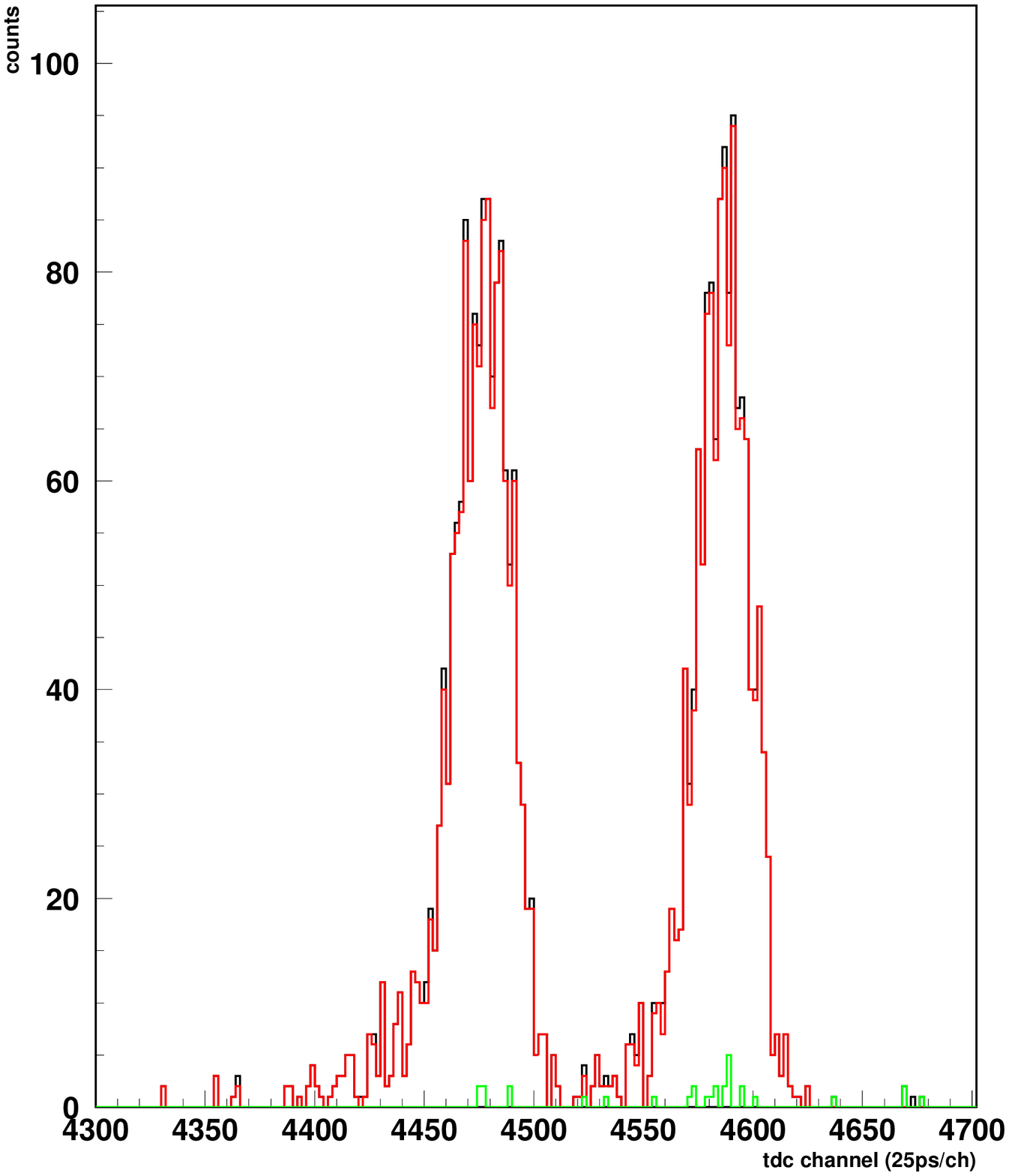}}
\end{tabular}
\caption{ADC and TDC spectra for MPPC+Sci-Fi setup mounted on DA$\Phi$NE collider (left picture); total spectra (black), signals in coincidence with Kaons (red) and with MIPs (green) are shown.}
\end{figure}

Using a Sci-Fi+MPPC setup detection of Kaons coming from DA$\Phi$NE is then possible with a time resolution of $\simeq \,300\,ps$.

\vfill  

\noindent{\bf References }
\begin{description}
\setlength\itemsep{-3pt}
\item{[1]} C. Curceanu et al., Eur. Phys. J. A 31 , (2007) , 537-539 
\end{description}


\setcounter{equation}{0} 
\setcounter{figure}{0}
\clearpage
\addcontentsline{toc}{section}{
{\bf Two- and three-body resonances 
in the $\bar{K}NN - \pi \Sigma N$ system}\\
N.V.~Shevchenko}

%

%



\titl{Two- and three-body resonances \\in the $\bar{K}NN - \pi \Sigma N$ system}

\name{
N.V.~Shevchenko$^{1}$
}

\adr{
$^1$ Nuclear Physics Institute, 25068 \v{R}e\v{z}, Czech Republic
}


One of the most important characteristics of a two-body interaction is
existence and properties of its bound states or resonances. When the
interaction is used in a few- or many-body calculation they also strongly
influence the properties of the few- or many-body system. We investigate
the dependence of the three-body resonance properties in the
$\bar{K}NN$ system on the different models of $\bar{K}N$ interaction,
providing one or two poles for $\Lambda(1405)$ resonance.

It is known, that $\bar{K}N$ interaction is strongly coupled to the
$\pi \Sigma$ channel through $\Lambda(1405)$ resonance. However, the nature
of the resonance is a question. A usual assumption is that $\Lambda(1405)$ is
a resonance in $\pi \Sigma$ and a quasi-bound state in $\bar{K}N$ channel.
There is also an assumption suggested by a chiral model, that the bump, which
is usually understood as $\Lambda(1405)$ resonance, is an effect of two poles.

In~[1] isospin symmetry breaking phenomenological $\bar{K}N-\pi \Sigma$
potentials were constructed in order to check whether it is possible to
reproduce all existing experimental data on $K^- p$ system with one-
and two-pole structures of $\Lambda(1405)$. It turned out, that
both versions of the potential can
reproduce existing data on $K^- p$ scattering and $K^- p$ atom level shift
equally well in such a way, that it is not possible
to draw conclusions about ``nature'' of $\Lambda(1405)$ resonance.

One possible way to clarify the situation is to perform a few- or many-body
calculation using two versions of $\bar{K}N-\pi \Sigma$ potential as
an input. Having this in mind, we repeated our calculations of
$\bar{K}NN-\pi \Sigma N$ system~[2, 3] searching for three-body poles in it.
As before, coupled-channel Faddeev equations in AGS form were solved.

Not only basic $\bar{K}N-\pi \Sigma$ interaction was changed during the
calculations, other two-body inputs were also improved. In particular,
we used new two-term potential
for $NN$ interaction~[4], which reproduce Argonne $v18$ $NN$ phase shifts,
having repulsion at short distances, and corresponding scattering length
and effective radius.

Also a new set of $\Sigma N (-\Lambda N)$ potential parameters was found
in order to reproduce existing experimental data on $\Sigma N$ and $\Lambda N$
scattering. The $I=3/2$ part of the interaction is a one-channel $\Sigma N$,
while $I=1/2$ $\Sigma N$ is connected with $\Lambda N$ channel. Due to this
a coupled-channel $I=1/2$  $\Sigma N - \Lambda N$ potential
was constructed first, then a corresponding one-channel optical
potential for $\Sigma N$ was derived.

Our preliminary results show, that while two versions of the
$\bar{K}N - \pi \Sigma$ interaction describe two-body experimental $K^- p$ data
indistinguishably well, the three-body results for the pole positions in
$\bar{K}NN - \pi \Sigma N$ system strongly depend on the two-body
$\bar{K}N - \pi \Sigma$ input. More detailed calculation is in progress.

\vspace{1mm}
\noindent
{\it Acknowledgments.} The work was supported by the Czech GA AVCR grant KJB100480801.

\vfill  

\noindent{\bf References }
\begin{description}
\setlength\itemsep{-3pt}
\item{[1]} J. R\'evai, N.V. Shevchenko, Phys. Rev. {\bf C 79} (2009) 035202.
\item{[2]} N.V. Shevchenko, A. Gal, J. Mare\v{s}, Phys. Rev. Lett. {\bf 98} (2007) 082301.
\item{[3]} N.V. Shevchenko, A. Gal, J. Mare\v{s}, J. R\'evai, Phys. Rev. {\bf C 76} (2007) 044004.
\item{[4]} P. Doleschall, {\it private communication}.
\end{description}


\setcounter{equation}{0} 
\setcounter{figure}{0}
\clearpage
\addcontentsline{toc}{section}{
{\bf The investigation of $\Lambda(1405)$ state 
in the stopped $K^-$ reaction on deuterium}\\
T.~Suzuki, J.~Esmaili, and Y. Akaishi}

%

%



\titl{The investigation of $\Lambda(1405)$ state in the stopped $K^-$ reaction on deuterium}

\name{
T.~Suzuki$^{1}$, J.~Esmaili$^{2,3}$, and Y. Akaishi$^{2,4}$
}

\adr{
$^1$ Department of Physics, The University of Tokyo, Tokyo 113-0033, Japan \\
$^2$ RIKEN Nishina Center, RIKEN, Saitama 351-0198, Japan \\
$^3$ Department of Physics, Isfahan University of Technology, Isfahan 84156-83111, Iran \\
$^4$ College of Science and Technology, Nihon University, Chiba  274-8501, Japan \\
}

Recently, intensive discussion of the problem of the deeply bound $\bar{K}$ nuclei reached to the reconsideration of $\Lambda(1405)$ state as the theoretical basis of the binding of $\bar{K}$ nuclei, and the old question of the nature of $\Lambda(1405)$ became a modern subject by the new interest.
 
 In contrast to the interpretation of $\Lambda(1405)$ as the $\bar{K}N$ quasi-bound state at 1405 MeV/$c^2$, a two-pole hypothesis, by which $\Lambda(1405)$  consists of two poles at 1420 and 1390 MeV/$c^2$ couple mainly with $\bar{K}N$ and $\Sigma \pi$ channels, respectively,  thus the less attractive $\bar{K}N$ interaction leads shallower binding of $\bar{K}$ nucleus, was proposed by several authors [1] [2].  
 
  On the other hand,  a very recent theoretical analysis has clarified that the $(\Sigma \pi)^0$ invariant mass spectra after $K^-$ absorption in $d$ do reflect resonant formation of $\Lambda(1405)$ (or $\Lambda(1420)$) and thus are capable of distinguishing different Ansatz's [3]. 
   
 We have discussed a new experimental proposal [4] by means of the stopped $K^-$ reaction on liquid deuterium at J-PARC K1.8BR beamline with E15 [5] /E17 [6] experimental devices, so as to give a new precision- and high-statistics-data of $(\Sigma \pi)^0$ mass spectra to examine the issue, $\Lambda(1405)$ or $\Lambda(1420)$, in the most reliable way, and thus to answer  the most fundamental questions of $\bar{K}N$ interaction and $\bar{K}$ nuclei.

\begin{figure}[h]
\centering
  \includegraphics[height=.15\textheight]{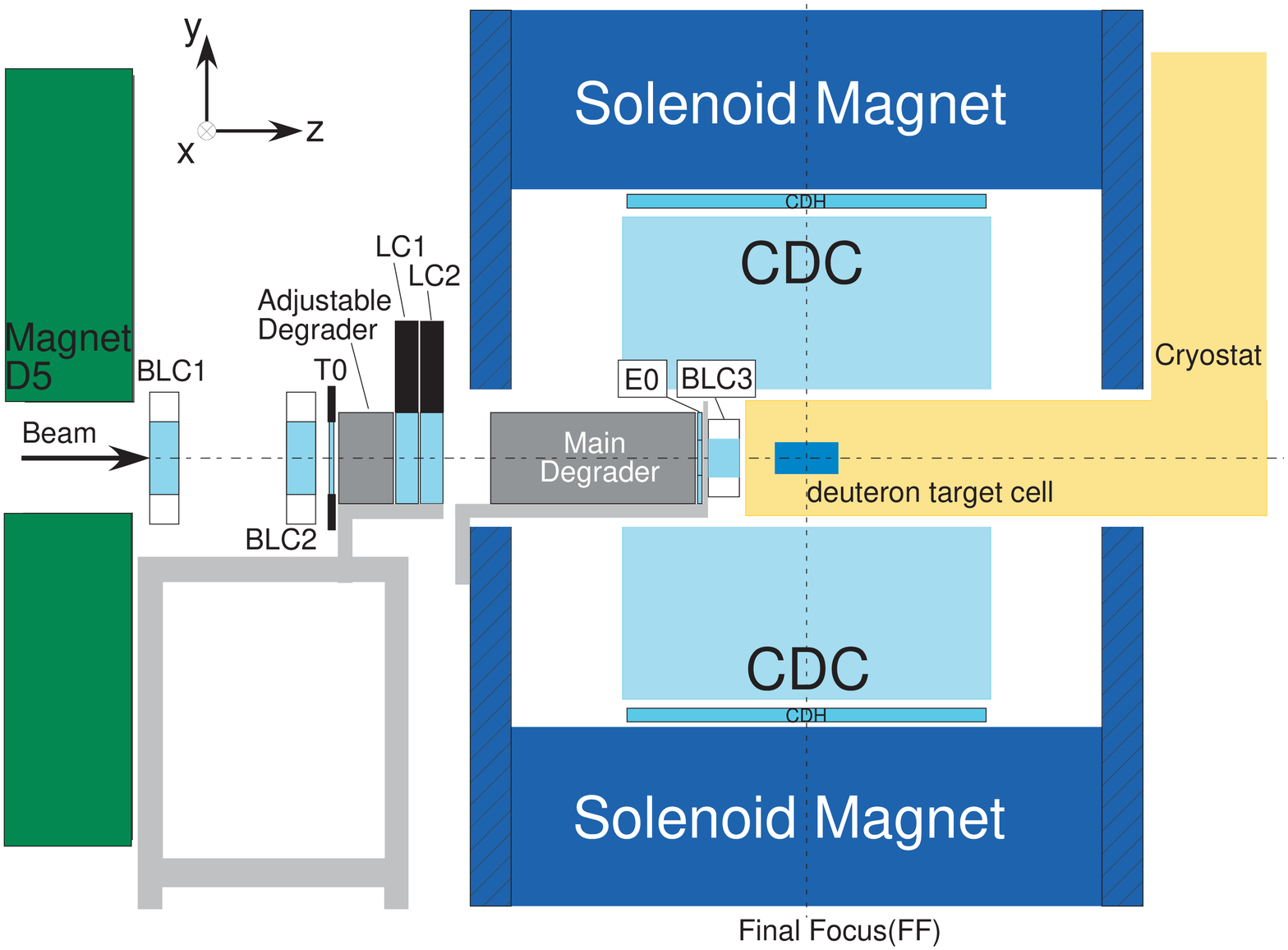}
  \hspace{15mm}
  \includegraphics[height=.15\textheight]{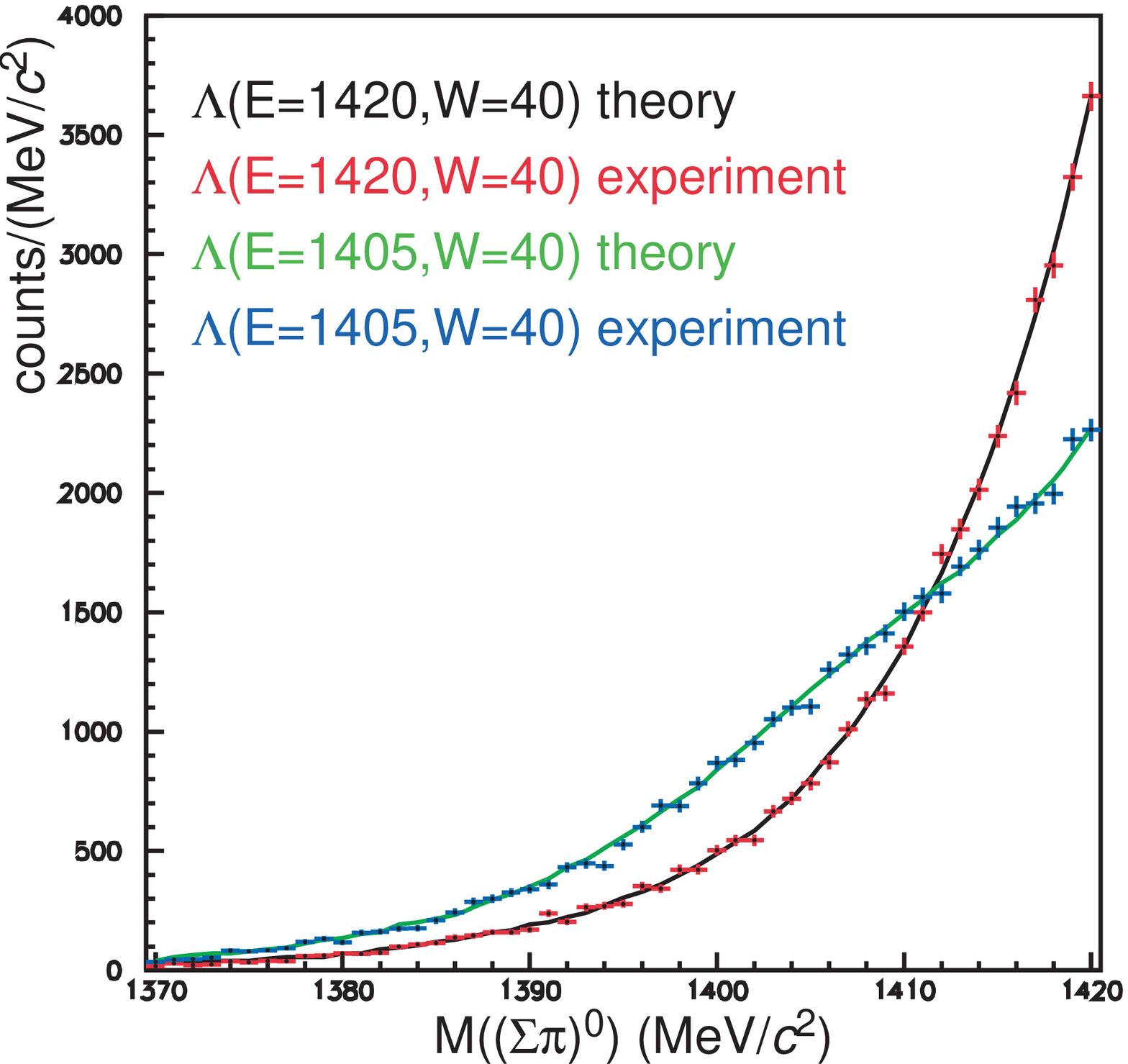}  
  \caption{Left: A schematic view of the proposed experiment. Right: Theoretical and expected $M(\Sigma \pi)^0$ spectra.}
\end{figure}

\vfill  

\noindent{\bf References }
\begin{description}
\setlength\itemsep{-3pt}
\item{[1]} V.K. Magas, E. Oset, and A. Ramos, Phys. Rev. Lett. {\bf 95} (2005) 052301.
\item{[2]} T. Hyodo and W. Weise, Phys. Rev. C  {\bf 77} (2008) 035204.
\item{[3]} J.~Esmaili, Y. Akaishi, and T. Yamazaki, ArXiv:0909.2573.
\item{[4]} T. Suzuki {\it et. al}, Proposal for J-PARC 50 GeV Proton Synchrotron (J-PARC P30), 2009.
\item{[5]} M. Iwasaki, T. Nagae {\it et. al}, Proposal for J-PARC 50 GeV Proton Synchrotron (J-PARC E15), 2006.
\item{[6]} R.S. Hayano, H. Outa {\it et. al}, Proposal for J-PARC 50 GeV Proton Synchrotron (J-PARC E17), 2006.
\end{description}


\setcounter{equation}{0} 
\setcounter{figure}{0}
\clearpage
\addcontentsline{toc}{section}{
{\bf Analysis of the K$^-$ He interactions in the KLOE drift chamber}\\
O.~Vazquez Doce$^1 $ on behalf of the AMADEUS Collaboration}

%

%



\titl{Analysis of the K$^-$ He interactions in the KLOE drift chamber}

\name{
O.~Vazquez Doce$^1 $ on behalf of the AMADEUS Collaboration
}

\adr{
$^1$ LNF-INFN, Enrico Fermi 40, 00044, Frascati (Rome), Italy \\
}


The AMADEUS experiment [1] at the Da$\Phi$ne accelerator of the Frascati National Laboratories (Italy) of INFN,
will perform,
for the first time,
full-acceptance studies of
antikaon interaction in light nuclei,
with a complete experimental program for
the case of the kaonic clusters.
Studying the absorption of antikaon by the nucleus will provide
information
concerning
the
$\bar{K}N$ interaction
and
the modification of the kaon mass
in the nuclear medium.

A preliminar study of these kind of hadronic interactions is being done
by the AMADEUS collaboration by analyzing the existent KLOE data [2].

The KLOE drift chamber [3] contains mainly helium, and a Monte Carlo study
shows that 0.1\% of the K$^-$ flying through the chamber should be stopped in the gas,
giving an unique scenario to study the hadronic interactions in
such an "active target".

Preliminary results of the analysis of a sample of the 2005 KLOE data
(corresponding to an integrated luminosity of $\sim 1.1 fb^{-1}$)
has shown the capabilities in performing nuclear physics measurements
with the KLOE detector. 
The strategy is focused on the identification of possible specific decay
products of the kaonic nuclear clusters:
specifically into channels containing the $\Lambda$(1116) hyperon,
present in most of the expected decay channels of the bound states.
An excellent result has been already achieved with a precise determination of the
lambda mass, the statistical error is below 3 KeV. The measurement shows an excellent mass resolution
FWHM $\sim$ 700 KeV/c$^2$,
found in the reconstruction of the
decay of $\Lambda$
into proton and negative pion [4].

Vertices produced by these lambda particles with protons or deuterons are searched
along the K$^-$ (tracked or extrapolated) decay path, or along the lambda path extrapolated backwards,
as direct signals of the formations of these clusters, or absorptions of K$^-$
by the nucleons of the gas nuclei.
Also neutral vertices are seached for, as the expected resulting from the formation
of a lambda(1405) decaying to neutral particles, $\Sigma^0 \pi^0$. In this
case the excellent performance of the electromagnetic calorimeter and its resolution for
the detection of photons is crucial.

In conclusion, a selection of thousands of $\Lambda$(1115) baryons
has been made from $\sim 1.1 fb^{-1}$ of KLOE data,
allowing to investigate diferent kind of reaction products from the interaction
of K$^-$ in the drift chamber.
The number and the quality of the signal
opens the door for studies of many hadronic physics hot topic items,
proving KLOE to be
a powerful instrument for performing
very interesting physics in the
strange nuclear and hadronic physics sectors too.

\vfill  

\noindent{\bf References }
\begin{description}
\setlength\itemsep{-3pt}
\item{[1]} The AMADEUS collaboration, LNF preprint, LNF-07/24(IR) (2007)
\item{[2]} O. Vazquez Doce, presentation on 38th LNF Scientific Committee, http://www.lnf.infn.it/committee/ (2009).
\item{[3]} M. Adinolfi et al., Nucl. Instr. Meth. A 488, 51-73 (2002)
\item{[4]} M. Cargnelli, C. Petrascu, O. Vazquez Doce and KLOE K-charged group, KLOE memo 337 (2007)
\end{description}


\setcounter{equation}{0} 
\setcounter{figure}{0}
\clearpage
\addcontentsline{toc}{section}{
{\bf Antikaon Interactions with Nucleons and Nuclei}\\
W.~Weise}




\titl{Antikaon Interactions with Nucleons and Nuclei}

\name{Wolfram Weise

}

\adr{
Physik-Department, Technische Universit\"at M\"unchen, D-85747 Garching, Germany
}


Precision measurements of kaonic hydrogen and their analysis extracting the real and imaginary parts of the  $K^-p$ scattering length set important quantitative constraints for chiral SU(3) dynamics. The best data so far have been obtained at LNF (the DEAR experiment) and earlier at KEK (the PS-E228 experiment). The  $K^-p$ scattering lengths deduced from these measurements are not fully consistent:
$a(K^-p) = -0.47\,(\pm\, 0.10) + i\,0.30\, (\pm\,0.17) ~[\mbox{fm}]$ [DEAR] and 
$a(K^-p) = -0.78\,(\pm\, 0.18) + i\,0.49\, (\pm\,0.37) ~[\mbox{fm}]$ [KEK].
Theoretical analyses of the kaonic hydrogen energy shift ($\Delta E$) and width ($\Gamma$), based on chiral SU(3) dynamics, have been performed in Refs. [1,2]. The calculations favour slightly the earlier KEK data. It is thus important to resolve this issue at the higher level of precision reached with the
SIDDHARTA experiment at LNF. High-precision $\bar{K}N$ threshold data and accurate $\pi\Sigma$ mass spectra are crucial in order to guide subthreshold extrapolations of  the antikaon-nucleon interaction into domains relevant for possible $\bar{K}$-nuclear quasibound states. 

Extrapolations of the $\bar{K}N \leftrightarrow \pi\Sigma$ coupled-channels dynamics into regions below $K^- p$ threshold can be tested by examining in detail the shapes and locations of the three $\pi^+\Sigma^-,~\pi^-\Sigma^+$ and $\pi^0\Sigma^0$ invariant mass distributions. They differ primarily due to the $I=1$ component of the amplitude as it interferes with the dominant $I=0$ part. There is not just a single $\pi\Sigma$ mass spectrum determining uniqely the position and width of the $\Lambda(1405)$. The $\pi\Sigma$ mass distributions depend on the process considered. 

Concerning the quest for quasibound $K^-pp$ systems, the present status of the theory can be summarized as follows. Two basic strategies have been employed: i) three-body calculations solving Faddeev equations with separable interactions [3], and ii) variational calculations using phenomenlogical input [4] or $\bar{K}N$ effective interactions based on chiral SU(3) dynamics [5]. Even though all input interactions in these calculations have been tuned to reproduce threshold $\bar{K}N$ observables or the $\Lambda(1405)$, one encounters a broad band of binding energies ranging between about 20 and 80 MeV, while the decay widths cover values between 40 and 110 MeV. 

A necessary (though not sufficient) condition for reliable subthreshold extrapolations is a controlled theoretical framework. Chiral SU(3) effective field theory combined with coupled-channel methods provides such a framework, but it requires a sufficiently large and accurate empirical data base in order to proceed.

While unambiguous conclusions about quasibound antikaon-nuclear systems can at present not yet be drawn, further progress is expected to come from detailed investigations of exclusive final states following $K^-$ absorption and photon- or hadron-induced $K^+$ production on nuclei, in order to constrain the underlying coupled-channel dynamics.

\vfill  

\noindent{\bf References }
\begin{description}
\setlength\itemsep{-3pt}
\item{[1]} B. Borasoy, R. Nissler and W. Weise, Eur. Phys. J {\bf A 25} (2005) 79; Phys. Rev. Lett. {\bf 94} (2005) 213401.  
\item{[2]} R. Nissler, PhD thesis, Univ. of Bonn (2007); B. Borasoy, U.-G. Mei{\ss}ner and R. Nissler, Phys. Rev. {\bf C 74} (2006) 055201.  
\item{[3]} N.V. Shevchenko, A. Gal and J. Mares, Phys. Rev. Lett. {\bf 98} (2007) 082301; N.V. Shevchenko, A. Gal, J. Mares and J. R\'evai, Phys. Rev. {\bf C 76} (2007) 044004; Y. Ikeda and T. Sato, Phys. Rev. {\bf C 76} (2007) 035203,  Phys. Rev. {\bf C 79} (2009) 035201.
\item{[4]} T. Yamazaki and Y. Akaishi, Phys. Lett. {\bf B 535} (2002) 70; Phys. Rev. {\bf C 76} (2007) 04520; S. Wycech and A.M. Green, Phys. Rev.  {\bf C 79} (2009) 014001.
\item{[5]} T. Hyodo and W. Weise, Phys. Rev. {\bf C 77} (2008) 03524; A. Dot\'e, T. Hyodo and W. Weise, Nucl. Phys. {\bf A 804} (2008) 197, Phys. Rev.  {\bf C 79} (2009) 014003.
\end{description}


\setcounter{equation}{0} 
\setcounter{figure}{0}
\clearpage
\addcontentsline{toc}{section}{
{\bf Extraction of the $\overline{ K}N$ subthreshold amplitudes
from $K^{-}$  atoms}\\
S.~Wycech}

%





\titl{Extraction of the $\overline{ K}N$ subthreshold amplitudes
from $K^{-}$  atoms}

\name{S.~Wycech
}

\adr{ Soltan Institute for Nuclear Studies, Warsaw, Poland
  }


The K  mesic atoms offer a chance to look directly into
subthreshold $ K^-N$ scattering amplitudes. This region is of
interest as there are conflicting models of the $\Lambda(1405)$
resonance and  related amplitudes. In $K^-$ atoms, the meson
interacts with bound nucleons and the energy in the $ K^-N$ system
is negative  as a result of the binding and $ K^-N$ recoil
energies. Most of the atomic data involve heavy nuclei. These are
not useful for the purpose since many nucleons participate in
interactions and nuclear effects are sizable.  Light nuclei are
convenient as the nucleon binding energies in $^1$H, $^2$H,
$^3$H,$^3$He and $^4$He span the 0- 20 MeV region in a
quasi-continuous way. With systematic experiments one can  extract
the $ K^-N$ scattering amplitudes below the $ K^-N$ threshold down
to about - 35 MeV. At this moment one has  only $K^1$H [1] and
$K^4$He [2] data. One should look with more care into usually
neglected "upper level" widths  in heavier $K$ atoms. In
particular one notices stronger absorption in nuclei with large
proton binding (C). There is a possibility to extract Im $a(K^-n)$
from large nuclei (Pb,U). Some old unprecise measurements are 
worthy of repetition.

A similar extraction has been performed  with the light
antiprotonic atoms [3]. Absorptive  $\overline{p}-N$ amplitudes
shown in the figure indicate existence of  S and P-wave
subthreshold quasi-bound states which find additional evidence in
other experiments [3,4].
\begin{figure}[h]
\centering
  \includegraphics[height=.25\textheight]{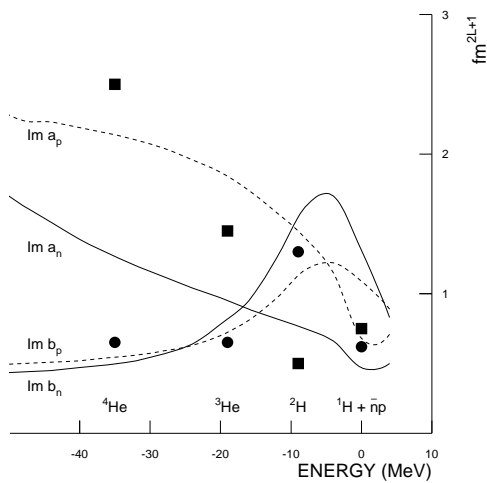}
  \caption{ The absorptive $\overline{p}-N$ scattering lengths (Im a, squares)  and scattering volumes (Im b, blobs)
  extracted from light atoms. The lines present Paris-model calculations.}
\end{figure}

\vfill  

\noindent{\bf References }
\begin{description}
\setlength\itemsep{-3pt}
\item{[1]} G. Beer et al. Phys.Rev.Lett \textbf{94 }(2009) 212302.
\item{[2]} S. Okada et al. Phys.Lett. \textbf{B653} (2007) 387.
\item{[3]} S. Wycech et al. Phys.Rev.\textbf{C76}(2007) 014311.
\item{[4]} J-P. Dedonder et al. Phys. \textbf{C80}(2009) 0452207.
\end{description}


\setcounter{equation}{0} 
\setcounter{figure}{0}
\clearpage
\addcontentsline{toc}{section}{
{\bf Experimental confirmation of the $\Lambda(1405)$ Ansatz
 from resonant formation  of a $K^-p$ quasi-bound state
 in $K^-$ absorption by $d$, $^3$He and $^4$He}\\
Toshimitsu Yamazaki}

%





\titl{Experimental confirmation of the $\Lambda(1405)$ Ansatz\\
 from resonant formation
  of a $K^-p$ quasi-bound state\\ in $K^-$ absorption by $d$, $^3$He and $^4$He
 }

\name{Toshimitsu Yamazaki$^{1,2}$
} 
\adr{
$^{1}$ Department of Physics, University of Tokyo, Tokyo, 116-0033 Japan 
$^{2}$ RIKEN Nishina Center, Wako, Saitama, 351-0198 Japan}

Where is the position of the $I=0$ $L=0$ $K^-p$ quasi-bound state? This question is directly connected to the strength of the s-wave $I=0$ $\bar{K} N$ interaction. Traditionally, the $\Lambda(1405)$ resonance is identified to this state, and a strongly attractive $\bar{K} N$ interaction is indicated [1]. Recently, another theoretical framework including a double-pole hypothesis has been proposed, claiming that the $K^- p$ bound state is located around 1420 MeV [2,3]. We call such a hypothetical state ``$\Lambda^*(1420)$". It is vitally important to distinguish between $\Lambda(1405)$ and $\Lambda^*(1420)$ experimentally, but there seem to be lots of confusing statements concerning the strategy as to how valid experimental evidence can be obtained.

In our recent work [4] we have shown that the $\Sigma\pi$ invariant-mass spectra ($M_{\Sigma\pi}$) in stopped-$K^-$ absorption in $^4$He, $^3$He and $d$ are governed by the spectator momentum distributions of $t$, $d$ and $n$, respectively, and do reflect the resonant formation of a quasi-bound $K^-p$ state, contrary to the past interpretation in terms of the non-resonant direct-capture process. Thus, the issue of the location of the $K^-p$ resonance state can be examined experimentally by a quantitative comparison of an observed $M_{\Sigma\pi}$ spectrum with predicted theoretical distributions including resonant formation. We made a $\chi^2$ analysis of  old bubble-chamber data of $M_{\Sigma\pi}$($^4$He) by varying the mass ($M$) and width ($\Gamma$) of an assumed resonance and found a significant minimum in the $M-\Gamma$ contour presentation of $\chi^2$ at
\begin{equation}
M = 1405.5^{+1.4}_{-1.0}~{\rm MeV}/c^2~ {\rm and}~ \Gamma = 25.6 ^{+4}_{-3}~{\rm MeV}.
\end{equation}
Thus, the $\Lambda^*(1420)$ Ansatz is excluded by more than 99.99\% confidence. However, the $\Lambda(1405)$ signal does not appear as a separate peak, but as a small component of the steeply falling tail, and this discrimination seems to be delicate. One would hope to have a more clear-cut case.  

More recently, we have pointed out [5] that the use of a deuteron target in the reaction,
\begin{eqnarray}
{\rm stopped-}K^- + d &\rightarrow& X + n,\\
&X& \rightarrow \Sigma + \pi,
\end{eqnarray}
can provide a more decisive conclusion. Since the deuteron wavefunction is composed of low- and high-momentum components, the dominant ``quasi-free (QF)'' shape of $M_{\Sigma\pi}$ is narrow, whereas its tail, resulting from the high-momentum component of $d$, extends to the region where a resonant formation of $\Lambda(1405)$ (but not of $\Lambda^*(1420)$) can be revealed as a separate peak. We investigate this problem in detail.\\

 


\vfill  

\noindent{\bf References }
\begin{description}
\setlength\itemsep{-3pt}
\item{[1]} Y. Akaishi, T. Yamazaki, Phys. Rev. C {\bf65} (2002) 044005.
\item{[2]} D. Jido, J. A. Oller, E. Oset, A. Ramos and U.-G. Mei{\ss}ner, Nucl. Phys. A {\bf725} (2003) 181.
\item{[3]}  T. Hyodo and W. Weise, Phys. Rev. C {\bf77} (2008) 035204.
\item{[4]}  J. Esmaili,Y. Akaishi and T. Yamazaki, arXiv:0906.0505v1.
\item{[5]}  J. Esmaili,Y. Akaishi and T. Yamazaki, arXiv:0909.2573.
\end{description}


\setcounter{equation}{0} 
\setcounter{figure}{0}
\clearpage
\addcontentsline{toc}{section}{
{\bf Indication of a strongly bound dense $K^-pp$ state
formed in the $pp \rightarrow p \Lambda K^+$ reaction at 2.85 GeV}\\
 T.~Yamazaki, M.~Maggiora, P.~Kienle, K.~Suzuki 
{\it on behalf of the DISTO collaboration}}

%





\titl{Indication of a strongly bound dense $K^-pp$ state\\
formed in the $pp \rightarrow p \Lambda K^+$ reaction at 2.85 GeV}

\name{
 T.~Yamazaki$^{1,2}$, M.~Maggiora$^3$, P.~Kienle$^{4,5}$, K.~Suzuki$^{4}$,\\  {\it on behalf of the DISTO collaboration} 
}

\adr{
$^{1}$ Department of Physics, University of Tokyo, Tokyo, 116-0033 Japan 
$^{2}$ RIKEN Nishina Center, Wako, Saitama, 351-0198 Japan 
$^3$ Dipartimento di Fisica Generale ``A. Avogadro'' and INFN, Torino, Italy 
$^{4}$ Stefan Meyer Institute for Subatomic Physics, Austrian Academy of Sciences, Vienna, Austria 
$^{5}$ Excellence Cluster Universe, Technische Universit\"at M\"unchen, Garching, Germany}

Recently, it was predicted that a strongly bound $K^-pp$ system with a short $p$-$p$ distance [1,2] can be formed in a $p+p \rightarrow p + \Lambda^* + K^+$ reaction with an enormously large sticking probability between $\Lambda^* \equiv \Lambda(1405)$ and $p$ due to the short range and high  momentum transfer of the $pp$ reaction [3,4]. Here, we report that existing data of a DISTO experiment show an evidence for this exotic formation. Preliminary reports have been published [5,6].

We have analyzed data of the DISTO experiment on the exclusive $pp \rightarrow p \Lambda K^+$ reaction at 2.85 GeV to search for a strongly bound $K^-pp ~(\equiv X)$ state to be formed in the $pp \rightarrow K^+ + X$ reaction. The observed spectra of the $K^+$ missing-mass and the $p \Lambda$ invariant-mass with high transverse momenta of $p$ and $K^+$ revealed a broad distinct peak with a mass $M_X = 2265 \pm 2~(stat) \pm 5~(syst)$ MeV/$c^2$ and a width $\Gamma_X = 118 \pm 8~(stat) \pm 5~(syst)$ MeV.

The $X$ production rate is found to be as much as the $\Lambda(1405)$ production rate. Such a large formation is theoretically possible {\it only when the $p$-$p$} (or $\Lambda^*$-$p$) {\it distance in $X$ is shorter than 1.7 fm}, whereas the average $N$-$N$ distance in ordinary nuclei is 2.2 fm [3,4]. No candidate other than $K^-pp$ for $X$ with such large formation is predicted so far. The dominance of the formation of the observed $X$ at high  momentum transfer gives direct evidence for the compactness and thus high density of the produced $K^-pp$ cluster.

The mass of $X$ corresponds to a binding energy $B_K = 105 \pm 2~(stat)~\pm 5~(syst)$ MeV for $X = K^-pp$. 
The observed binding energy is close to the mass $M(p \Lambda) \sim$ 2255 MeV/$c^2$ of the $K^-pp$ candidate observed by FINUDA [7]. It is larger than the original prediction, thus suggesting additional effects to be investigated [4]. The large width seems to be understood by the dense nature of $K^-pp$. 
The theoretical claims for shallow $\bar{K}$ binding [8,9] do not seem to be in agreement with the present observation.\\


\vfill  

\noindent{\bf References }
\begin{description}
\setlength\itemsep{-3pt}
\item{1} Y. Akaishi and T. Yamazaki, Phys. Rev. {\bf C 65}, 044005 (2002).
\item{2} T. Yamazaki and Y. Akaishi, Phys. Lett. {\bf B 535}, 70 (2002).
\item{3} T. Yamazaki and Y. Akaishi, Proc. Jpn Acad. {\bf B 83}, 144 (2007). 
\item{4} T. Yamazaki and Y. Akaishi, Phys. Rev. {\bf C 76}, 045201 (2007).
\item{5} T. Yamazaki {\it et al.}, Hyperfine Interactions {\bf 193}, 181 (2009). 
\item{6} M. Maggiora {\it et al.}, Hyp-X Proceedings; arXiv:0912.5116 [hep-ex].
\item{7} M. Agnello {\it et al.}, Phys. Rev. Lett. {\bf 94}, 212303 (2005).
\item{8} D. Jido, J.A. Oller, E. Oset, A. Ramos and U.-G. Meissner, Nucl. Phys. {\bf A725}, 181 (2003); V.K. Magas, E. Oset and A. Ramos, Phys. Rev. Lett. {\bf 95}, 052301  (2005).
\item{9} T. Hyodo and W. Weise, Phys. Rev. {\bf C 77}, 035204 (2008); A.~Dot$\acute{\rm e}$, T. Hyodo and W. Weise, Phys. Rev. {\bf C 79}, 014003 (2009).
 \end{description}


\setcounter{equation}{0} 
\setcounter{figure}{0}
\clearpage
\addcontentsline{toc}{section}{
{\bf The AMADEUS project -- precision studies of the low-energy 
antikaon nucleus/nucleon interaction}\\
J. Zmeskal$^{1}$ for the AMADEUS Collaboration}

%





\titl{The AMADEUS project -- precision studies of the low-energy 
antikaon nucleus/nucleon interaction}

\name{
J. Zmeskal$^{1}$ for the AMADEUS Collaboration
}

\adr{
$^1$ Stefan-Meyer-Institut f\"{u}r subatomare Physik, Vienna, Austria\\
}

The planned series of measurements with AMADEUS (Antikaon Matter At DA$\Phi$NE:
Experiments with Unravelling Spectroscopy) [1] will provide high precision data 
for a better understanding of the strong interaction in the low-energy regime 
due to the study of antikaon nucleus/nucleon dynamics. 
To achieve these goals AMADEUS will make use of the KLOE detector system at LNF, 
which is ideally suited for these measurements due to its large central 
drift chamber (CDC), which provides excellent charge particle identification and 
tracking. In addition, an almost $4 \pi$ calorimeter for the detection of neutral 
particles is surrounding the CDC. R\&D work has already started to construct a 
dedicated target and trigger system for optimizations on the kaon stopping efficiency 
and for further improvements of the signal to background ratio (Fig. 1).
\begin{wrapfigure}{r}[.4\width]{13.5cm}
\fbox{\includegraphics[scale=0.30,clip]{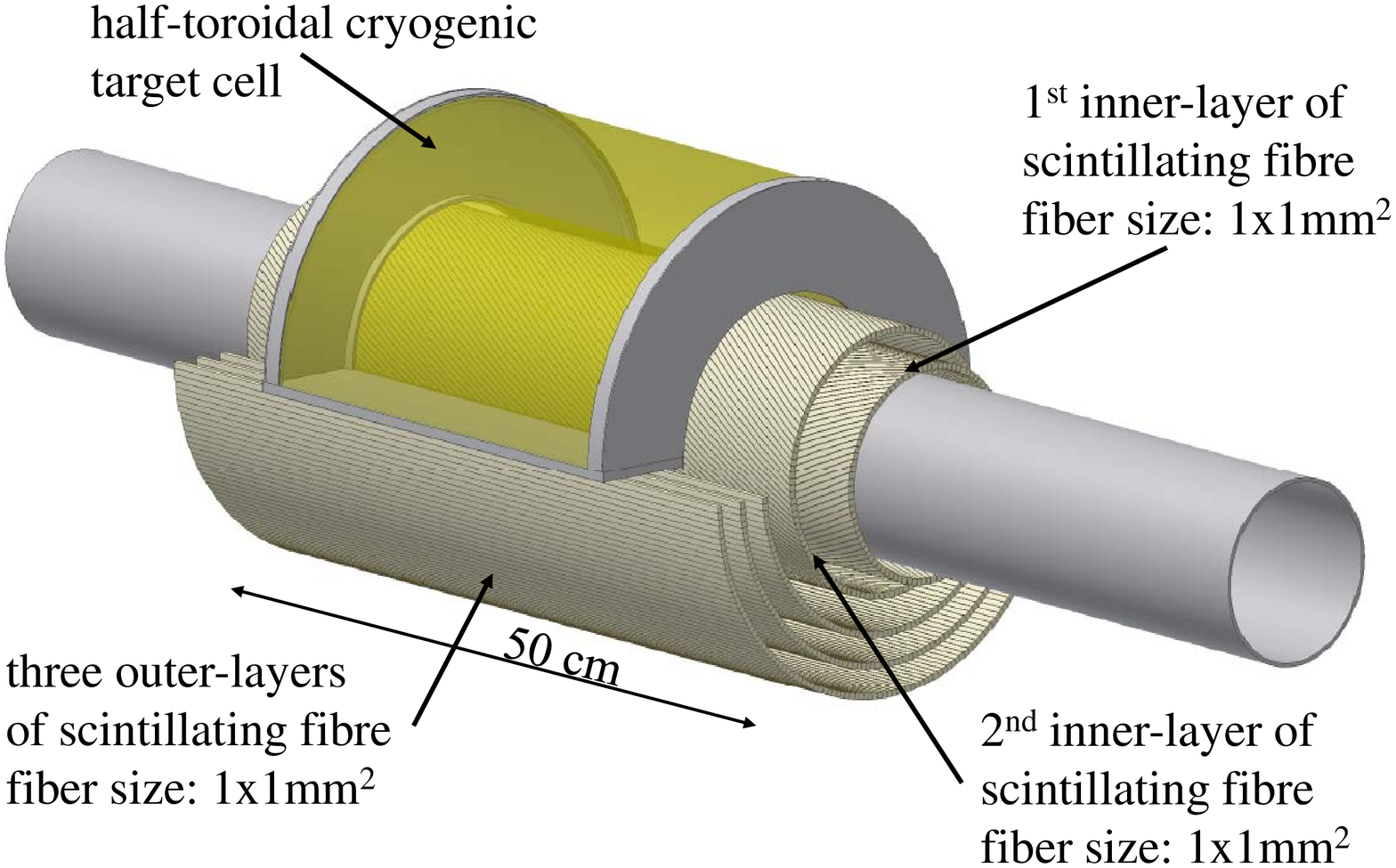}}
\newline
\small Fig. 1: A sketch of the AMADEUS target and 
\newline detector system within KLOE.
\end{wrapfigure}

The scientific case of AMADEUS deals with one of the most important, yet unsolved, 
problems in hadron physics: {\bf how the hadron masses are generated, how the hadron 
interactions change in the nuclear medium and what is the structure of cold dense 
hadronic matter.} With AMADEUS these problems will be attacked by detailed studies of 
the antikaon nucleus/nucleon interaction at low-energy and as well a dedicated search 
is planned to look for an experimental proof of the existence of antikaon-mediated 
deeply bound nuclear systems [2]. If they exist,  antikaon bound nuclear states 
($\bar{K}$-nuclear clusters) will indeed offer ideal conditions for investigating 
the way in which the spontaneous and explicit chiral symmetry breaking pattern of
low-energy QCD changes in the nuclear environment. Moreover, AMADEUS will measure
low-energy charged kaon cross sections on $^{3}$He and $^{4}$He and high statistic
data sets will be available to study resonance states like $\Lambda(1405)$ or 
$\Sigma(1385)$ to understand their structure and also their behaviour in a nuclear medium. 

\vfill  

\noindent{\bf References }
\begin{description}
\setlength\itemsep{-3pt}
\item{[1]} AMADEUS-LOI, http://www.lnf.infn.it  or  http://www.oeaw.ac.at/smi.
\item{[2]} J. Zmeskal, Prog. Part. Nucl. Phys. 61 (2008) 512.
\end{description}


\setcounter{equation}{0} 
\setcounter{figure}{0}
\clearpage

\addcontentsline{toc}{section}{
{\bf List of Participants} }

 
 
 
\titl{List of Participants}
\small
\begin{tabular}{l|l|l}
\hline
Family name & Given name& Institution\\
\hline \hline
Akaishi    & Yoshinori & RIKEN/Nihon University  (Japan) \\
Aslanyan   & Petros    & Joint Institute for Nuclear Research, LHEP (Russia) \\
Berger     & Martin    & TU-M\"{u}nchen (Germany) \\
Bosnar     & Damir     & Univ. Zagreb  (Croatia) \\
Chen       & Jia-Chii  & TU-M\"{u}nchen (Germany) \\
Cieply     & Ales      & NPI, Rez (Czech Republic) \\
Curceanu   & Catalina Oana & LNF, INFN  (Italy) \\
Donoval    & Jan       & NPI Rez  (Czech republic) \\
Epple      & Eliane    & TU-M\"{u}nchen (Germany) \\
Fabbietti  & Laura    &  TU-M\"{u}nchen (Germany) \\
Faber      & Manfried & TU-Wien (Austria) \\
Fayfman    & Mark     & Kurchatov Institute (Russia) \\
Filippi    & Alessandra & INFN Torino (Italy) \\
Friedman   & Eli      & Racah Institute of Physics (Israel) \\
Gal        & Avraham  & Hebrew University (Israel) \\
Gazda      & Daniel  & NPI, Rez  (Czech Republic) \\
Guaraldo   & Carlo   & LNF, INFN (Italy)  \\
Hartmann   & Olaf    & SMI Vienna (Austria) \\
Herrmann   & Norbert & Univ. Heidelberg (Germany) \\
Ishiwatari & Tomoichi & SMI, Vienna (Austria) \\
Ivanov     & Andrei & TU-Wien, (Austria) \\
Iwasaki    & Masahiko & RIKEN (Japan) \\
Kalantari  & Seyed Zafarollah & Isfahan Univ. of Tech. (IUT) (Iran)\\
Kienle     & Paul    & TU-Munich (Germany) \\
Krejcirik  & Vojtech & NPI, ASCR, Rez (Czech Republic) \\
Magas      & Volodymyr & University of Barcelona (Spain) \\
Mares      & Jiri & Nuclear Physics Institute,  Rez (Czech Republic) \\
MARTON     & Johann & SMI Vienna (Austria) \\
Outa       & Haruhiko & RIKEN (Japan)  \\
Pasqua     & Antonio & University of Manchester (United Kingdom) \\
Piano      & Stefano & INFN Trieste (Italy) \\
Pitschmann & Mario & TU Vienna (Austria) \\
Revai      & Janos & Research Institute for Particle and Nuclear Physics (Hungary) \\
Romero Vidal & Antonio & LNF, INFN (Italy) \\
Salvini    & Paola & INFN Pavia (Italy) \\
Sato       & Masaharu & Univ. of Tokyo (Japan) \\
Scordo     & Alessandro & LNF-INFN (Italy) \\
Shevchenko & Nina & NPI Rez (Czech republic) \\
Siebenson  & Johannes & TU-M\"{u}nchen (Germany) \\
Suzuki     & Takatoshi & Uni. of Tokyo (Japan) \\
Vazquez Doce & Oton & LNF, INFN (Italy) \\
Weise      & Wolfram & TU Munich (Germany)\\
Wuenschek  & Barbara & SMI  Vienna (Austria) \\
Wycech     & Slawomir & Soltan Institute for Nuclear Studies (Poland)\\
Yamazaki   & Toshimitsu & Univ. Tokyo (Japan) \\
Zmeskal    & Johann & SMI (Austria) \\
\hline
\end{tabular}

\setcounter{equation}{0} 
\setcounter{figure}{0}
\clearpage

\addcontentsline{toc}{section}{
{\bf Conference Photos} }

 
 
 
\titl{Conference Photos}
\begin{figure}[htbp]
\begin{center}
\includegraphics[width=7cm]{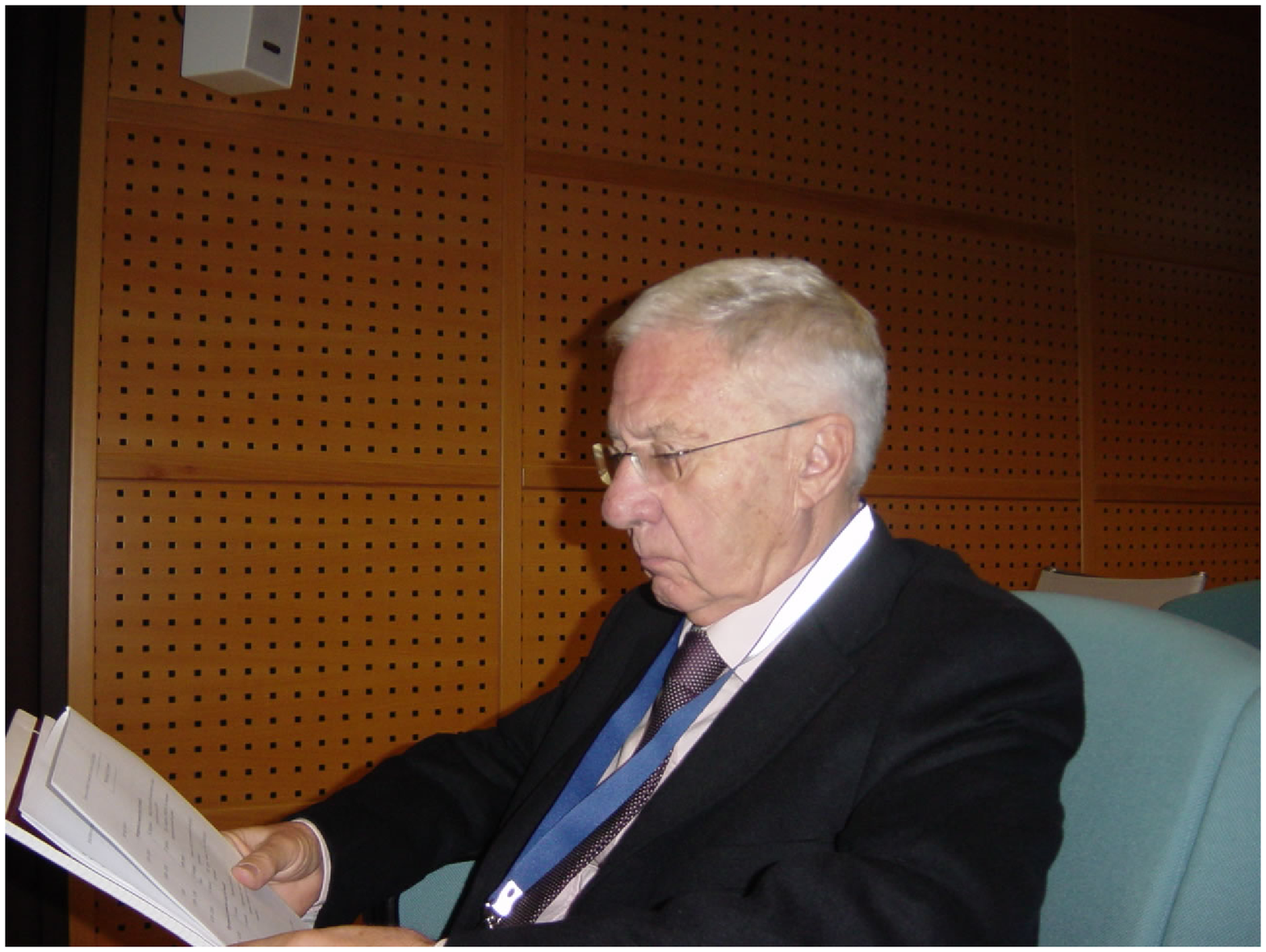}
\includegraphics[width=7cm]{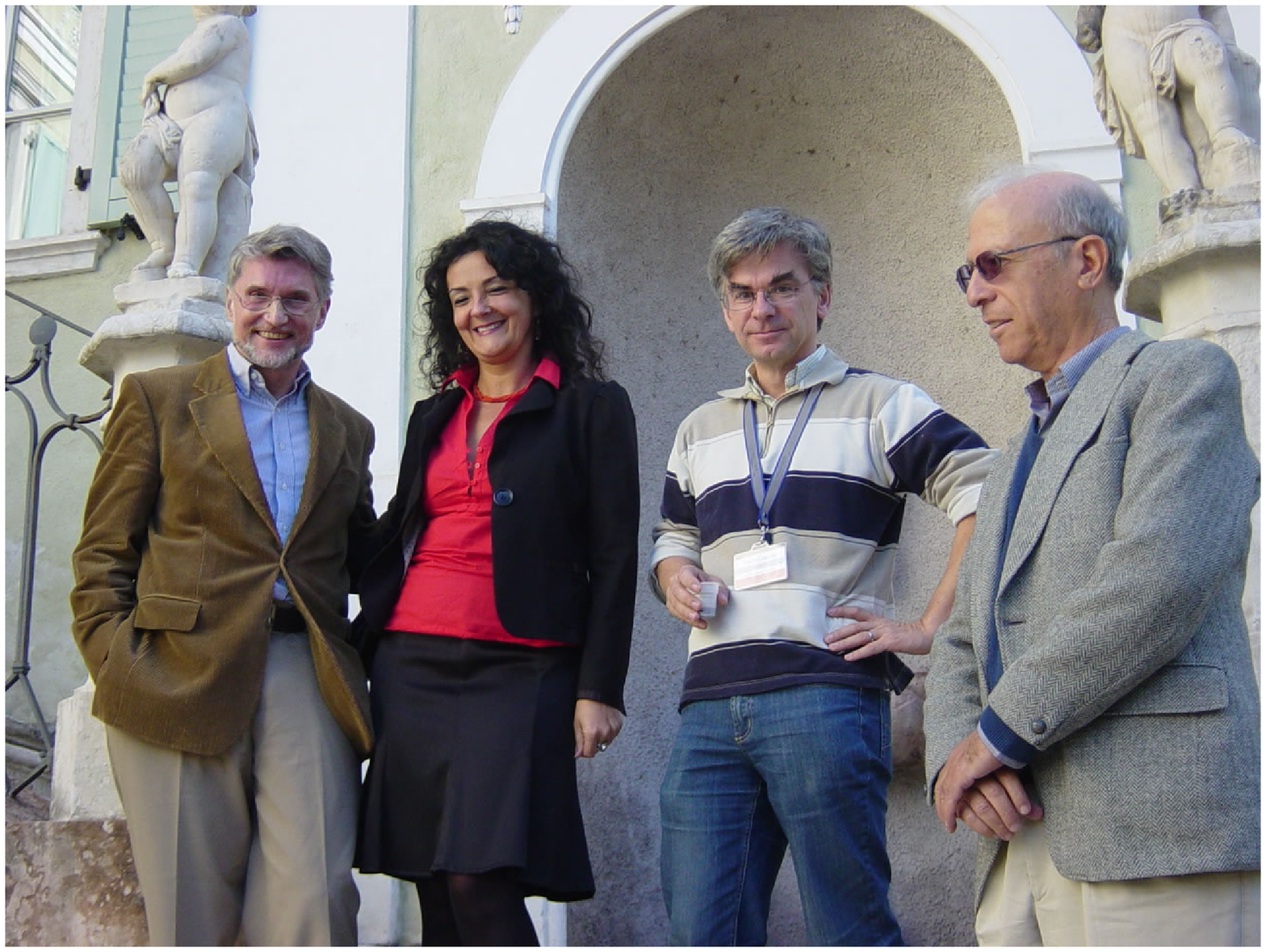}
\includegraphics[width=7cm]{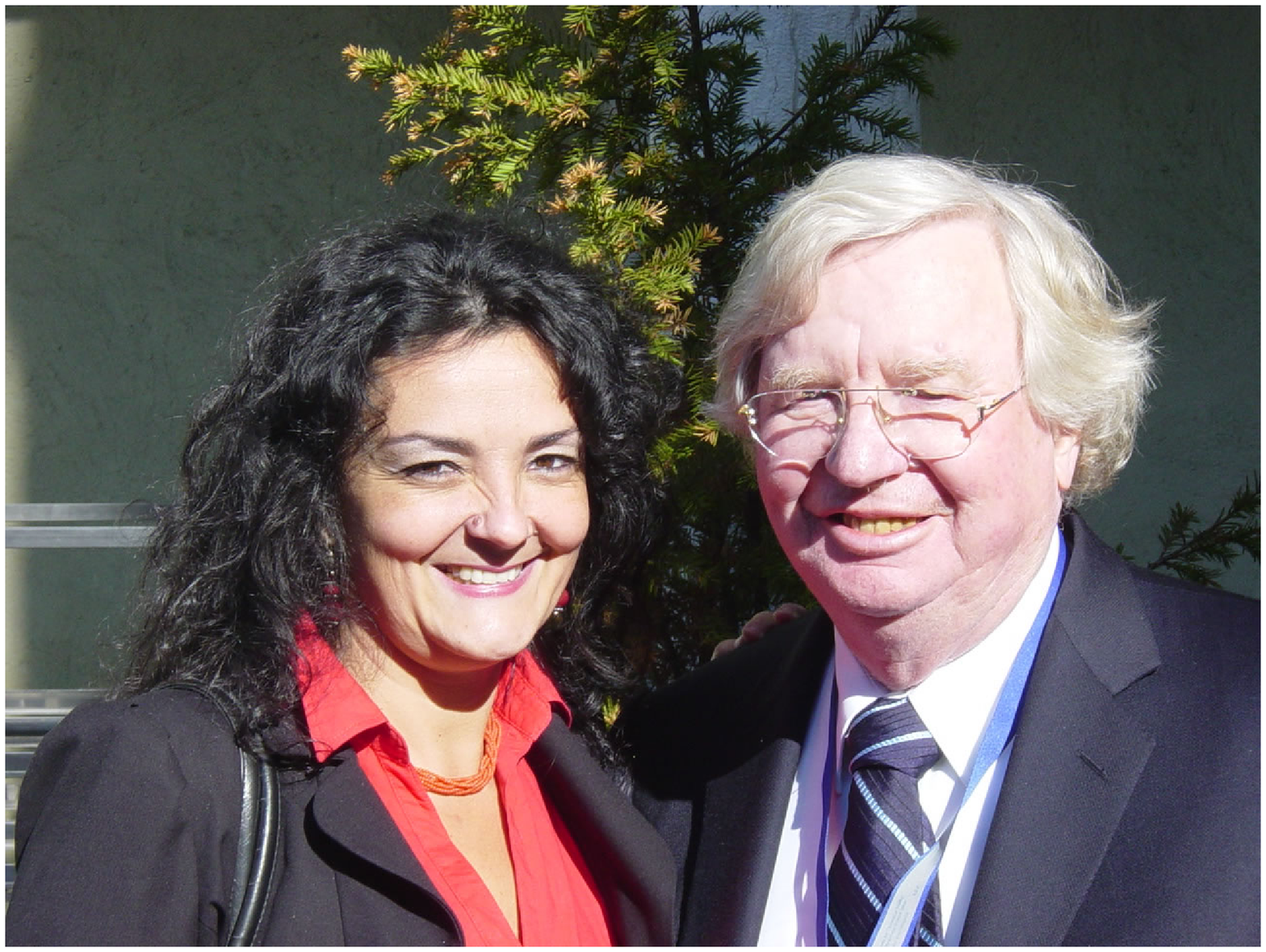}
\includegraphics[width=7cm]{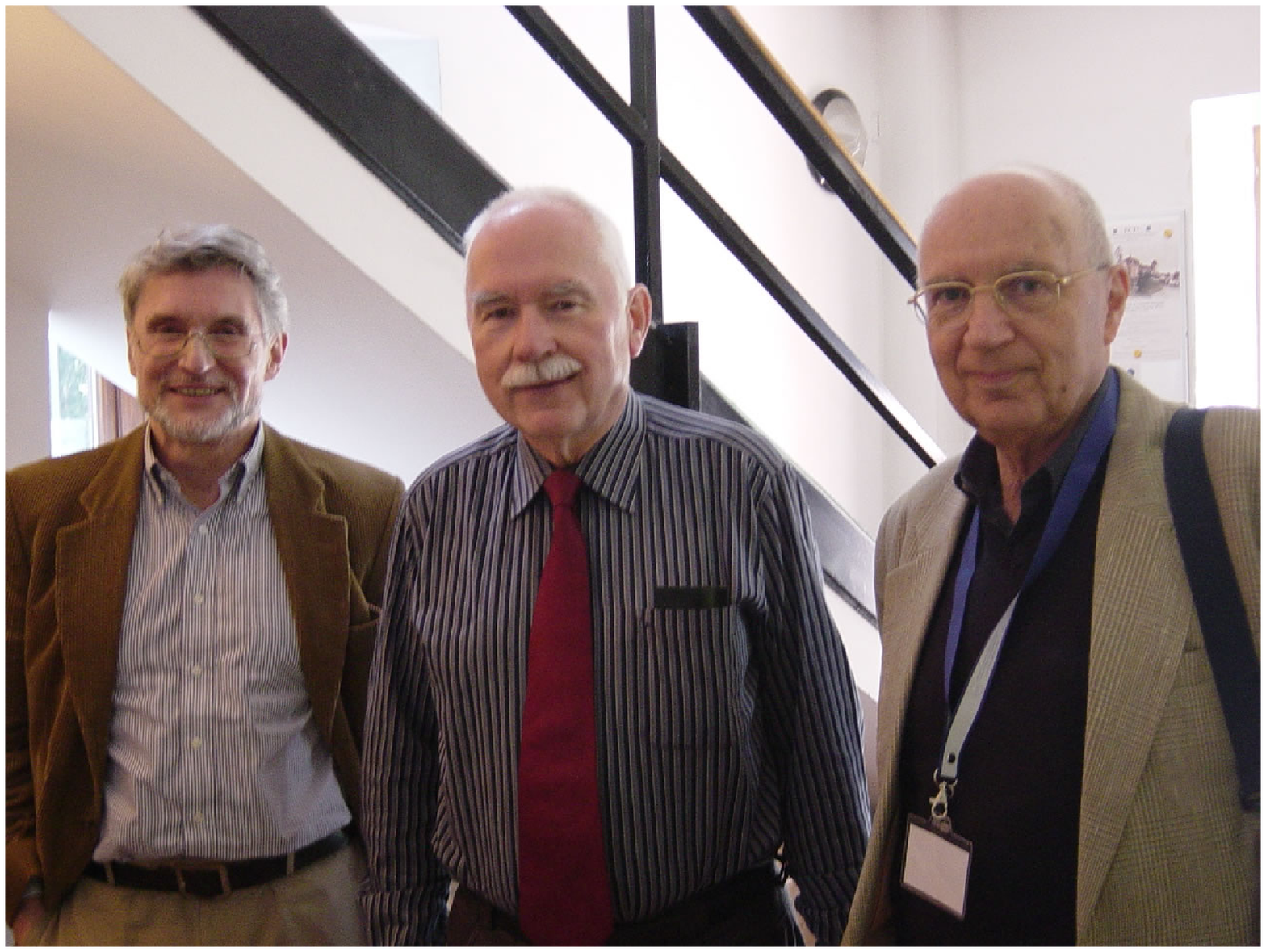}
\includegraphics[width=7cm]{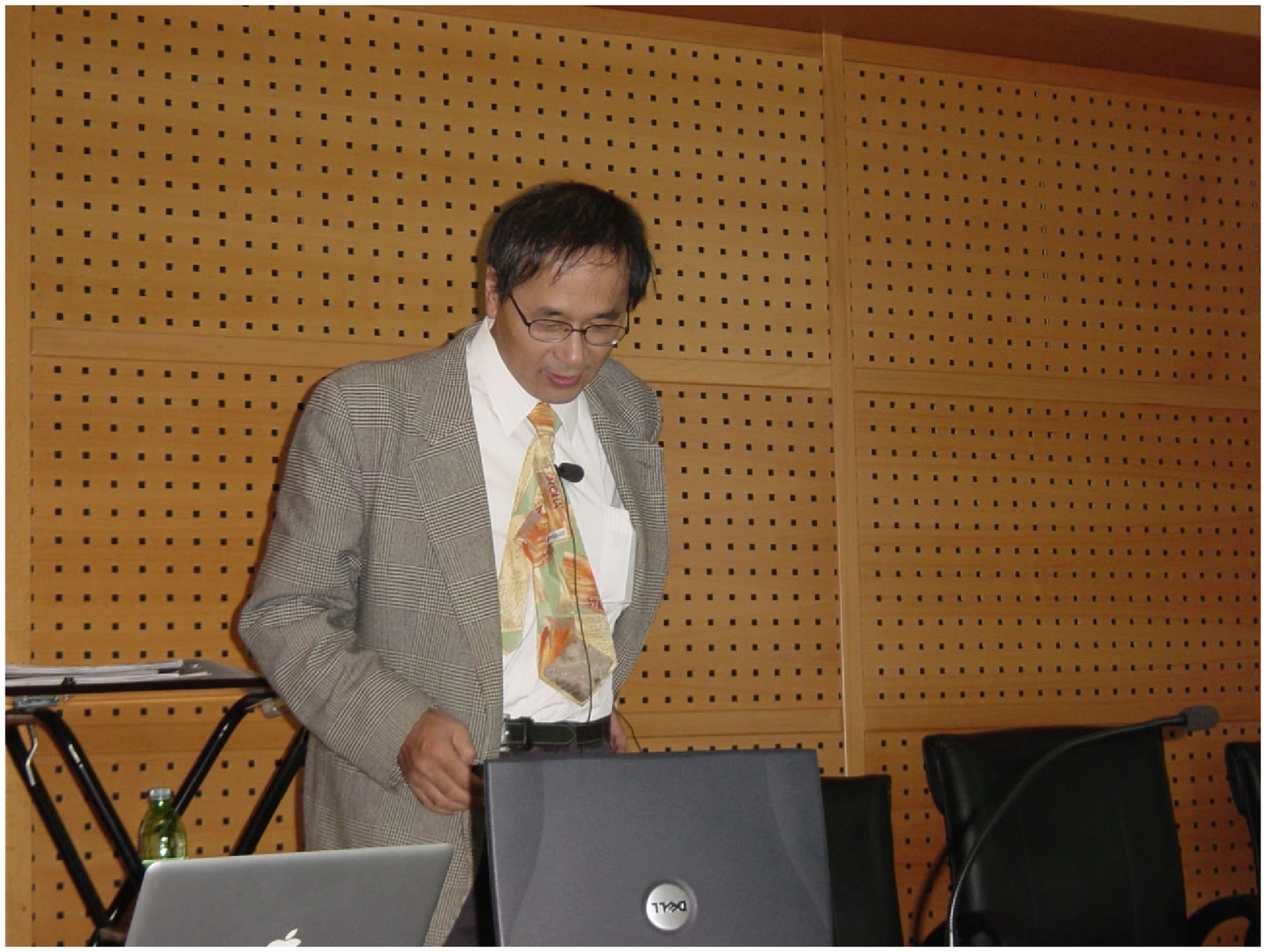}
\end{center}
\end{figure}
\begin{figure}[htbp]
\begin{center}
\includegraphics[width=7cm]{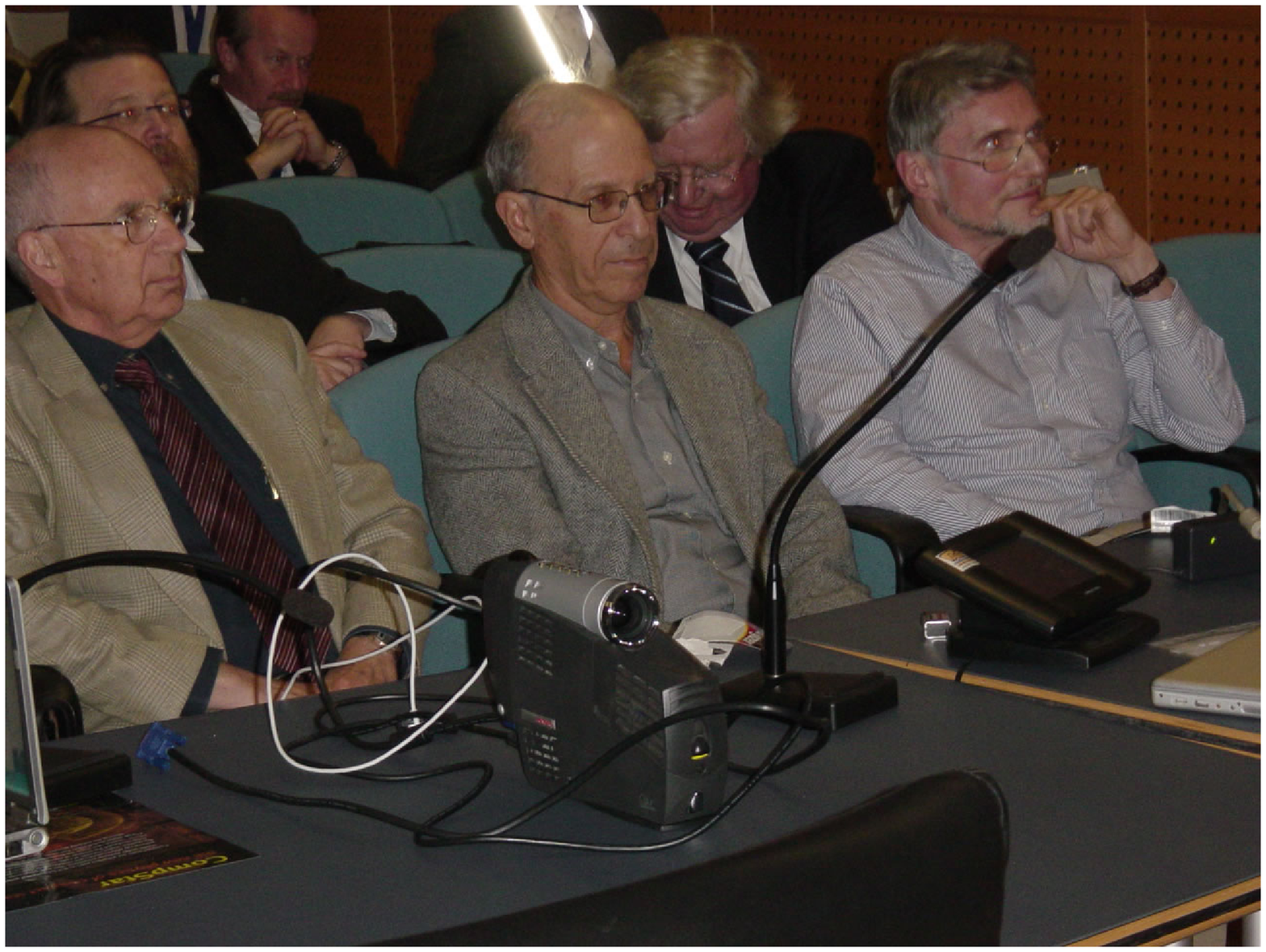}
\includegraphics[width=7cm]{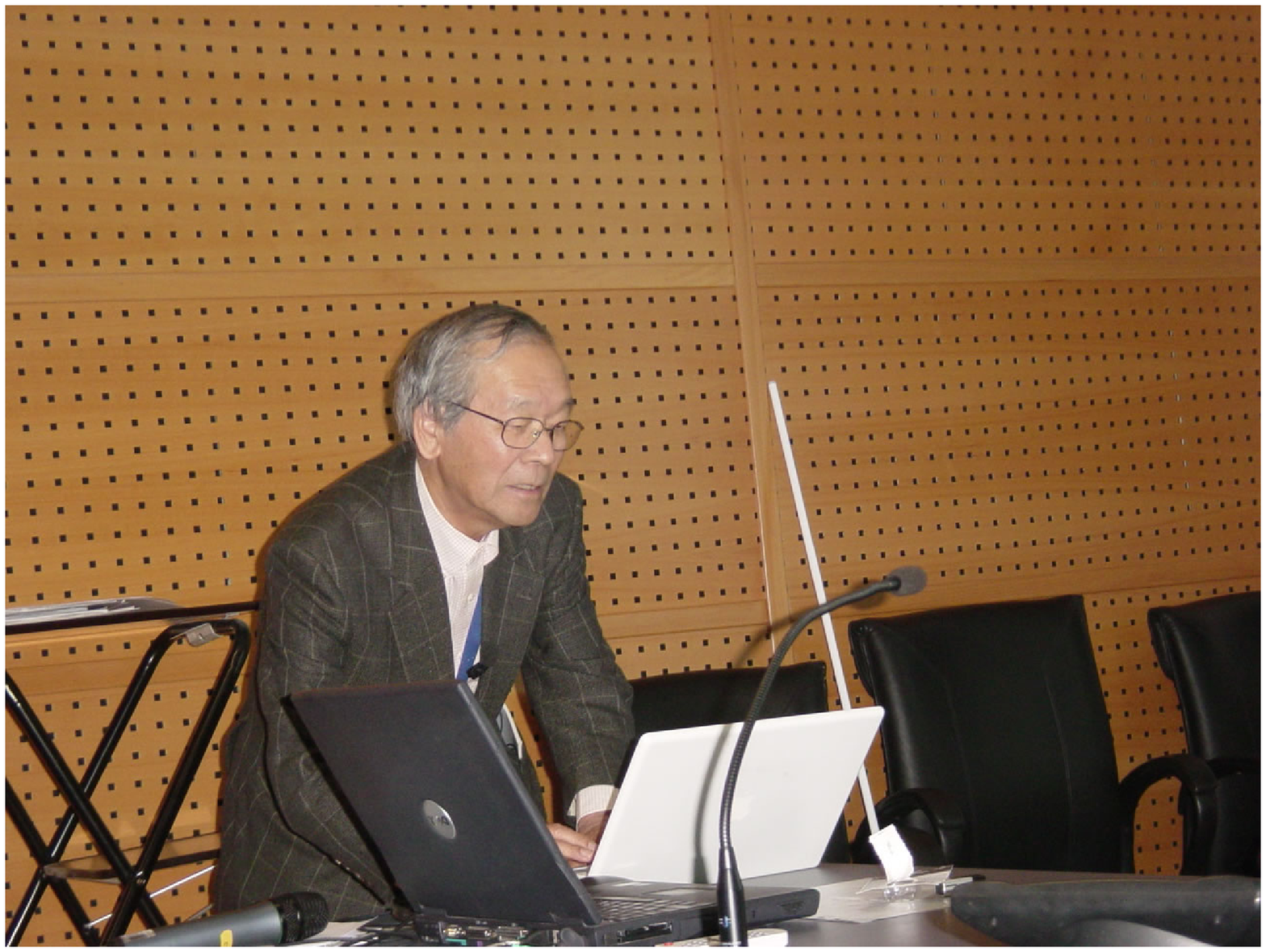}
\includegraphics[width=7cm]{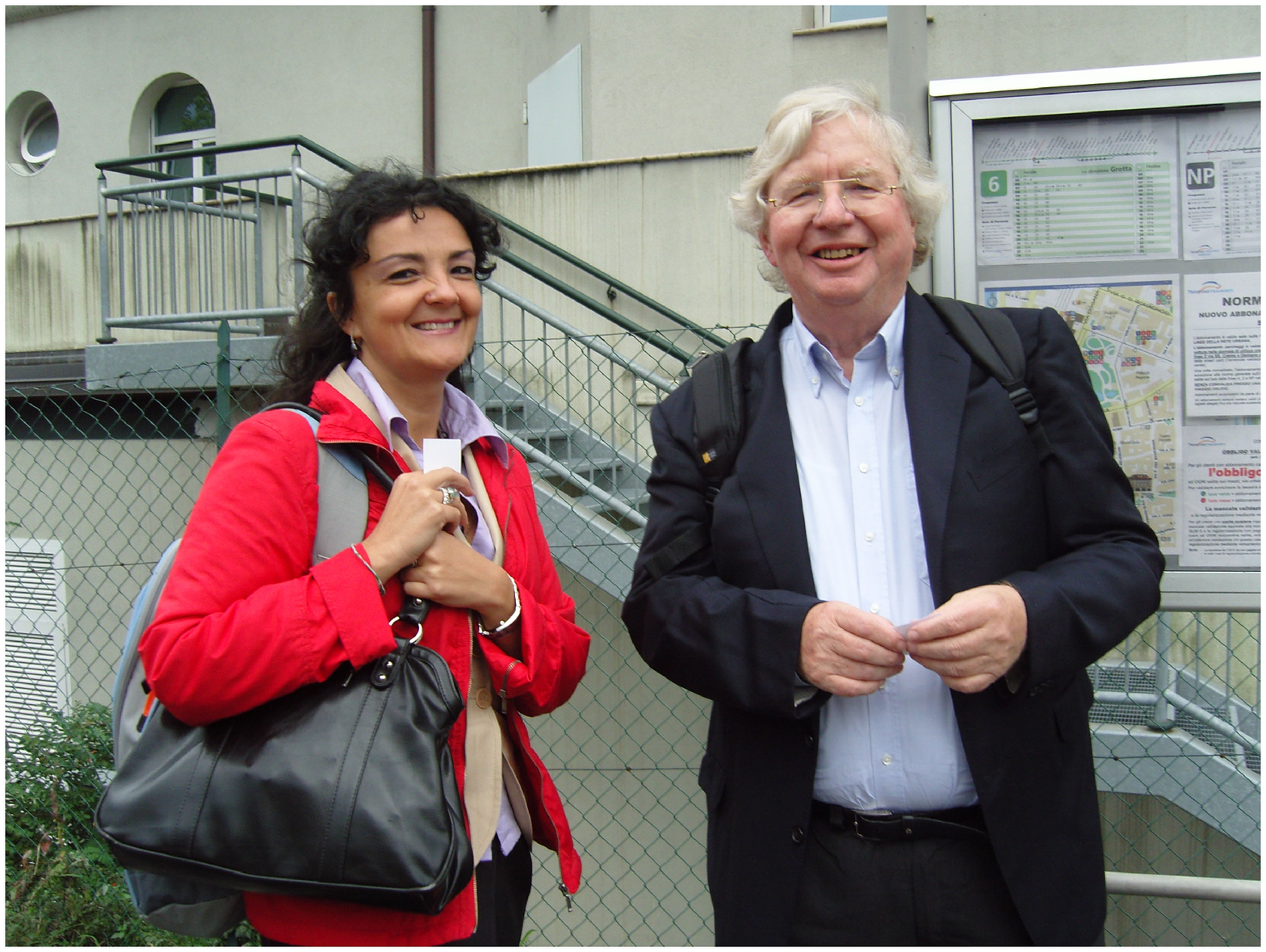}
\includegraphics[width=7cm]{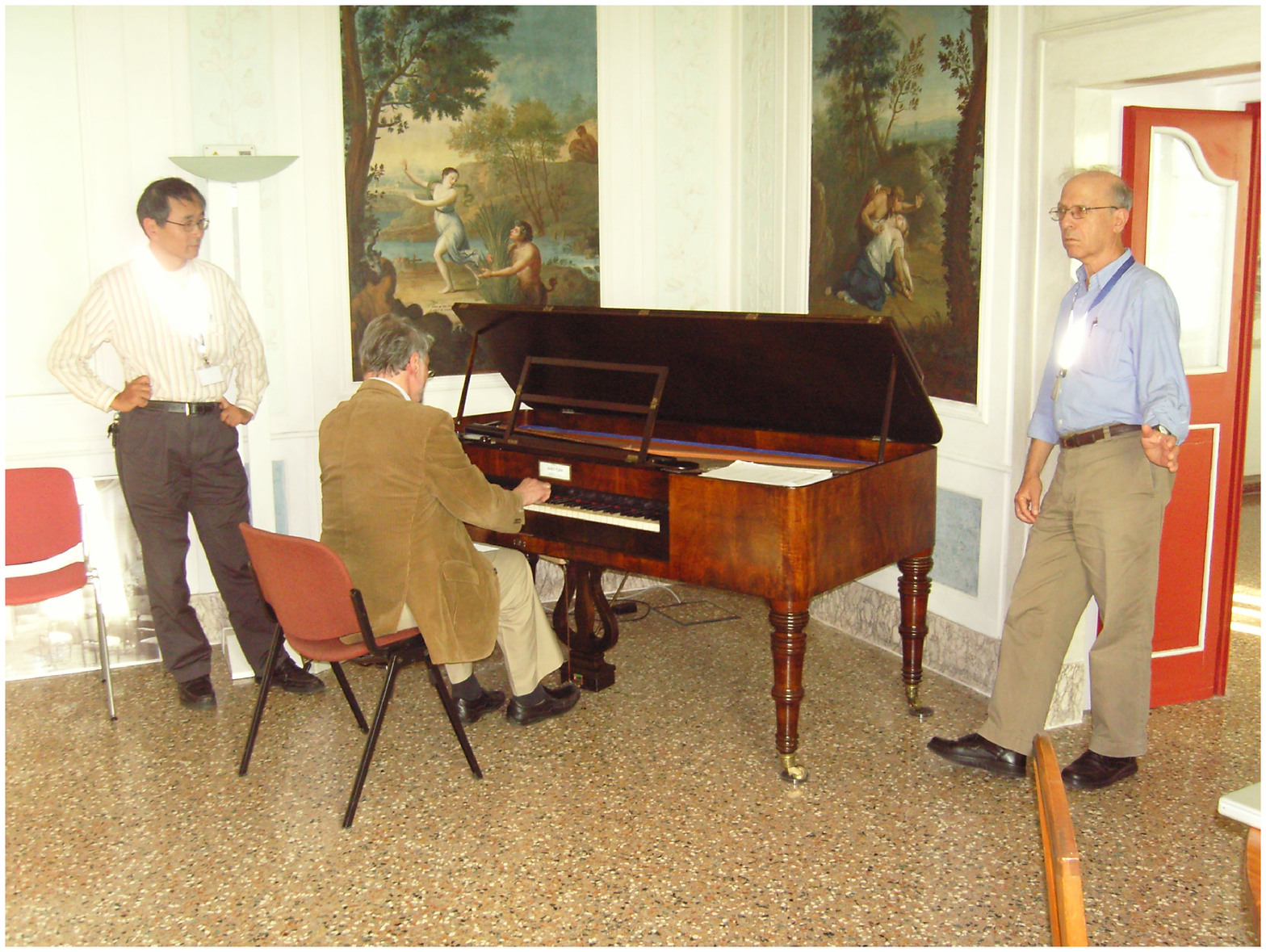}
\includegraphics[width=7cm]{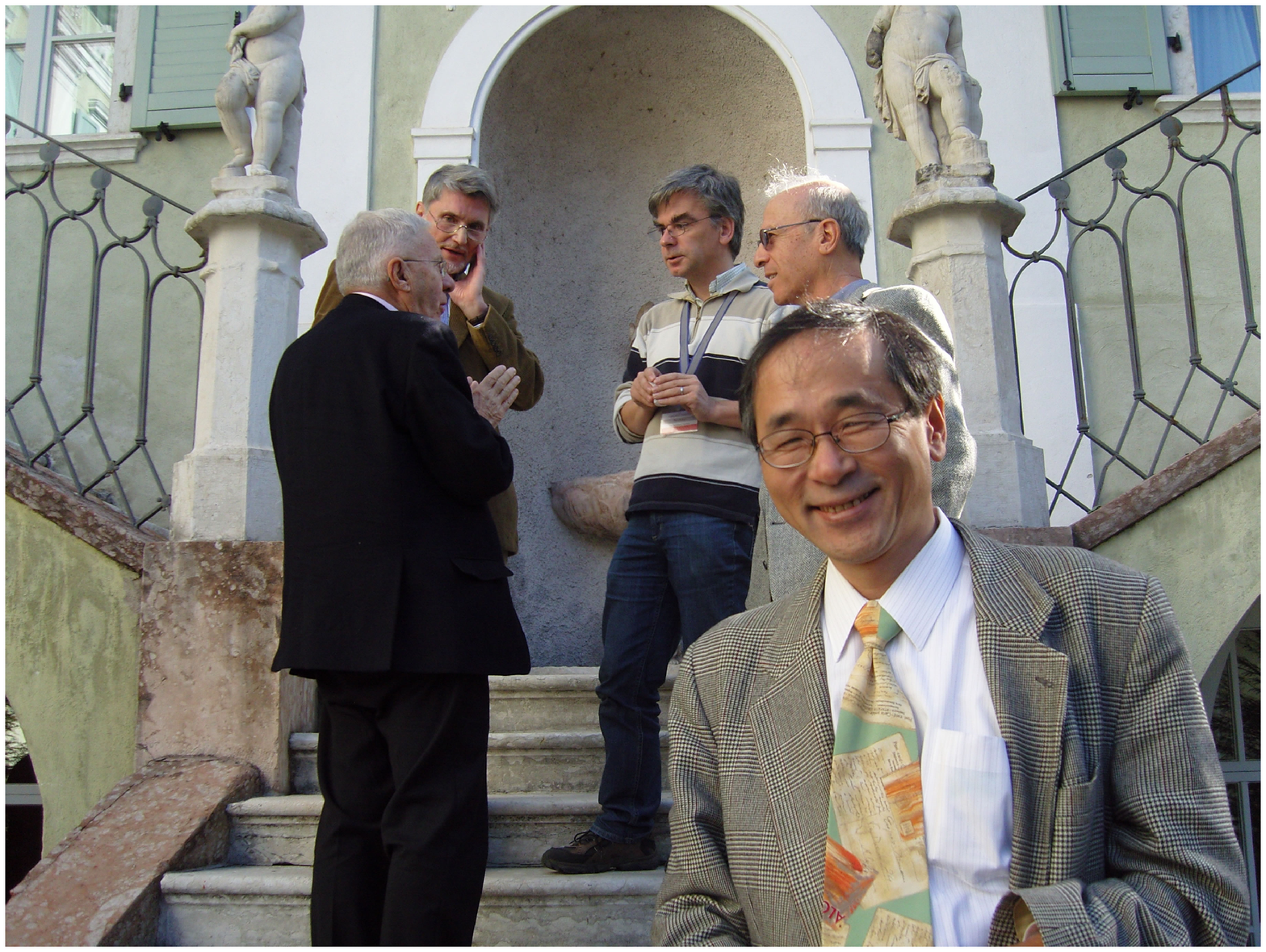}
\end{center}
\end{figure}


\setcounter{equation}{0} 
\setcounter{figure}{0}
\clearpage

\end{document}